\documentclass[letterpaper,11pt,fleqn]{article}
\pdfoutput=1
\usepackage{jheppub}

\usepackage{graphicx}
\usepackage[figuresright]{rotating}
\usepackage{bm,amsmath,amssymb}
\usepackage[mathscr]{eucal}
\usepackage{color}
\usepackage{multirow}

\newcommand{\abar}{\bar{\alpha}_s}

\newcommand{\eqn}[1]{Eq.~\eqref{#1}}
\newcommand{\beq}{\begin{equation}}
\newcommand{\eeq}{\end{equation}}
\newcommand{\nn}{\nonumber\\}

\newcommand{\rmd}{{\rm d}}
\newcommand{\rme}{{\rm e}}

\newcommand{\order}[1]{\mathcal{O}{(#1)}}
\newcommand{\rmI}{{\rm I}}

\newcommand{\mcal}{\mathcal}
\newcommand{\tf}{t_{\rm f}}

\newcommand{\kt}{k_\perp}
\newcommand{\bk}{\bm{k}}

\newcommand{\tbr}{t_{\rm br}}
\newcommand{\obr}{\omega_{\rm br}}

\newcommand{\amed}{{\alpha}_{s,\text{med}}}

\newcommand{\oc}{\omega_{\rm c}}
\newcommand{\zc}{z_{\text{cut}}}
\newcommand{\minus}{\!-\!}

\newcommand{\njets}{$N_\text{jets}$}

\def\thetacut{\theta_\text{cut}}

\makeatletter
\g@addto@macro\bfseries{\boldmath}
\makeatother

\title{Deciphering  the $z_g$ distribution in ultrarelativistic heavy ion collisions}
\author{P.~Caucal,}
\author{E.~Iancu,}
\author{and G.~Soyez}
\affiliation{Institut de Physique Th\'{e}orique, Universit\'{e} Paris-Saclay, CNRS, CEA, F-91191, Gif-sur-Yvette, France}

\emailAdd{paul.caucal@ipht.fr}
\emailAdd{edmond.iancu@ipht.fr}
\emailAdd{gregory.soyez@ipht.fr}

\abstract{
  Within perturbative QCD, we develop a new picture for the parton
  shower generated by a jet propagating through a dense quark-gluon
  plasma.
  This picture combines in a simple, factorised, way multiple medium-induced
  parton branchings and standard vacuum-like emissions, with the
  phase-space for the latter constrained
  by the presence of the medium.
  We implement this picture as a Monte Carlo generator that we use
  to study two phenomenologically important observables: the jet
  nuclear modification factor $R_{AA}$ and the $z_g$ distribution
  reflecting the jet substructure.
  In both cases, the outcome of our Monte Carlo simulations is in good
  agreement with the LHC measurements.
  We provide basic analytic
  calculations that help explaining the main features observed in the data.
  We find that the energy loss by the jet is increasing with the jet
  transverse momentum, due to a rise in the number of partonic sources
  via vacuum-like emissions.
  This is a key element in our description of both $R_{AA}$ and the 
  $z_g$ distribution.
  For the latter, we identify two main nuclear effects: incoherent
  jet energy loss and hard medium-induced emissions.
  As the jet transverse momentum increases,  we predict a qualitative change
   in the ratio between the $z_g$ distributions in PbPb and pp collisions: 
   from increasing at small $z_g$, this ratio becomes essentially flat, or
   even slightly decreasing.
   
  \vspace*{2.0cm}
  {\it We want to dedicate this paper to the 80$^\text{th}$ birthday of Al Mueller, who pioneered
  the studies of jet quenching in perturbative QCD and was also our collaborator on a previous
  paper which introduced the general physical picture that we further develop in this work.
  }
}

\begin{document}
\maketitle

\section{Introduction}\label{sec:intro}

``Jet quenching'' is a rather generic denomination for the ensemble of
nuclear modifications affecting a highly-energetic jet or hadron
propagating through the dense partonic medium created in
ultrarelativistic heavy ion collisions.
Its study at RHIC and at the LHC is one of the main sources of
information about the deconfined, quark-gluon plasma, phase of QCD.
It is associated with a rich and complex phenomenology which includes
a broad range of observables, from very inclusive one like the
``nuclear modification factor'' $R_{AA}$ (for hadrons or jets), to
more detailed ones like the jet shape, the jet fragmentation function
and jet substructure.
These observables are both delicate to measure (e.g.\ due to the
complexity of the background) and delicate to interpret theoretically
(due to the high density of partons and their interplay with various
collective, non-perturbative, phenomena).
In this context, it is difficult to build a full theoretical picture
of jet quenching from first principles, although a few specific
phenomena have been described using (or at least taking inspiration
from) perturbative QCD (pQCD). We refer for example to the recent reviews
\cite{Mehtar-Tani:2013pia,Blaizot:2015lma,Qin:2015srf}.

Jet substructure observables have attracted a lot of attention
recently, due to their potential to capture detailed aspects of the
dynamics of jet quenching (see e.g.~\cite{Andrews:2018jcm} for a
recent review and more references).
This paper focuses on the ``$z_g$
distribution''~\cite{Larkoski:2015lea}, where $z_g$ is the splitting
fraction of the first hard splitting selected by the Soft Drop
(SD) procedure \cite{Larkoski:2014wba}.
This observable may reveal the nuclear modification effects on the parton
splitting functions.
In nucleus-nucleus ($AA$) collisions experimental studies at the
LHC, by CF'S \cite{Sirunyan:2017bsd} and ALICE \cite{Acharya:2019djg},
reported significant nuclear effects in the $z_g$ distribution.
Expectations on the theoretical side are less obvious and the main
goal of this paper is to highlight the jet quenching mechanisms
controlling the $z_g$ distribution (see also
Refs.~\cite{Chien:2016led,Mehtar-Tani:2016aco,Chang:2017gkt,Milhano:2017nzm}).

Since the SD procedure clusters the constituents of the jet with the
Cambridge/Aachen algorithm~\cite{Dokshitzer:1997in,Wobisch:1998wt}
(see Sect.~\ref{zgdef} below for details), it is intrinsically built
with the expectation of {\it angular ordering} (AO) between successive
parton branchings.
This property follows from the colour coherence of parton
splittings~\cite{Dokshitzer:1991wu} and is at the heart of partonic
cascades in $pp$ collisions.
One reason why the $z_g$ distribution is difficult to describe
theoretically in $AA$ collisions is that angular ordering is expected
to be violated. This is certainly the case for the {\it medium-induced
  emissions} (MIEs), the parton branchings triggered by collisions
with the medium constituents
\cite{Baier:1996kr,Baier:1996sk,Zakharov:1996fv,Zakharov:1997uu,Baier:1998kq}. As
shown in
Refs.~\cite{MehtarTani:2010ma,MehtarTani:2011tz,CasalderreySolana:2011rz,Blaizot:2012fh,Blaizot:2013vha,Apolinario:2014csa},
these collisions wash out the quantum coherence between the daughter
partons produced by a MIE, thus suppressing the interference effects
responsible for angular ordering. {\it A priori}, the collisions can
also affect the partons produced via {\it vacuum-like emissions}
(VLEs) --- the standard bremsstrahlung triggered by the parton
virtualities --- occurring inside the medium.

The situation of VLEs has been clarified recently in
Ref.~\cite{Caucal:2018dla}, which demonstrated that the VLEs occurring
inside the medium  do still obey AO, essentially because they occur
too fast to be influenced by the collisions. The only effect of the
medium is to restrict the phase-space available for VLEs.

The same argument implies that the vacuum-like emissions can be
factorised in time from the medium-induced radiations, at least within
the limits of a leading double-logarithmic approximation (DLA) in
which the successive branchings are strongly ordered in both energies
and angles.
This picture, derived from first principles, allows for a Monte Carlo
implementation of the parton showers produced in $AA$ collisions.
In this picture, VLEs (satisfying angular ordering) and MIEs (for
which quantum coherence can be neglected) are factorised and are both
described by a Markovian process.

The phenomenological discussions in Ref.~\cite{Caucal:2018dla} only
included VLEs at DLA.
To have a chance to be realistic, several extensions are necessary.
First, it must include medium-induced radiation and transverse
momentum broadening. These higher twist effects carry particles
to large angles and are responsible for the energy loss by the jet.
Second, it must include the complete (leading-order) DGLAP splitting
functions for the VLEs, to have a more realistic description for the
energy flow in the cascade and to ensure energy conservation. This
also means that one must go beyond the DLA, by giving up the strong
ordering in energies at the emission vertices.

We will show in this paper that the factorised picture still holds in a single logarithmic 
approximation in which successive VLEs are strongly ordered in angles.
We will then provide a corresponding Monte Carlo implementation.
In this implementation, the medium-induced radiation will be treated
in the spirit of the effective theory developed in
Refs.~\cite{Blaizot:2012fh,Blaizot:2013hx,Blaizot:2013vha,Fister:2014zxa},
i.e.\ as a sequence of independent emissions occurring at a rate given
by the BDMPS-Z spectrum
\cite{Baier:1996kr,Baier:1996sk,Zakharov:1996fv,Zakharov:1997uu,Baier:1998kq}. This
is the right approximation for the relatively soft MIEs that we are
primarily interested in this work.

In this first study, our description of the medium will be
oversimplified: we assume a fixed ``brick'' of size $L$, the distance
travelled by the jet inside the medium, and characterised by a uniform
value for the jet quenching parameter $\hat q$, the rate for
transverse momentum broadening via elastic collisions. This
description can certainly be improved in the future, e.g. by including
the longitudinal expansion of the medium as a time-dependence in
$\hat q$.
Given these simplifying assumptions, we concentrate on observables
which, in our opinion, are mainly controlled by the parton showers and
which are mostly sensitive to global properties of the medium, like
the typical energy loss by a jet (which scales like $\hat q L^2$) or
the typical transverse momentum acquired via elastic collisions (the
``saturation momentum'' $Q_s^2\equiv \hat q L$).
At least in our theoretical approach, this is the case for observables
like the jet nuclear modification factor $R_{AA}$ and the
$z_g$ distribution. This is further supported by the good agreement
that we shall find between our results and the corresponding LHC data.

As a baseline for discussion, we will study the jet nuclear
modification factor $R_{AA}$.
Comparing our predictions to the ATLAS
measurements~\cite{Aaboud:2018twu} will allow us to calibrate our
medium parameters $\hat q$ and $L$ (and a coupling
$\alpha_{s,\text{med}}$).
We will find that the dependence of $R_{AA}$ on the jet transverse
momentum $p_T$
is controlled by the evolution of the parton multiplicity via VLEs
inside the medium. Each of these partons then acts as a source for
medium-induced radiation, enhancing the jet energy loss as $p_T$
increases.
We shall notably check that $R_{AA}$ is primarily controlled by the
energy scale $\omega_\text{br}\equiv \alpha_{s,\text{med}}^2\hat q L^2/2$
(the characteristic scale for multiple medium-induced branchings).

The $z_g$-distribution is an observable particularly suited for our
purposes. On one hand, it is associated with relatively hard
branchings, for which perturbative QCD is expected to be
applicable. On the other hand, it is sensitive to the dynamics of the
MIEs, that are probed both {\em directly} (especially at relatively
small values of $z_g$, where the SD procedure can select a MIE) and
{\em indirectly} (via the energy loss of the subjets produced by a
hard vacuum-like branching).

Our purpose is to provide a transparent physical interpretation and a
qualitative description of the relevant LHC
data~\cite{Sirunyan:2017bsd,Acharya:2019djg}. To that aim, we also construct
piecewise analytic approximations, whose results are eventually compared to our
numerical simulations.
In this process, two natural kinematic regimes will emerge, ``low
energy'' and ``high energy'', with the transition between them
occurring around $p_T=\hat q L^2/(2\zc )$. 
Here, $\zc$ is the lower limit on $z_g$ which is used by the SD algorithm, 
that we shall chose as $\zc=0.1$ in our explicit calculations,  
in compliance with the experimental
measurements in \cite{Sirunyan:2017bsd,Acharya:2019djg}. 
For a ``high energy jet'', the SD procedure can only select a
vacuum-like splitting, so the only nuclear effect is the (incoherent)
energy loss of the two subjets created by this splitting.
For ``low energy jets'',  both VLEs and MIEs can be captured by SD and
the contribution from the MIEs leads a significant rise in
the $z_g$ distribution at small $z_g$.

From this perspective, the current measurements of the $z_g$-distribution at
RHIC~\cite{Kauder:2017cvz} and the
LHC~\cite{Sirunyan:2017bsd,Acharya:2019djg} belong to the low energy
regime. Our predictions reproduce the trends seen in the LHC data, both
 qualitatively and semi-quantitatively (see the discussion in Sect.~\ref{sec:data},
 notably Figures~\ref{Fig:zgvar1} to \ref{Fig:nsd}).
We argue that the onset of the transition between the low- and
high-energy regimes is already visible in the current LHC data (in the
highest energy bin, with 300~GeV$\ < p_{T} < 500$~GeV) and that the
change in behaviour should become even more visible when further increasing $p_T$

Given the importance of the jet energy loss for both the
$z_g$-distribution and the nuclear modification factor $R_{AA}$, we
propose a new measurement to study the correlation between these two
observables. The idea is to measure the jet $R_{AA}$ as a function of
the jet $p_T$ in bins of $z_g$ or, even better, in bins of $\theta_g$
(the angular separation between the two subjets identified by SD).
We find (see Sect.~\ref{sec:beyond-zg} and, in particular
Fig.~\ref{Fig:zgRAA}), a larger suppression of $R_{AA}$, meaning a
larger energy loss, for 2-prong jets which have passed the SD criteria
than for the 1-prong jets which did not.

This paper is structured as follows: Sect.~\ref{sec:phys} 
describes the general physical picture and the underlying
approximations. We start with a brief summary of the argument for the
factorisation of the in-medium parton shower as originally formulated
in Ref.~\cite{Caucal:2018dla} and then explain the extension of this
argument beyond DLA.
Sect.~\ref{sec:MC} presents the Monte Carlo implementation of this
factorised picture.
We begin the discussion of our new results in Sect.~\ref{sec:eloss},
where we study the jet energy loss and present our predictions for the
jet $R_{AA}$.
The next two sections present an extensive study of the nuclear
effects on the $z_g$-distribution.
In Sect.~\ref{sec:zgmed}, we consider ``monochromatic'' jets generated
by a leading parton (gluon or quark) with a fixed energy $p_T$. To
uncover the physics underlying the $z_g$ distribution, we construct
analytic calculations that we compare to our Monte Carlo results.
In Sect.~\ref{sec:data}, we move to our phenomenological
predictions using a full matrix element for the production of the
leading partons. We compare our numerical results with the experimental
analyses of the LHC data~\cite{Sirunyan:2017bsd,Acharya:2019djg},
whenever applicable.
Finally, Sect.~\ref{sec:conc} presents our conclusions together with
open problems and perspectives.

\section{Parton shower in the medium: physical picture}
\label{sec:phys}

In this section, we describe our factorised pQCD picture for the
parton shower generated by an energetic parton propagating through a
homogeneous dense QCD medium of size $L$. We discuss the validity of
this picture beyond the double-logarithmic approximation originally
used in Ref.~\cite{Caucal:2018dla}.

\subsection{Basic considerations}

We aim at describing jets created in ultrarelativistic heavy ion
collisions and which propagate at nearly central rapidities (the most
interesting situation for the physics of jet quenching). For such
jets, one can identify the energy and the transverse momentum
(w.r.t. the collision axis), so in what follows we use these
notations interchangeably. In particular, the energy of the leading
parton (quark or gluon) initiating the jet will be interchangeably denoted by $E$ or
$p_{T0}$.

The leading parton is created with a time-like virtuality $Q^2\ll E^2$
via a hard $2\to 2$ partonic process. In the ``vacuum'' (i.e.\ in a
proton-proton collision), such a parton typically decays after a
time of the order of the  {\it formation time} $\tf\equiv 1/\Delta E$.
Using $\Delta E=\sqrt{E^2+Q^2}-E\simeq Q^2/2E$, we have
\beq\label{tvac}
 t_{\rm f}\,\simeq\,\frac{2E}{Q^2}\,\simeq\,\frac{2}{z(1-z)E\theta^2}\,\simeq\,\frac{2z(1-z)E}{\kt^2}\,,
\eeq
where $z$ and $\theta$ (assuming $\theta\ll 1$) are the
energy fraction and opening angle of the partonic decay and
$k_\perp\simeq z(1-z)\theta E$ is the (relative) transverse momentum
of any of the two daughter partons
w.r.t. the direction of the leading parton.\footnote{For more clarity, we use the subscript $T$ for momentum components transverse to the {\em collision axis} and the subscript $\perp$ for the components transverse to the {\it jet axis}, here identified with the original direction of the leading parton.}

The differential probability for vacuum-like branching is then given by the bremsstrahlung law,
\beq\label{brem}
\rmd^2\mcal{P}_{\text{vac}}\,=\,\frac{\alpha_s(\kt^2)}{2\pi}\,\frac{\rmd^2\bk_\perp}{\bk_\perp^2}\,P_{a\to bc}(z)\,\rmd z
\,=\,\frac{\alpha_s(\kt^2)}{2\pi}\,\frac{\rmd\theta^2}{\theta^2}\,P_{a\to bc}(z)\,\rmd z\,,
\eeq
where $P_{a\to bc}$ is the Altarelli-Parisi splitting function for the
branching of a parton of type $a$ into two partons of type $b$ and $c$
with energy fractions $z$ and respectively $1-z$. The second equality
above holds after averaging over the azimuthal angle $\phi$ (the
orientation of the 2-dimensional vector $\bk_\perp$).
For physics discussions, it is often helpful to consider the limit
$z\ll 1$ where the emitted gluon is soft. In this case,
$P(z)\simeq 2C_R/z$ --- with $C_R=C_F=(N_c^2-1)/(2N_c)$ for quarks and
$C_R=C_A=N_c$ for gluons --- and we can write
$\tf\simeq 2/(\omega\theta^2)$ and $\kt\simeq \omega\theta$, with
$\omega\equiv zE\ll E$.

In the presence of a medium, additional effects have to be taken into
account as high-energy partons traversing the medium suffer elastic
collisions and thus receive transverse kicks.
This has three main consequences: (i) they affect the available
phase-space for vacuum-like emissions, (ii) they trigger additional,
{\it medium--induced} emissions, and (iii) they yield a {\it
  broadening} of the transverse momentum of high-energy partons.
These three effects are discussed separately in the following subsections.

In what follows, we assume that the medium is sufficiently dense
to be weakly coupled so that the successive collisions are quasi-independent
from each other.
In the multiple soft scattering approximation and after travelling
through the medium along a time/distance $\Delta t$, the random kicks
yield a Gaussian broadening of the transverse momentum distribution with a width
$\langle \Delta \kt^2\rangle =\hat q \Delta t$. The jet quenching
parameter $\hat q$ in this relation is the average transverse momentum
squared transferred from the medium to a parton per unit
time.\footnote{Strictly speaking, this quantity is (logarithmically)
  sensitive to the ``hardness'' of the scattering, i.e.\ to the total
  transferred momentum $Q_s^2(L)=\hat q L$ (see
  e.g.~\cite{Liou:2013qya,Blaizot:2014bha,Iancu:2014kga}). This
  sensitivity is ignored in what follows.} This quantity is
proportional to the Casimir $C_R$ for the colour representation of the
parton and in what follows we shall keep the simple notation $\hat q$
for the case where the parton is a gluon. The corresponding quantity
for a quark reads $\hat q_F=(C_F/C_A)\hat q$.

\subsection{Factorisation of vacuum-like emissions in the presence of the medium}\label{sec:factorisation-vles}

We now discuss how interactions with the medium affect the way a
vacuum partonic cascade develops in the medium.
This mostly follows the picture emerging from our previous study,
Ref.~\cite{Caucal:2018dla}, valid in the double-logarithmic
approximation.
We summarise below the main physics ingredients behind this picture
and then turn to a few new ingredients, going beyond the strict
double-logarithmic approximation, that were added for the purpose of
this paper.

First note that the expression  $\tf\simeq 2\omega/\kt^2$ for the formation
time is a direct consequence of
the uncertainty principle --- it is the time after which the parent
parton and the emitted gluon lose their mutual
quantum coherence --- and hence also holds for medium-induced emissions.
While an emission occurring in the vacuum can have an arbitrary
$k_\perp$, in the medium $k_\perp$ cannot be smaller than $k_{\rm f}^2\equiv \hat q\tf$, the
momentum broadening accumulated via collisions over the formation
time.
This defines a clear boundary between vacuum-like emissions (VLEs) for
which $k_\perp\gg k_{\rm f}$, and medium-induced emissions (MIEs) for
which $k_\perp\simeq k_{\rm f}$. Converting this in formation times, a
VLE satisfies
\begin{equation}\label{tfvac} 
  \tf(\omega, \theta) =
  \frac{2}{\omega\theta^2}\ll\sqrt{\frac{2\omega}{\hat q}}\equiv
  t_{\rm med}(\omega),
  \qquad \text{i.e.}\quad \omega^3\theta^4\gg 2\hat{q},
\end{equation}
where the strong ordering is valid in the sense of the
double-logarithmic approximation and $t_\text{med}(\omega)$ is the typical
formation time of MIEs.

\begin{figure}[t] 
\centering
\hspace*{0.5cm}%
\raisebox{0.5cm}{\includegraphics[width=0.45\textwidth]{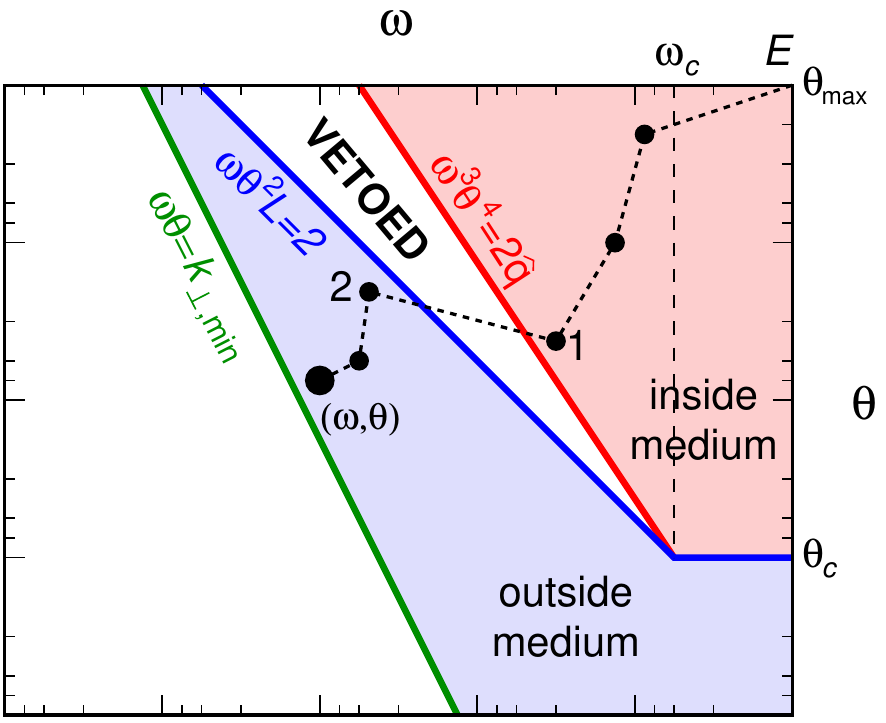}}\hfill%
\includegraphics[width=0.4\textwidth]{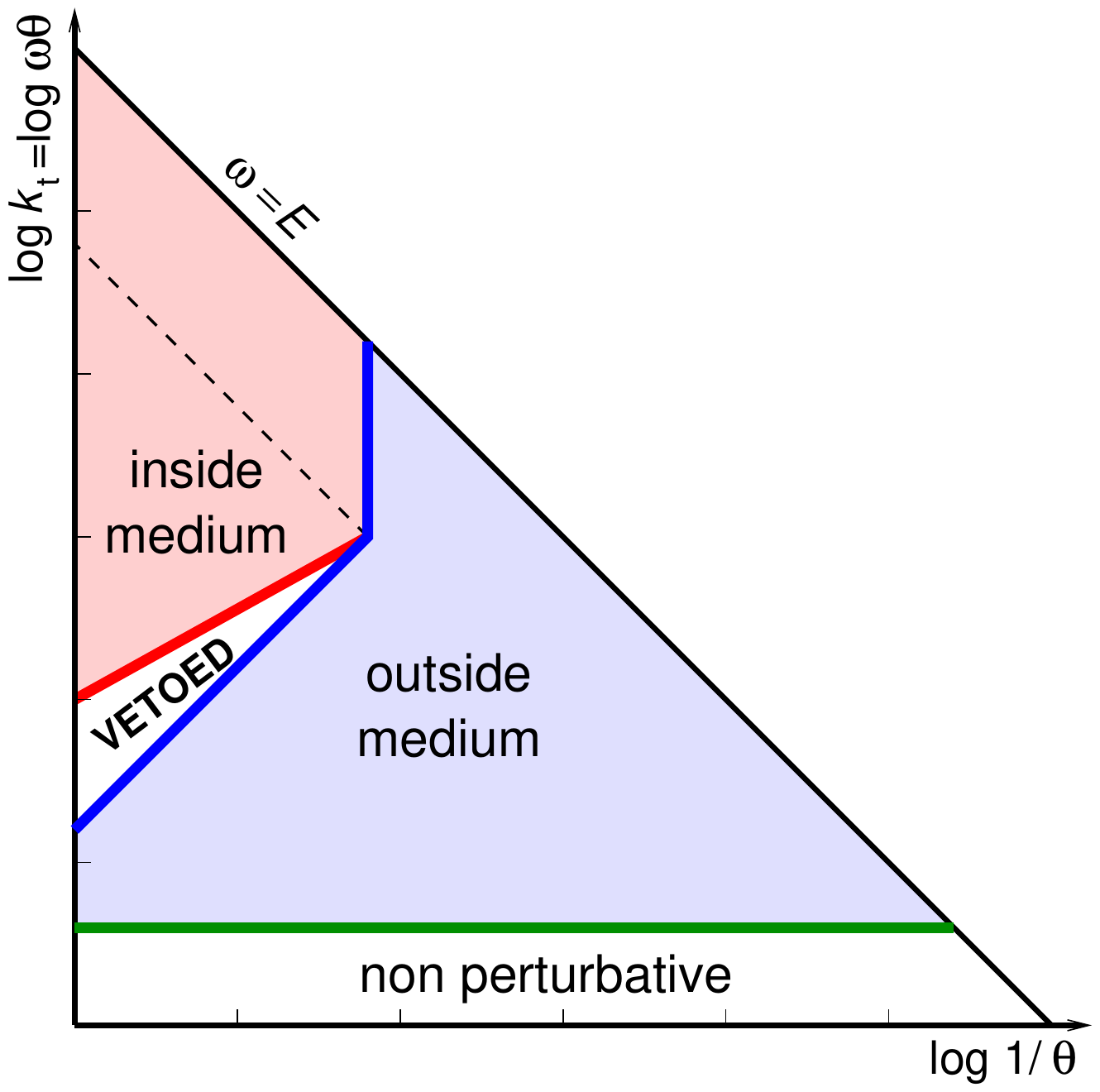}\hspace*{0.5cm}
\caption{\small The phase-space for vacuum-like gluon emissions by a jet propagating
through a dense QCD medium, in logarithmic units. In the left plot, the variables are the gluon energy $\omega$ and its emission angle $\theta$. In the right plot, we rather use the relative transverse momentum $\kt\simeq \omega\theta$ and the inverse of the angle $1/\theta$.}
 \label{Fig:LundPS}
\end{figure}

Eq.~(\ref{tfvac}) is the cornerstone on which the partonic cascade in
the medium is built. From this fundamental relation, the full physical
picture can be obtained based on a few additional observations:
\begin{itemize}
\item The formation time $t_{\rm med}(\omega)$ corresponding to a MIE
  must be shorter than $L$, implying an upper limit on the energy of the
  MIEs: $\omega\le \omega_c\equiv \hat q L^2/2$. This argument also shows that the
  constraint in \eqn{tfvac}  exists only for $\omega\le \omega_c$; more energetic emissions with $\omega > \omega_c$ are always vacuum-like, irrespective of their formation time (smaller or larger than $L$).
 \item VLEs can either occur inside the medium, in which case they
   satisfy~(\ref{tfvac}), or outside the medium, in which case $\tf(\omega, \theta)>L$. This implies that emissions with intermediate formation times, with $t_\text{med}\ll t_\text{f}(\omega, \theta)\ll L$, are forbidden. This gives the ``vetoed''  region  in Fig.~\ref{Fig:LundPS}.
\item The above one-emission picture can be generalised to the
  multiple emission of $N$ gluons with $\omega_n\ll\omega_{n-1}$ and
  $\theta_n\ll\theta_{n-1}$, $n=1,\dots, N$.
  The corresponding formation times are strongly increasing from one
  emission to the next one, $t_{{\rm f}, n}\gg t_{{\rm f}, n-1}$ (with
  $t_{{\rm f}, n}\equiv \tf(\omega_n, \theta_n)$).
  As a consequence if Eq.~(\ref{tfvac}) is satisfied by the last
  emitted gluon i.e.\ if $t_{{\rm f}, N}\ll t_{\rm med}(\omega_N)$,
  then it is automatically satisfied by all earlier emissions,
  $t_{{\rm f}, n}\ll t_{\rm med}(\omega_n)$.
\item To obtain the strong ordering above we have assumed that
  emissions were both soft and collinear. This is known as the
  double-logarithmic approximation where the emission
  probability~\eqref{brem} is enhanced by logarithms of both the
  energy and the emission angle. This approximation is at the heart of
  a large range of calculations in perturbative QCD.
  We briefly discuss below some new elements (and limitations) going
  beyond this approximation in the next section.
\item A key ingredient of the above generalisation to multiple gluon
  emissions is the fact that the in-medium partonic cascade
  preserves angular ordering, meaning that $\theta_n\ll\theta_{n-1}$.
  This is highly non-trivial as this is a subtle consequence of
  colour coherence for vacuum emissions and collisions in the
  medium which eventually wash out this coherence~\cite{Caucal:2018dla}.
  %
\item The characteristic time for colour decoherence
  is~\cite{MehtarTani:2010ma,MehtarTani:2011tz,CasalderreySolana:2011rz,MehtarTani:2011gf}
  \begin{equation}\label{tdec}
    t_{\rm coh}(\theta)\,=\,\Big(\frac{4}{\hat q\theta^2}\Big)^{1/3}.
  \end{equation}
  Hence, VLEs with emission angles $\theta\gg \theta_c\equiv 2/\sqrt{\hat q L^3}$ rapidly lose their colour coherence ($t_{\rm coh}(\theta)
  \ll L$) and act as independent sources for MIEs over a time $\Delta t\simeq L$. 
  Conversely, VLEs with $\theta< \theta_c$ remain coherent with their
  parent parton and are not discriminated by the medium: the
  associated pattern of MIEs is as if created by their parent. This
  explains why the emissions with $\theta<\theta_c$ are included in
  the ``outside'' region in Fig.~\ref{Fig:LundPS}.
\item Over the development of a vacuum-like cascade, the partons can
  also lose energy by emitting MIEs. Since the (relative) energy loss
  is suppressed as $(\tf(\omega, \theta)/t_{\rm med}(\omega))^2\ll 1$
  this effect can be neglected. In other words, even though the
  partons created via VLEs will ultimately lose energy via MIEs, this
  effect is negligible during the development of the vacuum-like
  cascade. Note that Eq.~(\ref{tfvac}) also means that broadening
  effects are negligible.
\item After being emitted at a time $t$, a parton propagates
  through the medium over a distance $L-t$. During this propagation it
  interacts with the medium. From the point of view of these
  interactions with the medium, we can safely set the formation time
  of the VLEs to 0, as a consequence of the strong ordering in
  formation time. Therefore, MIEs and transverse momentum broadening
  can occur at any time $0<t<L$.
  We describe these phenomena in the following sections.
\item Once the partons have travelled through the medium, undergoing
  MIEs and broadening, they are again allowed to fragment as a
  standard vacuum-like cascade {\it outside the medium}. Since the
  partons have lost their colour coherence during their traversal of
  the medium, the first emission in the outside-medium VLE cascade can
  violate angular ordering i.e.\ happen at any angle \cite{Caucal:2018dla}.
\end{itemize}

A very simple picture for the development of a partonic cascade in the
medium emerges from the above observations. The full cascade can be
factorised in three major steps, represented in Fig.~\ref{Fig:LundPS}:
\begin{enumerate}
\item {\it in-medium vacuum-like cascade}: an angular-ordered
  vacuum-like cascade governed by the standard DGLAP splitting
  functions occurs inside the medium up to
  $\tf(\omega, \theta) = t_{\rm med}(\omega)$. During this process,
  the only effect of the medium is to set the constraint \eqref{tfvac} on the formation time;
\item {\it medium-induced emissions and broadening}: every parton
  resulting from the in-medium cascade travels through the medium,
  possible emitting (a cascade of) MIEs and acquiring momentum
  broadening (see discussions below for details);
\item {\it outside-medium vacuum-like cascade}: each parton exiting
  the medium at the end of the previous step initiates a new
  vacuum-like cascade outside the medium, down to a non-perturbative
  cut-off scale. The first emission in this cascade can happen at an
  arbitrary angle.
\end{enumerate}

Before moving to new considerations specific to this paper, let us
provide a few additional comments of use for our later discussions.
First of all, the condition \eqref{tfvac} can be reformulated as a
lower limit on the emission angle $\theta$ for a given energy,
$\theta\gg \theta_{0}(\omega)\equiv \left({2\hat
    q}/{\omega^3}\right)^{\!\frac{1}{4}}$, or as a condition on the
energy of the emission at fixed angle $\theta$:
$\omega>\omega_{0}(\theta)\equiv \left({2\hat
    q}/{\theta^4}\right)^{\!\frac{1}{3}}$.
Physically, $\theta_{0}(\omega)$ is the formation angle of a MIE,
i.e.\ the value of the emission angle at the time of formation.

Furthermore, one can easily compute the in-medium parton multiplicity
at a given energy and angle in the above double-logarithmic
approximation. Each emission is enhanced by a double logarithm and the
corresponding contribution to the double differential gluon
distribution at a given point $(\omega,\theta)$ in phase-space is,
using a fixed-coupling approximation
\begin{equation}\label{DLA}
  \frac{\rmd^2 N_{\rm DLA}^\text{(in-medium)}}{\rmd\omega\rmd\theta^2}
  =\,\frac{\abar}{\omega\,\theta^2}
\sum_{n\ge 0}\,\abar^{n} \left[\frac{1}{n!}\ln^n\frac{E}{\omega}\right]
 \left[\frac{1}{n!}\ln^n\frac{\theta_{\rm max}^2}{\theta^2}\right]
 =\frac{\abar}{\omega\,\theta^2}\,\rmI_0\left(2\sqrt{\abar 
\ln\frac{E}{\omega}\,\ln\frac{\theta_{\rm max}^2}{\theta^2}}\right), 
\end{equation}
where $\abar=\tfrac{\alpha_s N_c}{\pi}$ and $\theta_\text{max}$ is the maximal angle allowed for the emissions. The term with $n=0$ in this series represents the direct emission of the gluon $(\omega,\theta)$ by the leading parton, while a term with $n\ge 1$ describes a sequence of $n$ intermediate emissions acting as additional sources for the final gluon.

Finally, note that whereas it provides a reasonable (first) estimate
for the gluon multiplicity at small $\omega\ll E$, the  above
double-logarithmic approximation (DLA) proposed in
Ref.~\cite{Caucal:2018dla}  is not appropriate for observables
which are sensitive to the energy loss, or for quantitative studies
of the multiplicity.
This is due to two main reasons:
the DLA for the VLEs uses only the singular part of the splitting
functions, i.e.\ $P(z)\simeq 2C_R/z$, and the energy of the parent
parton is unmodified after the emission so the energy is not
conserved at the splitting vertices.
In the next section we show that this can easily be fixed by
accounting for the full splitting function.
Second, MIEs, which represent the main mechanism for jet energy loss
in the medium, were neglected at DLA. However, the factorisation
between VLEs and MIEs (described above) which was rigorously proven at
DLA~\cite{Caucal:2018dla} is still valid beyond, typically in the {\it
  single logarithmic approximation} in which one only enforces a
strong ordering of the successive emissions angles.
One can therefore include MIEs and broadening effects in the above
picture, as described in the text below.
This is the picture that we adopt throughout this paper.

\subsection{A single-logarithmic approximation with angular ordering}

Now that we have recalled the basic picture for the development of
parton showers in the presence of a medium, we show that
several subleading corrections, beyond DLA, can easily be taken into account.

The validity of our factorisation between VLEs and MIEs relies on
strong inequalities between the formation times. Clearly, these
inequalities do still hold if the strong ordering refers only to the
emission angles ($\theta_n\ll\theta_{n-1}$), but not also to the
parton energies (as is the case beyond the soft limit $z\ll
1$). There is nevertheless some loss of accuracy w.r.t.\ a strict
single logarithmic approximation associated with the uncertainties in
the boundaries of the vetoed region in phase-space. Notably the
condition  $\tf(\omega, \theta)=t_{\rm med}(\omega)$ defining the
upper boundary is unambiguous only at DLA. For a generic splitting
fraction $z$, the formation times also depend upon the energy $xE$ of
the parent parton and not just upon the energy $\omega=xzE$ of the
soft daughter gluon. For a generic $1\to 2$ splitting where
the ``vacuum-like'' formation time $\tf\equiv \tf(x, z, \theta)$ is
given by \eqn{tvac} with $E\to xE$. The corresponding ``medium''
formation time $t_{\rm med}(x,z)$ is different for different partonic
channels. For example, for a $g\to gg$ splitting, it reads
(see e.g. \cite{Blaizot:2012fh,Blaizot:2013vha,Apolinario:2014csa})
\beq\label{tmed}
t_{\rm med}(x,z)\big|_{g\to gg}\,=\,\sqrt{\frac{2z(1-z)xE}{ \hat
    q_{\text{eff}}(z)}}\
   \overset{z\ll 1}{\approx}\ \sqrt{\frac{2\omega}{\hat q}}\,,\qquad \hat q_{\text{eff}}(z)\equiv\hat q \big[1-z(1-z)\big],
\eeq
with $\omega=zxE$. One could in principle use these more accurate
estimates for $\tf$ and $t_{\rm med}$ in  \eqn{tfvac}. One would then
need to deal with the difficulty that the evolution phase-space
depends explicitly on $xE$, $z$ and $\theta$ and not just on $\omega$
and $\theta$. The corresponding generalisation of \eqn{tfvac} would
also be different for different partonic channels. Last but not least,
the distinction between VLEs and MIEs according to their formation
times only holds so long as the {\it strong} inequality $\tf\ll t_{\rm
  med}$ is satisfied, meaning that the precise form of the boundary
could also be sensitive to subleading corrections. In practice, our
strategy to deal with this ambiguity (notably, in the Monte-Carlo
simulations) is to stick to the simpler form of the boundary in
\eqn{tfvac}, but study the sensitivity of our results to variations in
$\hat q$, effectively mimicking the  $z$-dependence of $t_{\rm med}(x,z)$ in \eqn{tmed} when $z\sim \order{1}$. Similarly, the ambiguity in the definition of the other boundary of the ``vetoed'' region, which corresponds to $\tf=L$, can be numerically studied by considering variations in $L$.

\subsection{Medium-induced radiation}

We now turn to a discussion of the medium-induced radiation and the
associated energy loss.  We start by observing that all the partons
created via VLEs with emission angles $\theta\gg \theta_c$ act as
quasi-independent sources for MIEs. Such partons have very short
formation times, $\tf(\omega, \theta)\ll L$, so after formation they
propagate through the medium over a distance of the order of the
medium size $L$. Since their coherence time $t_{\rm coh}(\theta)$
(cf. \eqn{tdec}) is much smaller that $L$, they rapidly lose colour
coherence during this propagation.
Each parton therefore acts as independent sources for medium-induced
radiation with energies up to $\omega_c=\hat q L^2/2$ (such that
$t_{\rm med}(\omega_c)=L$).

In practice, the emissions which control the energy loss {\it by the
  jet} , i.e.\ emissions at large angles, are dominated by much softer
gluons, with energies $\omega\ll \omega_c$ and short formation times
$t_{\rm med}(\omega)\ll L$~\cite{Blaizot:2013hx,Fister:2014zxa}.
Since these soft emissions occur with a probability of order one, one
must include multiple branching to all orders.
With this kinematics, this multiple-branching problem can be treated as a
classical Markovian
process~\cite{CasalderreySolana:2011rz,Blaizot:2012fh,Blaizot:2013hx,Blaizot:2013vha,Apolinario:2014csa,Arnold:2015qya,Arnold:2016kek}. This
stems from the following two observations: \texttt{(i)} the time
interval between two MIEs of comparable energies is much larger than
their respective formation time, meaning hat successive MIEs do not
overlap with each other, and \texttt{(ii)} interference effects can be
neglected since, in a medium-induced branching, the colour coherence
between the daughter partons is lost during formation. This last point
also implies that successive MIEs do not obey angular ordering. The
evolution ``time'' of the Markovian process is the physical time $t$
at which the MIEs occur in the medium (with $0 < t < L$).

The differential probability for one emission is given by the BDMPS-Z
spectrum for energies $\omega\ll \omega_c$
\cite{Baier:1996kr,Baier:1996sk,Zakharov:1996fv,Zakharov:1997uu,Baier:1998kq}
(see also 
\cite{Wiedemann:2000za,Wiedemann:2000tf,Arnold:2001ba,Arnold:2001ms,Arnold:2002ja}
for related developments).
For definiteness, consider the $g\to gg$ splitting of a gluon of
energy $xE$ (see e.g.~\cite{Mehtar-Tani:2018zba} for the other
partonic channels). The differential splitting rate integrated over
the emission angles (or, equivalently, over the transverse momentum
$\bk_\perp$) reads
\begin{equation}
\label{BDMPSZrate}
 \frac{\rmd^2\Gamma_{\rm med}}{\rmd z \rmd t}=\frac{\alpha_sP_{g\to gg}(z)}{2\pi}\sqrt{\frac{\hat{q}_{\text{eff}}(z)}{z(1-z)xE}}=\frac{\alpha_sP_{g\to gg}(z)}{\sqrt{2}\pi}\frac{1}{t_{\rm med}(x,z)}
 \end{equation}
where it is understood that $z(1-z)xE \ll\omega_c$. The second rewriting above uses \eqn{tmed} and makes it clear that the splitting rate is proportional to the inverse formation time. 

It is interesting to consider the emission probability (integrated
over a time/distance $L$), i.e.\ the BDMPS-Z spectrum for a single
soft emission, in the limit of a soft splitting,\footnote{Here, the
  concepts of ``soft splitting'' ($z\ll 1$) and ``soft emitted
  gluons'' ($\omega\ll \omega_c$) are not equivalent. For example a
  soft parent gluon (with $xE\ll \omega_c$) splits in soft daughter
  gluons for any value of $z$, including the symmetric (or
  ``democratic'') case $z\sim 1/2$.} $z\ll 1$. We find\footnote{ This
  is strictly valid when $\omega \ll\omega_c$. For larger energies
  $\omega \gg\omega_c$, the BDMPS-Z spectrum decreases as
  $(\omega_c/\omega)^2$. Eq.~\eqref{BDMPSZspec} can be used up to
  $\omega \sim\omega_c$ for parametric estimates.}
\beq
 \label{BDMPSZspec}
 \omega\frac{\rmd\mathcal{P}_{\rm med}}{\rmd\omega}=\abar\sqrt{\frac{2\omega_c}{\omega}}\,\Theta(\omega_c-\omega),
 \eeq
where we have used $P_{g\to gg}(z)\simeq 2N_c/z$, $\abar=\alpha_s N_c/\pi$, $\omega= zxE$, and $\omega_c=\hat q L^2/2$.

So long as its r.h.s. is strictly smaller than one, \eqn{BDMPSZspec}
expresses the probability to emit a gluon with energy $\omega$ by an
energetic gluon propagating through the medium along a distance
$L$. For $\omega \sim\omega_c$, this probability is of
$\order{\abar}$, showing that hard MIEs are rare. On the contrary, the
emission probability becomes of $\order{1}$ when
$\omega\sim\obr\equiv \abar^2\oc$ at which point multiple branchings
become important and the single-emission spectrum~\eqref{BDMPSZspec}
is no longer appropriate.

For gluons with $\omega\lesssim\omega_\text{br}$, we can
use~(\ref{BDMPSZrate}) to estimate the typical time interval between
successive branchings. It is given by the condition
\beq\label{tbr}
\omega\frac{\rmd^2\Gamma_{\rm med}}{\rmd \omega \rmd
  t}\,\tbr(\omega)\,\sim\,1 \qquad \Longrightarrow \qquad
\tbr(\omega)\,\simeq\,\frac{1}{\abar}\,t_{\rm med}(\omega)\,.
\eeq
The fact that this is parametrically larger than the formation time
$t_{\rm med}(\omega)$ justifies the picture of independent
emissions. Note that $\tbr(\omega)< L$ when $\omega < \obr$.

The hard but rare emissions with  $\omega \sim\omega_c$ control the
average {\it energy lost by the leading parton}, as can be seen by
integrating \eqn{BDMPSZspec} over $\omega$: the integral is dominated
by its upper limit $\omega_c$ and yields, parametrically,
$\varepsilon_{\rm LP}\sim\abar\omega_c$. However, a hard emission with
$\omega \sim\omega_c$ makes a small angle $\theta\sim\theta_c\ll R$
w.r.t.\ the jet axis and hence remains inside the jet. To estimate the
{\it energy lost by the jet}, one must instead consider the emissions
which are soft enough to be deviated outside the jet via elastic
collisions. For this, we need a more quantitative understanding of the {\it transverse momentum broadening}.

\subsection{Transverse momentum broadening}
\label{sec:TMB}

We first consider a MIE with $\obr\ll\omega\ll\oc$, i.e.\ soft but not
too soft, emitted at a time $t$. In this case, multiple emissions can
be neglected and after being emitted, the gluon propagates through the medium over a
distance $L-t$. Via elastic collisions, this gluon accumulates a
transverse momentum $k_{\rm f}^2\simeq \hat q\tf$ during its formation
and an additional $\Delta \kt^2=\hat q (L-t)$ after formation. Given
that $\tf\ll L$ and $L-t\sim L$, we have
$k_{\rm f}^2\ll \Delta \kt^2$. We can therefore neglect the momentum
broadening during formation and assume that the MIE is produced
collinearly with its source. Accordingly, the final transverse
momentum distribution of this gluon can be approximated by
\begin{equation}
\label{ktbroad}
 \frac{\rmd \mathcal{P}_{\rm broad}}{\rmd^{2}\bk_\perp}=\,\frac{1}{\pi\hat{q}(L-t)}\exp\left\{-\frac{\kt^2}{\hat{q}(L-t)}\right\}\,.
\end{equation}
The maximal average transverse momentum that can be acquired in this
way is $Q_s^2\equiv \hat q L$ (realised for $t=0$). Using
$\kt\simeq\omega\theta$ and averaging over the azimuthal angle one can
rewrite~\eqref{ktbroad} as
\begin{equation}
\label{thetabroad}
 \frac{\rmd \mathcal{P}_{\rm broad}}{\rmd\theta}\simeq
 \frac{2\omega^2\theta}{\hat{q}(L-t)}\exp\left\{-\frac{\omega^2\theta^2}{\hat{q}(L-t)}\right\}.
\end{equation}
This strictly applies to soft emitted gluons ($\omega\ll\oc$) and small angles
($\theta\ll 1$), but it is formally normalised by integrating over all
values $0<\theta<\infty$. 
The above equations describe the Gaussian
broadening via multiple soft scattering, but neglect the possibility
to acquire large transverse momentum $\kt^2\gg \hat q (L-t)$ via a
single hard scattering. 

We now move to even softer MIEs with energies $\omega\lesssim \obr$ which
have a finite lifetime $\tbr(\omega)< L$ (and much shorter formation times $\tf\ll \tbr$,
cf.\ \eqn{tbr}). The momentum/angular broadening is therefore
obtained by replacing $L-t\to \tbr(\omega)$ in
Eqs.~\eqref{ktbroad}--\eqref{thetabroad} yielding a typical deviation
angle
\beq
\theta(\omega)\simeq \frac{\sqrt{\hat q  \tbr(\omega)}}{\omega} =
\frac{1}{\sqrt{\abar}} \left(\frac{2\hat q}{\omega^3}\right)^{1/4}
\gtrsim \theta(\obr)\simeq \frac{\theta_c}{\abar^2}
\,,
\eeq
This angle is large and typically larger than the jet opening angle
$R$.
Furthermore, gluons with energies $\omega\lesssim \obr$ have a
probability of $\order{1}$ to undergo a democratic branching. They
therefore split over and over again in many soft quanta at large
angles until they thermalise in the plasma via elastic
collisions~\cite{Iancu:2015uja}. 
This implies in particular that the whole energy carried by primary
emissions with $\omega\lesssim \obr$ is eventually lost by the
jet. (See also the next subsection.)

\subsection{Energy loss by the jet: medium-induced emissions only}
\label{sec:energyloss-mies}

So far, we have argued that the energy lost by the jet via
medium-induced radiation at large angles ($\theta > R$) is controlled
by multiple branchings (resummed to all orders) and the
associated characteristic scale is $\obr=\abar^2\hat q L^2/2$. To
facilitate the physical interpretation of our subsequent Mote-Carlo
results it is useful to first recall a few more specific analytic results
related to energy
loss~\cite{Blaizot:2013hx,Fister:2014zxa,Blaizot:2014ula,Blaizot:2014rla,Escobedo:2016jbm,Escobedo:2016vba}.

These previous studies only addressed the case of jets generated via
medium-induced emissions {\it alone},\footnote{Formally, the leading
  parton was assumed to be on its mass shell when produced in the
  plasma.} and with a simplified version of the BDMPS-Z kernel where
only the poles at $z=0$ and $z=1$ in the splitting rate from
\eqn{BDMPSZrate} were kept. This is indeed sufficient to capture the most
salient features of the full dynamics while allowing for a good degree
of physical insight.

First, it has been shown~\cite{Blaizot:2013hx} that for a leading
parton of energy $E<\omega_c$, the energy lost by a jet via multiple
soft branchings at large angles is given by (with $\upsilon_0=2\pi$)
\beq\label{Eflow}
\varepsilon_{\rm flow}(E)\,=\,E\big(1-
\rme^{-\upsilon_0\frac{\obr}{E}}\big).
\eeq
This contribution is independent of the jet radius $R$.

To this ``turbulent'' component of the jet energy loss, one must add
the (average) energy taken away by semi-hard gluons whose energies are
larger than $\omega_\text{br}$, yet small enough for the associated propagation angles $\theta\sim
Q_s/\omega$ to be larger than $R$.
The (average) semi-hard contribution to the energy loss is therefore
obtained by integrating the emission spectrum over $\omega$ up to
$\bar\omega\equiv c_*Q_s/R$, with $c_*$ a number smaller than
one:\footnote{We will see later, cf.~\eqn{pbroad}, that an average value for it is $c_*=\sqrt{\pi}/3$.}
\beq\label{Espec}
\varepsilon_{\rm spec}(E,R)\,\simeq \int_0^{\bar\omega}\rmd\omega\,
\abar\sqrt{\frac{2\omega_c}{\omega}}\,\rme^{-\upsilon_0\frac{\obr}{E}}\,=\,
2\abar\oc\sqrt{\frac{2c_*\theta_c}{R}}
\,\rme^{-\upsilon_0\frac{\obr}{E}}\,,
\eeq
where we have also used $Q_s=\theta_c\oc$. Note that the above
expression uses the BDMPS-Z spectrum dressed by multiple branchings
computed under the same assumptions as~\eqn{Eflow} yielding an extra
exponential factor~\cite{Blaizot:2013hx}.
The $R$-dependence here is easy to understand: with increasing $R$,
more and more semi-hard emissions are captured inside the jet
so the energy loss is decreasing.
Not that when $R$ becomes as small as $\theta_c$, all the MIEs are
leaving the jet and the average energy loss by the jet coincides with
that of the leading parton. In practice, one usually has $R\gg\theta_c$ and therefore $\varepsilon_{\rm spec}\ll \varepsilon_{\rm LP}$.

The total energy lost by the jet under the present assumptions is
$\varepsilon_{\text{MIE}}=\varepsilon_{\rm flow}+\varepsilon_{\rm
  spec}$, where the subscript ``MIE'' indicates that for the time
being only MIEs are included.
Eqs.~\eqref{Eflow}--\eqref{Espec} exhibit some general features which
go beyond the approximations required for their derivation (see~\cite{Blaizot:2013hx,Fister:2014zxa}):
\begin{itemize}
\item For jets with energies $E\gg \upsilon\obr$, the jet energy loss
  via MIEs becomes independent of $E$:
  \beq\label{ElossHigh}
  \varepsilon_{\text{MIE}}(R)\,\simeq\,\upsilon\obr+2\abar\oc\sqrt{\frac{2c_*\theta_c}{R}}\qquad\mbox{when}\quad
  E\gg \upsilon\obr\,.\eeq
  The parameter $\upsilon$, equal to $\upsilon_0=2\pi$ for the simplified branching kernel considered in~\cite{Blaizot:2013hx}, is smaller for the full splitting rate: for jets with $E < \oc$, one finds $\upsilon\simeq 4.96$~\cite{Baier:2000sb,Blaizot:2013hx}, whereas in the high-energy limit $E \gg \oc$,   Ref.~\cite{Fister:2014zxa} reported  $\upsilon\simeq 3.8$ for $\abar=0.25$.
\item Jets with $E\lesssim \upsilon\obr$ lose their whole energy via democratic branchings:  $\varepsilon_{\text{MIE}}\simeq \varepsilon_{\rm flow}\simeq E$.
\item For a large jet radius
  $R\gtrsim \theta_c/\abar^2$, the flow component dominates over the
  spectrum component,
  $\varepsilon_{\rm flow}\gg\varepsilon_{\rm spec}$, for any energy
  $E$, and the energy loss
  $\varepsilon_{\text{MIE}}\simeq \varepsilon_{\rm flow}$ becomes
  independent of $R$.
\end{itemize}

\subsection{Energy loss by the jet:  full parton shower}
\label{sec:energyloss-full}

We can now consider the generalisation of the above results to the
full parton showers, including both VLEs and MIEs. In our sequential
picture, in which the two kind of emissions are factorised in time,
each of the VLE inside the medium act as an independent source of MIEs
and hence the energy loss by the full jet can be computed by
convoluting the distribution of partonic sources created by the VLEs
in the medium with the energy loss via MIEs by any of these sources.
Assuming that all the in-medium VLEs are collinear with the jet axis
(which is the case in the collinear picture described in
Sect.~\ref{sec:factorisation-vles}), the energy lost by the {\it
  full} jet is computed as
\begin{align} \label{logeloss}
 \mathcal{E}_{\rm{jet}}(E,R)&\simeq  \int_{\omega_0(R)}^{E} \rmd\omega\, \frac{\rmd N_{\text{VLE}}}
 {\rmd \omega}\,\varepsilon_{\text{MIE}}(\omega,R)\,,\nn
 &\simeq\varepsilon_{\text{MIE}}(E,R) + 2\abar\int_{\theta_c}^R\frac{\rmd\theta}{\theta}
  \int_{\omega_0(\theta)}^{E}\frac{\rmd\omega}{\omega}\,\rmI_0\left(2\sqrt{\abar 
\ln\frac{E}{\omega}\,\ln\frac{R^2}{\theta^2}}\right)\varepsilon_{\text{MIE}}(\omega,R)\,,
\end{align}
 where ${\rmd N_{\text{VLE}}}/ {\rmd \omega}$ is the energy
 distribution of the partons created via VLEs inside the medium (cf.\
 Fig.~\ref{Fig:LundPS}). The second line follows after using the DLA
 result for the gluon multiplicity, \eqn{DLA}. This is of course a
 rough approximation which overestimates the number of sources, but it
 remains useful to get a physical insight. Similarly, for qualitative purposes, one can use the simple estimate for $\varepsilon_{\text{MIE}}(\omega,R)$ given by the sum of Eqs.~\eqref{Eflow}--\eqref{Espec}.

\section{Parton shower in the medium: Monte-Carlo implementation}
\label{sec:MC}

The factorised picture for parton showers in the medium developed in
the last section (see in particular
Sect.~\ref{sec:factorisation-vles}), is well-suited for an
implementation as a Monte Carlo generator.
In this section we describe the main lines of this implementation and
its limitations. We also provide the details for the simulations done
throughout this paper.

\paragraph{Generic kinematic.}
We will represent the massless 4-vectors corresponding to emissions
using their transverse momentum $p_{Ti}$, their rapidity $y_i$ and
their azimuth $\phi_i$.
Since our physical picture is valid in the collinear limit, we will
often neglect differences between physical emission angles $\theta$
and distances $\Delta R=\sqrt{\Delta y^2+\Delta\phi^2}$ in the
rapidity-azimuth plane.
All showers are considered to be initiated by a single parton of given
transverse momentum $p_{T0}$, rapidity $y_0$ and azimuth $\phi_0$, and
of a given flavour (quark or gluon).

\paragraph{Vacuum shower.}
Still working in the collinear limit we will generate our partonic
cascades using an angular-ordered approach, starting from an initial
opening angle $\theta_\text{max}$. The initial parton can thus be seen
as having $\theta=\theta_\text{max}$ and a relative transverse
momentum $k_{\perp}=p_{T0}\theta_\text{max}$.
To regulate the soft divergence of the splitting functions, we
introduce a minimal (relative) transverse momentum cut-off
$k_{\perp,\text{min}}$. This corresponds to the transition towards the
non-perturbative physics of hadronisation (see Fig.~\ref{Fig:LundPS}).
Note that for a particle of transverse momentum $p_T$, the condition
$\kt >k_{\perp,\text{min}}$ imposes a minimal angle for the
next emission: $\theta>\theta_\text{min}=k_{\perp,\text{min}}/p_T$.

The shower is generated using the Sudakov veto algorithm.
More precisely, if the previous emission happened at an angle
$\theta_0$ and with relative transverse momentum $k_{\perp 0}$
(i.e.\ with transverse momentum\footnote{For the purposes of the
  subsequent discussion, $p_{T0}$ denotes the transverse
  momentum of a {\it generic} parent parton, which is not necessarily
  the {\it leading} parton.}  $p_{T0}=k_{\perp 0}/\theta_0$), the next
emission is generated with coordinates $\theta$, $k_\perp$ (and hence
with $p_{T}=k_{\perp}/\theta$), using the following procedure.
We first generate the angle $\theta$ according to the Sudakov factor
\begin{align}\label{eq:next-angle-sudakov}
  \Delta_i(\theta|\theta_0,k_{\perp 0})
  & = \exp\bigg[
    -\int_\theta^{\theta_0}\frac{\rmd\theta'}{\theta'}\int_{k_{\perp,\text{min}}}^{k_{\perp 0}} \frac{\rmd k_\perp'}{k_\perp'}\frac{2\alpha_s(k_\perp')C_i}{\pi}
    \bigg]\\
  & = \exp\bigg[
  -\frac{C_i}{\pi\beta_0}\log\Big(\frac{1-2\alpha_s\beta_0\log(k_{\perp 0}/M_Z)}{1-2\alpha_s\beta_0\log(k_{\perp,\text{min}}/M_Z)}\Big)
  \log\Big(\frac{\theta_0}{\theta}\Big)\bigg],
\end{align}
where $i=(\text{g,\,q})$ is a flavour index, $C_\text{g}\equiv C_A$, $C_\text{q}\equiv C_F$, and we use 5 flavours of  massless quarks.
To obtain the second line, we used a 1-loop running coupling
$\alpha_s(k_\perp)=\tfrac{\alpha_s}{1-2\alpha_s\beta_0\log(k_\perp/M_Z)}$ with
$\alpha_s\equiv\alpha_s(M_Z)$ the running coupling at the $Z$ mass,
fixed to 0.1265, and $\beta_0=(11C_A-2n_f)/(12\pi)$ the 1-loop QCD
$\beta$ function  (with $n_f =5$).
The $k_\perp$ of the emission is then generated between $k_{\perp,\text{min}}$
and $k_{\perp 0}$ following the distribution
$\tfrac{\rmd k_\perp}{k_\perp}\alpha_s(k_\perp)$.
This procedure neglects finite effects in the splitting function and
momentum conservation as the splitting fraction
$z'\equiv\tfrac{p_T'}{p_{T0}}
=\tfrac{k_\perp'}{k_{\perp 0}}\tfrac{\theta_0}{\theta'}$
 associated with the emission of the gluon $\theta'$ and $k_\perp'$ in
(\ref{eq:next-angle-sudakov}) can take values larger than one.
This is simply taken into account by vetoing emissions with $z>1$ and
accepting those with $z\le 1$ with a probability
$\tfrac{z}{2C_i}P_i(z)$ with $P_i(z)$ the targeted splitting
function.\footnote{A similar trick allows us to select between the
  $g\to gg$ and $g\to q\bar q$ channels for gluon splitting.}
If any of these vetoes fails, we set $\theta\to \theta_0$ and $k_\perp\to
k_{\perp 0}$ and reiterate the procedure.
After a successful parton branching, both daughter partons are further
showered.
The procedure stops when the generated angle $\theta$ is smaller than
the minimal angle $\theta_\text{min}$.

To fully specify the procedure we still need to specify how to
reconstruct the daughter partons from the parent.
For this, we use $z\equiv \tfrac{p_T}{p_{T0}}=\tfrac{k_\perp}{k_{\perp 0}}\tfrac{\theta_0}{\theta}
$ and also generate an azimuthal angle
$\varphi$ around the parent parton, randomly chosen between 0 and $2\pi$.
We then write
\begin{equation}\label{eq:splitting-kinematics}
  \text{parent: }(p_{T0},y_0,\phi_0)
  \longrightarrow \begin{cases}
    \text{daughter 1: }\big(z\,p_{T0},y_0+(1-z)\theta\cos(\varphi),\phi_0+(1-z)\theta\sin(\varphi)\big)\\
    \text{daughter 2: }\big((1-z)p_{T0},y_0-z\,\theta\cos(\varphi),\phi_0-z\,\theta\sin(\varphi)\big)
  \end{cases}.
\end{equation}

\paragraph{Medium shower: MIEs.}
The cascade of MIEs is better described using an ordering in emission
time.
For simplicity, we assume a uniform medium of fixed length $L$ and
adopt a fixed-coupling approximation for the interactions between the
hard partons and the medium, with no feedback.

The rate for the emission of a MIE per unit time is given by the
kernel\footnote{All the expressions here are given for a pure-gluon
  cascade. In practice, our implementation includes all the flavour
  channels using the expressions from
  Ref.~\cite{Mehtar-Tani:2018zba}.} (see e.g.~\cite{Blaizot:2013vha})
\begin{equation}\label{eq:mie-evolution-kernel}
  \mathcal{K}(z|p_T)
  \equiv \frac{\alpha_s}{2\pi}
  \sqrt{\frac{\hat q[1-z(1-z)]}{z(1-z)p_T}} P_{gg}(z)
  \overset{z\ll 1}{\approx} \frac{\alpha_sC_A}{\pi}
  \sqrt{\frac{\hat{q}}{p_T}} \frac{1}{[z(1-z)]^{3/2}},
\end{equation}
where $p_T$ is the transverse momentum of the parton that splits and
$z$ the splitting fraction. Throughout this paper, the QCD coupling
occurring in \eqn{eq:mie-evolution-kernel} will be assumed to be fixed and
treated as a free parameter, to be often denoted as $\amed$ for more clarity.
This Markovian process can be simulated from $t=0$ to $t=L$ using a
Sudakov veto algorithm as for the vacuum-like shower. From a time
$t_0$, we first generate the next splitting time $t$ according to a
Sudakov factor which integrates~\eqref{eq:mie-evolution-kernel} over
time between $t_0$ and $t$ and over $z$ between some cut-off
$z_\text{min}$ and $1-z_\text{min}$. We then generate $z$ according
to~\eqref{eq:mie-evolution-kernel}. Both steps are done using the
simplified kernel in the limit $z\ll 1$ and a veto with probability
$\tfrac{z}{2C_A} P_{gg}(z)$ is applied to get the full splitting rate.
In practice, we have set $z_\text{min}=10^{-5}$ and checked that this
choice has no influence on our final results.

In the cascade described above, all the splittings are considered to
be exactly collinear. The angular pattern is generated afterwards via
transverse momentum broadening, cf.\ Sect.~\ref{sec:TMB}. 
For this, we go over the whole cascade and, for each parton, we generate
an opening angle $\theta$ and azimuthal angle $\varphi$ according to
the two-dimensional Gaussian distribution Eq.~(\ref{thetabroad}) where
$L-t$ is replaced by the lifetime $\Delta t$ of the parton in the cascade.
Once we have the transverse momenta and angles of each parton in the
cascade, we use~(\ref{eq:splitting-kinematics}) to reconstruct the
kinematics.
Partons which acquire an angle larger than $\theta_\text{max}$ via
broadening are discarded together with their descendants.

\paragraph{Medium shower: global picture.}
The in-medium shower is generated in three stages, according to the
factorisation discussed in Sect.~\ref{sec:factorisation-vles}.
The first step is to generate in-medium VLEs. This is done exactly as
for the full vacuum shower except that each emission is further tested
for the in-medium conditions $k_\perp^3\theta>2\hat{q}$ and
$\theta>\theta_c$. If any of these two conditions fails, the emission
is vetoed.
The second step is to generate MIEs for each of the partons obtained
at the end of the first step, following the procedure described above.
The third step is to generate the VLEs outside the medium.
For this, each parton at the end of the MIE cascade is taken and
showered outside the medium. This uses again the vacuum shower,
starting from an angle $\theta_\text{max}$ since decoherence washes
out angular ordering for the first emission outside the medium. Each
emission which satisfies either $k_\perp\theta<2/L$ or $\theta<\theta_c$
is kept, the others are vetoed.

\paragraph{Final-state reconstruction.}
The full parton shower can be converted to 4-vectors suited for any
analysis. Final-state (undecayed) partons are taken massless with a
kinematics taken straightforwardly from
Eq.~(\ref{eq:splitting-kinematics}). If needed, the 4-vectors of the
other partons in the shower are obtained by adding the 4-momenta of
their daughter partons. This requires traversing the full shower
backwards.

Whenever an observable requires to cluster the particles into jets and
manipulate them, we use the {\tt FastJet} program
(v3.3.2)~\cite{Cacciari:2011ma} and the tools in {\tt fjcontrib}.
In particular, the initial jet clustering is always done using the
anti-$k_\perp$ algorithm~\cite{Cacciari:2008gp} with $R=0.4$ unless
explicitly mentioned otherwise.

\paragraph{Limitations.}
The Monte Carlo generator that is described above is of course very
simplistic and has a series of limitations. We list them here for the
sake of completeness.
First of all, we only generate a partonic
cascade, neglecting non-perturbative effects like hadronisation. Even
if one can hope that these effects are limited --- especially at large
$p_T$ ---, our description remains incomplete and, for example,
track-based observables are not easily described in our current framework.
Additionally, our partonic cascade only includes final-state
radiation. Including initial-state radiation goes beyond our collinear
picture and is left for future work. This would be needed, for
example, to describe the transverse momentum pattern of jets recoiling
against a high-energy photon.

Our description of the medium is also simplified: several effects like
medium expansion, density non-uniformities and fluctuations, and the
medium geometry are neglected.
For the observables discussed in this paper, this can to a large extend
be hidden into an adjustment of the few parameters we have left, but we
would have to include all these effects to claim a full in-medium
generator.

\begin{table}
  \centering
  \begin{tabular}{|l|ccc|ccc|}
    \hline
    & \multicolumn{3}{|c|}{parameters} 
    & \multicolumn{3}{|c|}{physics constants} \\
    \cline{2-7}
    Description
    & $\hat{q}$ [GeV$^2$/fm] & $L$ [fm] & $\amed$
    & $\theta_c$ & $\omega_c$ [GeV] & $\omega_\text{br}$ [GeV] \\
    \hline
    default
    & 1.5   & 4     & 0.24  & 0.0408 &  60 & 3.456 \\
    \hline
    & 1.5   & 3     & 0.35  & 0.0629 & 33.75 & 4.134 \\
    similar $R_{AA}$
    & 2     & 3     & 0.29  & 0.0544 & 45    & 3.784 \\
    & 2     & 4     & 0.2   & 0.0354 & 80    & 3.200 \\
    \hline
    \multirow{2}{*}{vary $\theta_c$}
    & 0.667 & 6     & 0.24  & 0.0333 & 60    & 3.456 \\
    & 3.375 & 2.667 & 0.24  & 0.05   & 60    & 3.456 \\
    \hline
    \multirow{2}{*}{vary $\omega_c$}
    & 0.444 & 6     & 0.294 & 0.0408 & 40    & 3.456 \\
    & 5.063 & 2.667 & 0.196 & 0.0408 & 90    & 3.456 \\
    \hline
    \multirow{2}{*}{vary $\omega_\text{br}$}
    & 1.5   & 4     & 0.196 & 0.0408 & 60    & 2.304 \\
    & 1.5   & 4     & 0.294 & 0.0408 & 60    & 5.184 \\
    \hline
  \end{tabular}
  \caption{\small Table of medium parameters used in this paper. The default
    set of parameters is given in the first line. The next 3 lines
    are parameters which give a similar prediction for
    $R_\text{AA}$. The last 6 lines are up and down variations of
    $\theta_c^2$, $\omega_c$ and $\omega_\text{br}$ by 50\%, keeping
    the other two physics constants fixed.}\label{tab:parameters}
\end{table}

\paragraph{Choices of parameters.}
The implementation of in-medium  partonic cascades described above has
5 free parameters: two unphysical ones, $\theta_\text{max}$ and
$k_{\perp,\text{min}}$, essentially regulating the soft and collinear
divergences, and three physical parameters, $\hat{q}$, $L$ and
$\amed$, describing the medium.
In our phenomenological studies, we will make sure that our results
are not affected by variations of the unphysical parameters and we
will study their sensitivity to variations of the medium parameters.

The different sets of parameters we have used are listed in
Table~\ref{tab:parameters}. The first line is our default setup. It
has been chosen to give a reasonable description of the $R_{AA}$ ratio
measured by the ATLAS collaboration (see Sect.~\ref{sec:RAA} below).
The next 3 sets are variants which give a similarly good description of
$R_{AA}$ and can thus be used to test if other observables bring an
additional sensitivity to the medium parameters compared to $R_{AA}$.
The last 6 lines are variations that will be used to probe which
physical scales, amongst $\theta_c$, $\omega_c$ and
$\omega_\text{br}$, influence a given observable.

\begin{figure}[t] 
  \centering
  \includegraphics[width=0.48\textwidth,page=1]{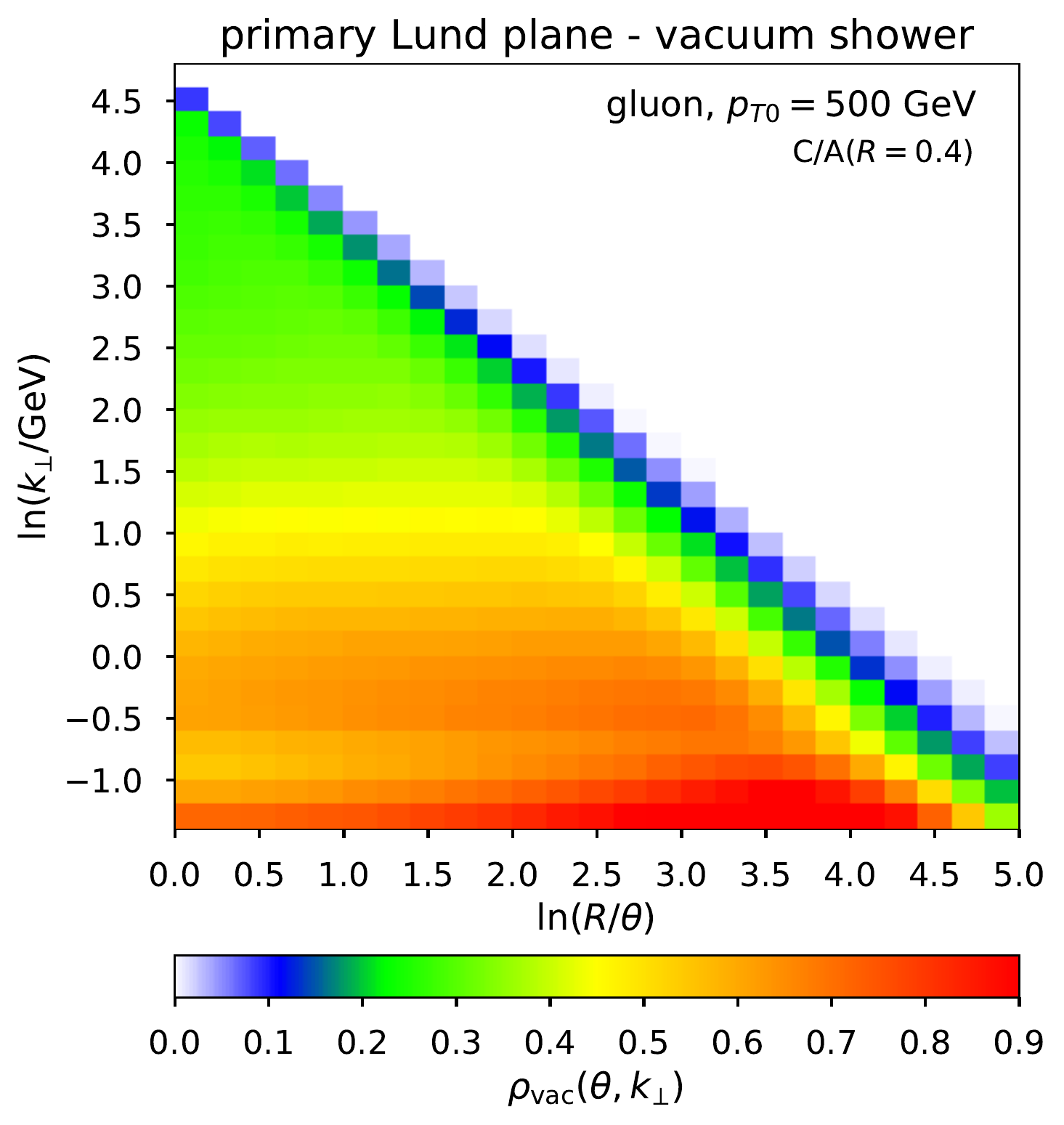}\hfill%
  \includegraphics[width=0.48\textwidth,page=2]{./FIGS/basic-lund-images.pdf}
 \caption{\small plot of the primary Lund plane density $\rho$ for the two
   main showers in our Monte Carlo: the vacuum shower (left) and the
   medium-induced shower (right). In both cases, the shower is
   initiated by a 500~GeV gluon. }
 \label{Fig:Lund1}
\end{figure}

\begin{figure}[t] 
  \centering
  \includegraphics[width=0.48\textwidth,page=5]{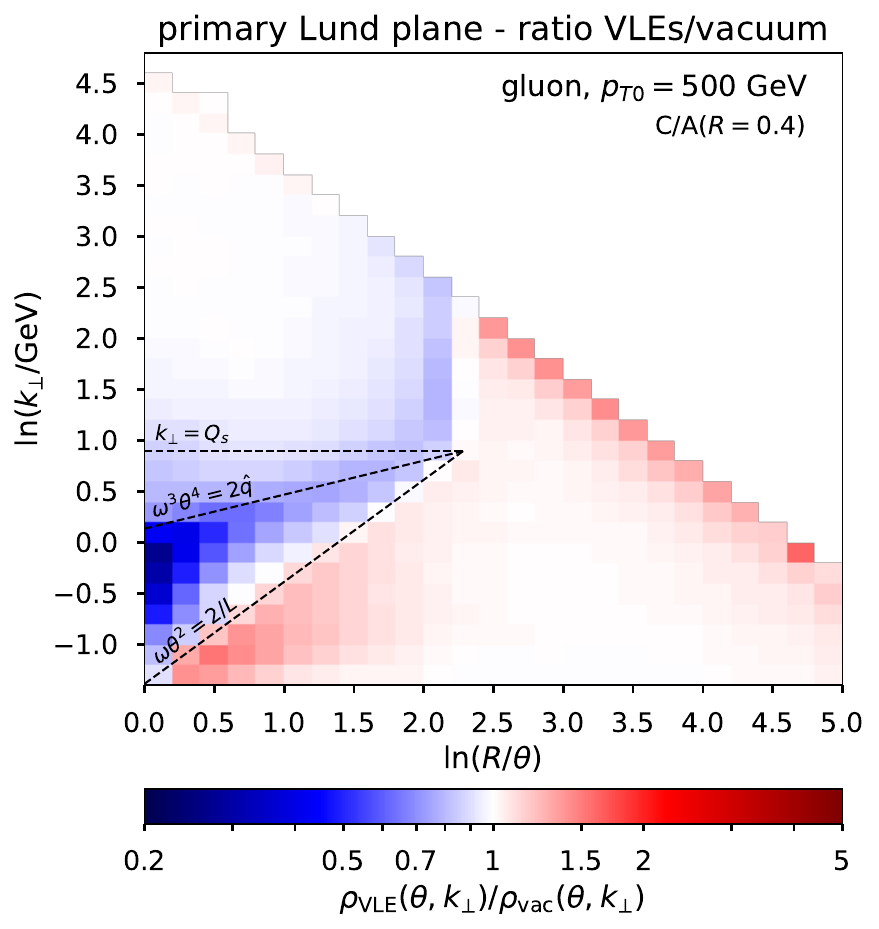}\hfill%
  \includegraphics[width=0.48\textwidth,page=6]{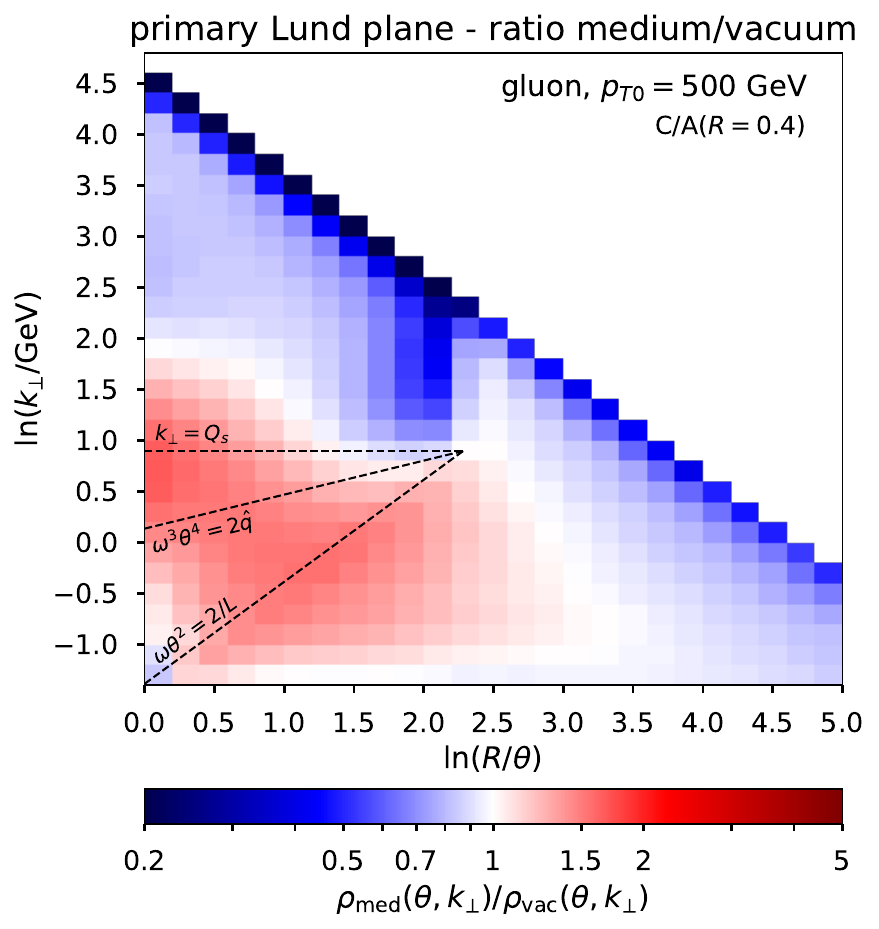}
 \caption{\small Effect of the medium on the primary Lund plane
   density $\rho$. The left plot show the ratio of the density
   obtained with VLEs, taking into account the kinematic constraints
   from the medium, to the vacuum reference. The right plot is the
   ratio of the full shower over the vacuum density.}
 \label{Fig:Lund2}
\end{figure}

\paragraph{Illustration of the behaviour.}
To illustrate the properties of our in-medium parton shower, we
consider the primary Lund plane density $\rho(\theta, k_\perp)$.
In a nutshell, this uses an iterative declustering procedure,
following the hardest branch at each step, to measure the density of
emissions  at an angle $\theta$ and with
relative transverse momentum $k_\perp$ from the hard subjet (see
Ref.~\cite{Dreyer:2018nbf}).
It can be viewed as a representation of the emissions from the leading
parton in a jet.

Fig.~\ref{Fig:Lund1} shows the primary Lund plane for the two main
showers introduced above: the (genuine) vacuum-shower (left) ---
our reference for the calibration of the nuclear effects --- and the
medium-induced shower (right).
The vacuum shower shows the expected pattern of a density which is
mostly constant at a fixed $k_\perp$, increases due to the running of
the strong coupling with decreasing $k_\perp$, and vanishes when
approaching the kinematic limit $z\to 1$.
The right plot of Fig.~\ref{Fig:Lund1} highlights that MIEs have a
typical transverse momentum $k_\perp\lesssim Q_s$
(cf.~(\ref{ktbroad})) and a density which increases at small $z$.

Fig.~\ref{Fig:Lund2} then shows the effect of our factorised picture
for the parton shower in a dense plasma. In the left plot, we have
neglected MIEs and only included the VLEs both inside and outside the
medium. The plot shows the ratio of the resulting density to the
vacuum density.
The vetoed region is clearly visible on the plot. The small density
reduction in the in-medium region and the small increase in the
outside region, especially at large angles, can both  be attributed to the
fact that the first emission outside the medium can violate angular
ordering and be emitted at any angle.

Finally, the right plot of Fig.~\ref{Fig:Lund2} shows the ratio of the
density $\rho_\text{med}$ for the full shower to the corresponding
vacuum density. In this case, one clearly see a region of enhanced
emissions corresponding to MIEs, as well as a decrease at large $p_T$
due to energy loss.

\section{Energy loss by the jet and the nuclear modification factor}
\label{sec:eloss}

Here we present our Monte Carlo results for the jet nuclear modification
factor $R_{AA}$. We first discuss the case of a monochromatic
leading parton, for which we compute the jet energy loss, and then
turn to $R_{AA}$ itself, using a Born-level jet spectrum for the
hard process producing the leading parton.

\subsection{The average energy loss by the jet}

To study the jet energy loss we start with a single hard parton of
transverse momentum $p_{T0}$ and shower it with the Monte Carlo
including either MIEs only, or both VLEs and MIEs.
The jet energy loss is defined as the difference between the energy of
the initial parton and the energy of the reconstructed jet. To avoid
artificial effects related to emissions with an angle $\theta$ between the jet
radius $R$ and the maximal opening angle $\theta_\text{max}$ of the
Monte Carlo, we have set $\theta_\text{max}=R$. Furthermore, for the
case where both VLEs and MIEs are included, we have subtracted the
pure-vacuum contribution (which comes from clustering and other edge
effects and is anyway small for $\theta_\text{max}=R$).

\begin{figure}[t] 
  \centering
  \includegraphics[width=0.48\textwidth,page=1]{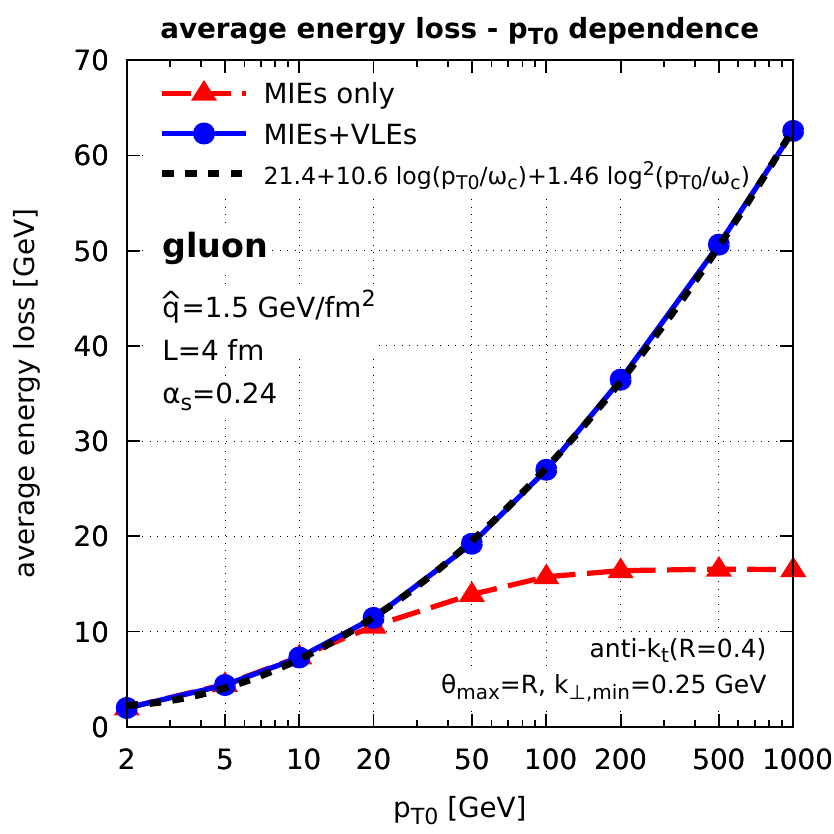}\hfill%
  \includegraphics[width=0.48\textwidth,page=1]{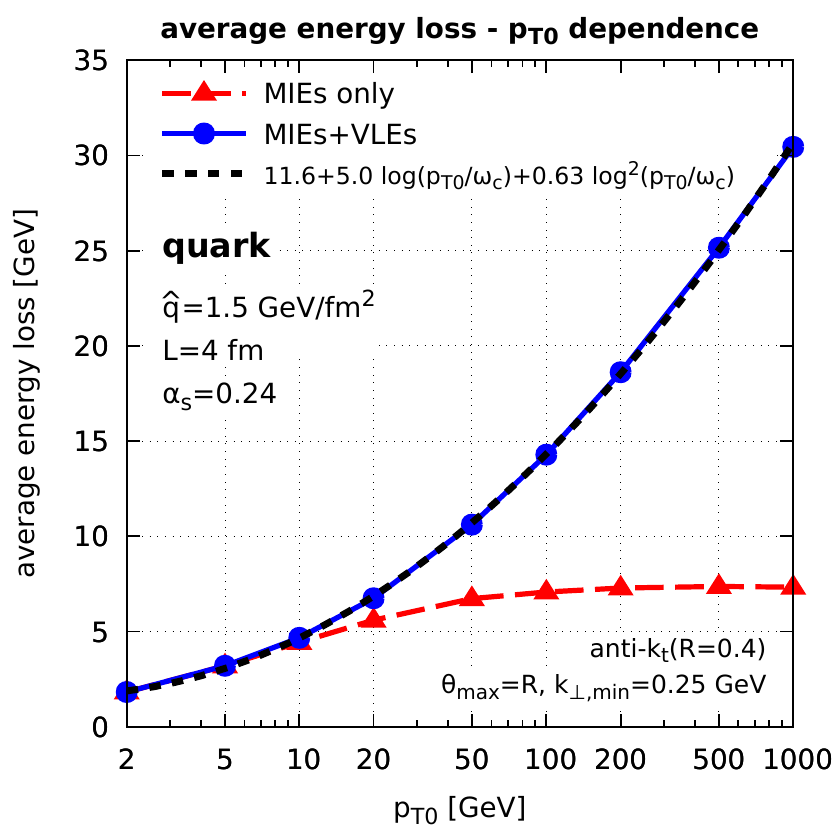}
 \caption{\small The MC results for the average energy loss by a gluon-initiate jet (left), respectively,
 a quark-initiated one (right), is displayed as a function of the initial energy $p_{T0}$ of the leading parton,
 for two scenarios for the jet evolution: jets with MIEs only (triangles) and full showers with both MIEs and VLEs (circles). The dashed line shows the quadratic fit to the energy loss by the full parton shower.}
 \label{Fig:jeteloss-v-pt}
\end{figure}

\begin{figure}[h] 
  \centering
  \includegraphics[width=0.48\textwidth,page=2]{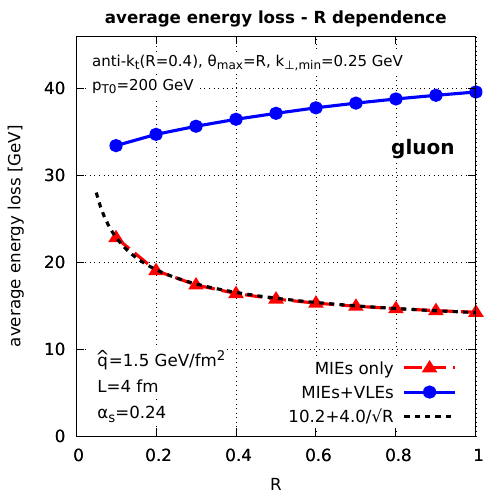}\hfill%
  \includegraphics[width=0.48\textwidth,page=2]{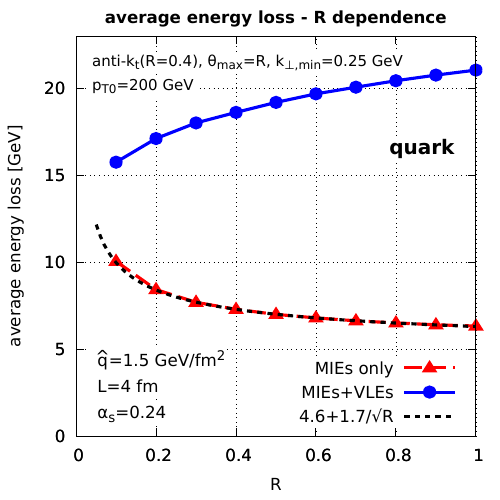}
 \caption{\small The MC results for the average energy loss by the jet are displayed
 as a function of the jet angular opening $R$, for the same choices as in Fig.~\ref{Fig:jeteloss-v-pt}.  We also show in dashed line a fit (inspired by the theoretical estimate in \eqn{Espec}) for the jet built with MIEs alone.}
 \label{Fig:jeteloss-v-R}
\end{figure}

Our results for the energy loss are shown in
Fig~\ref{Fig:jeteloss-v-pt} as a function of $E\equiv p_{T0}$ and in
Fig.~\ref{Fig:jeteloss-v-R} as a function of $R$, for both gluon- and
quark-initiated jets. For these plots we have used the default values
for the medium parameters (see the first line of
Table~\ref{tab:parameters}).
Overall, we see a good qualitative agreement with the features expected
from the theoretical discussion in sections~\ref{sec:energyloss-mies}
and~\ref{sec:energyloss-full}.

For the jets involving MIEs only, we see that the energy loss first
increases with $p_{T0}$ and then saturates, as predicted by
Eqs.~\eqref{Eflow}--\eqref{Espec}. Also, as a function of $R$ for
fixed $p_{T0}=200$~GeV, it decreases according to the expected
$1/\sqrt{R}$ behaviour, cf.\ \eqn{Espec}.
The agreement is even quantitatively decent. Indeed, with the physical
parameters given in Table~\ref{tab:parameters}, the fitted
$R$-dependence for gluon-initiated jets,
$\varepsilon_{\text{MIE}}(R)\simeq \epsilon_0+\epsilon_1/\sqrt{R}$,
with $ \epsilon_0=10.2$~GeV and $ \epsilon_1=4.0$~GeV, corresponds to
the prediction in~\eqn{ElossHigh} provided one chooses
$ \upsilon\simeq 2.95$ and $c_*\simeq 0.38$, which are both
reasonable.

Consider now the full parton showers, with both VLEs and
MIEs. Although we do not have accurate-enough analytic results to
compare with (only the DLA estimate~\eqn{logeloss}), the curves
``MI+VLEs'' in Figs.~\ref{Fig:jeteloss-v-pt}
and~\ref{Fig:jeteloss-v-R} show the expected trend: the energy loss
increases with both $p_{T0}$ and $R$, due to the rise in the
phase-space for VLEs. For later use, we have fitted the dependence on
$p_{T0}$ with a quadratic polynomial in $\ln(p_{T0}/\omega_c)$ and the
resulting coefficients are shown on Fig.~\ref{Fig:jeteloss-v-pt}.

It is also striking from Figs.~\ref{Fig:jeteloss-v-pt}
and~\ref{Fig:jeteloss-v-R}, that the average energy loss obeys a
surprisingly good scaling with the Casimir colour factor of the leading
parton: the energy loss by the quark jet is to a good approximation
equal to $C_F/C_A=4/9$ times the energy loss by the gluon jet. Such a
scaling, natural in the case of a single-gluon emission, is very
non-trivial in the presence of multiple branchings.
Let us first give an argument explaining this scaling for the case of
MIEs alone.
The main observation 
(see~\cite{Blaizot:2013hx,Fister:2014zxa,Mehtar-Tani:2018zba} for more
details and for numerical results) is that the small-$x$ gluon distribution
within a jet initiated by a parton of colour representation $R$
develops a scaling behaviour with $1/x$, known as a Kolmogorov-Zakharov (or ``turbulent'') fixed
point.
For large-enough initial jet energy $E\equiv p_{T0}$ this scaling spectrum is identical to the
BDMPS-Z spectrum created by a single emission and reads
\beq\label{gluonR}
x\frac{\rmd G_R}{\rmd x}\,\simeq\,\frac{\alpha_s C_R}{\pi}
\sqrt{\frac{\hat q L^2}{xE}}\qquad\mbox{for  \ $x\ll 1$ \ and}\ \ E\gg \obr^{(R)}
\equiv \frac{\alpha_s^2}{\pi^2} \,C_A C_R\,  \frac{\hat q L^2}{2}\,,
\eeq
where $\hat q$ is the {\it gluonic} jet quenching parameter,
proportional to $C_A$, since~\eqn{gluonR} refers to the emission of
soft gluons. This spectrum is directly proportional to the Casimir
$C_R$ of the leading parton as expected.
Note that the scale $\obr^{(R)}$ which appears in the
validity condition of~\eqn{gluonR} involves the product $C_A C_R$ of
two Casimir factors, one for each power of $\alpha_s$.
One actually gets a factor $\alpha_s C_R/\pi$ associated with the
emission from the leading parton, whereas the other coupling
$\alpha_s C_A/\pi$ refers to the turbulent energy flux of the emitted
gluons and carried away at large angles.

From~\eqref{gluonR} it is easy to show the energy loss~\eqn{ElossHigh}
of a jet initiated by a parton in an arbitrary colour representation
$R$ scales linearly with $C_R$: the first term in~\eqn{ElossHigh} is
proportional to $C_R$ as it is proportional to $\obr^{(R)}$ and the
second term is also proportional to $C_R$ in the general case, as shown in
\eqn{gluonR}. All the other factors only refer to
gluons and are independent of $R$.
This justifies the Casimir scaling visible in
Figs.~\ref{Fig:jeteloss-v-pt} and \ref{Fig:jeteloss-v-R} for the
cascades with MIEs only.

For the full cascades including VLEs, the linear dependence on $C_R$
can be argued based on~\eqn{logeloss}, assuming $p_{T0}\gg \obr$. The
first term in the r.h.s.\ of~\eqn{logeloss} is the energy lost by the
leading parton and is by itself proportional to $C_R$, as just
argued. The second term in \eqn{logeloss}, which refers to the
additional ``sources'' created via VLEs, one can assume that most of
these ``sources'' are gluons, so they all lose energy (via MIEs) in
the same way; the only reference to the colour Casimir of the leading
parton is thus in the overall number of sources, which is indeed
proportional to $\alpha_s C_R/\pi$ (for a gluon-initiated jet, this is
the factor $\abar$ in front of the integral in
\eqn{logeloss}).\footnote{The factor $\bar\alpha$ in the $I_0$ is
  associated to the further  fragmentation of the gluons emitted from
  the main parton and therefore remains $\alpha_sC_A/\pi$
  independently of the leading parton. \label{foot:CR}}

\subsection{The nuclear modification factor $R_{AA}$}
\label{sec:RAA}

We now consider the physically more interesting jet nuclear modification factor $R_{AA}$,
which is directly measured in the experiments. 
In order to compute this quantity, we have considered a sample of Born-level
$2\to 2$ partonic hard scatterings.\footnote{We have used the same hard-scattering
  spectrum for both the pp baseline and the PbPb sample. This
  typically means that we neglect the effects of nuclear PDF which are
  most likely small for the observables studied in this paper.} For
each event, both final partons are showered using our Monte Carlo. Jets are
reconstructed using the anti-$k_\perp$ algorithm~\cite{Cacciari:2008gp} as
implemented in FastJet v3.3.2~\cite{Cacciari:2011ma}.
All the cuts are applied following the ATLAS measurement from
Ref.~\cite{Aaboud:2018twu}.

\begin{figure}[t] 
  \centering
  \includegraphics[width=0.48\textwidth]{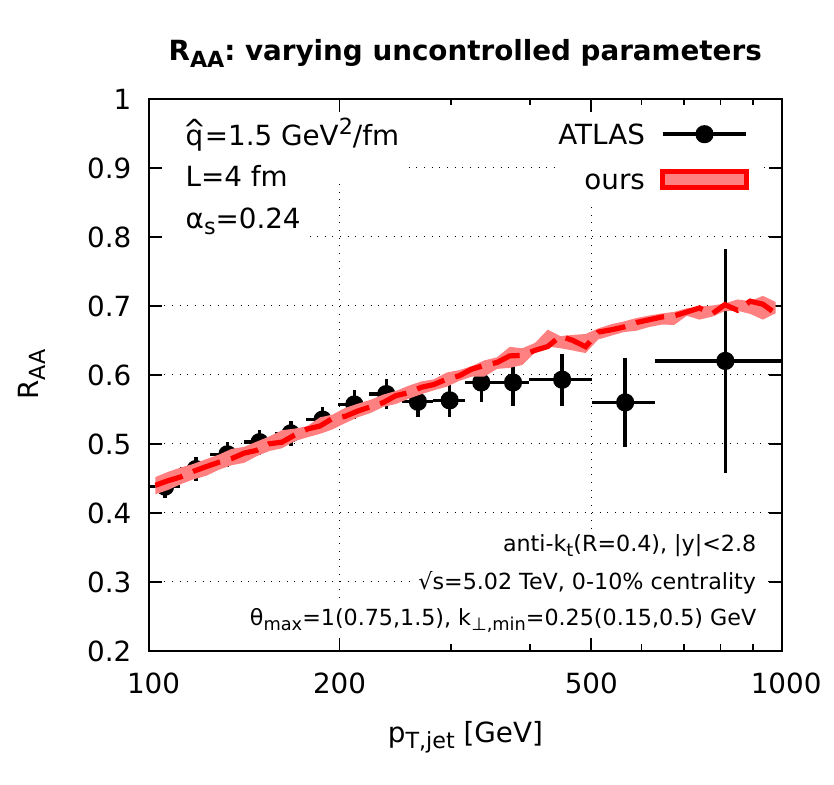}\hfill
  \includegraphics[width=0.48\textwidth]{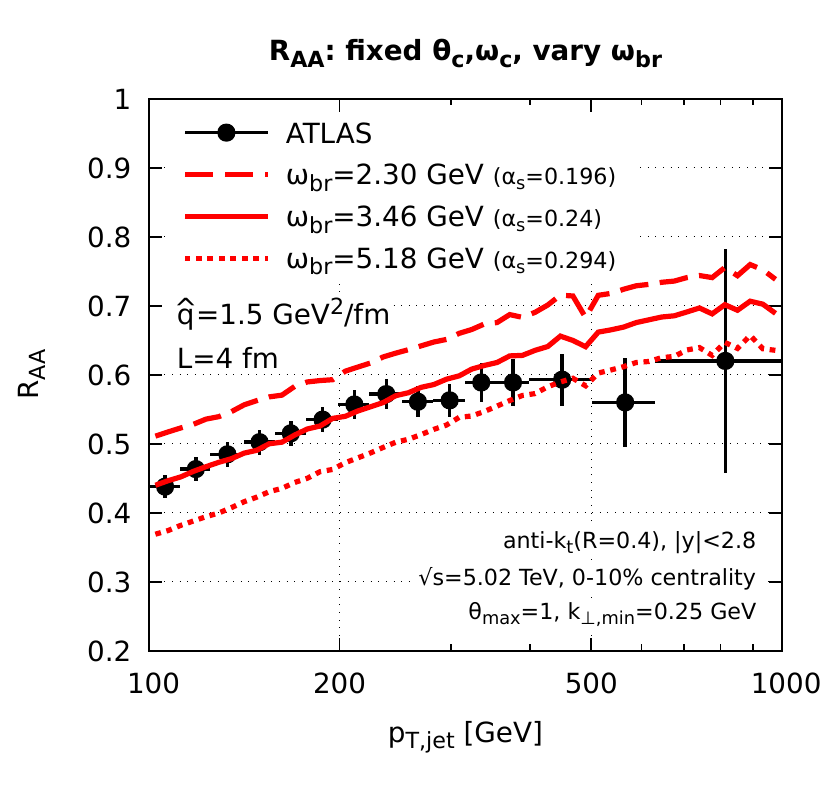}
  \caption{\small Our MC predictions for $R_\text{AA}$ are compared with the results
  of an experimental analysis by ATLAS \cite{Aaboud:2018twu} (shown as dots with error bars).
  Left: the sensitivity of our results to changes in the kinematic cuts $\theta_{\rm max}$ and $k_{\perp,\text{min}}$. Right: the effect of varying $\obr$ (by $\pm 50\%$) at fixed values for $\oc$ and $\theta_c$.}
  \label{Fig:RAA}
\end{figure}

\begin{figure}[t] 
  \centering
  \includegraphics[width=0.48\textwidth]{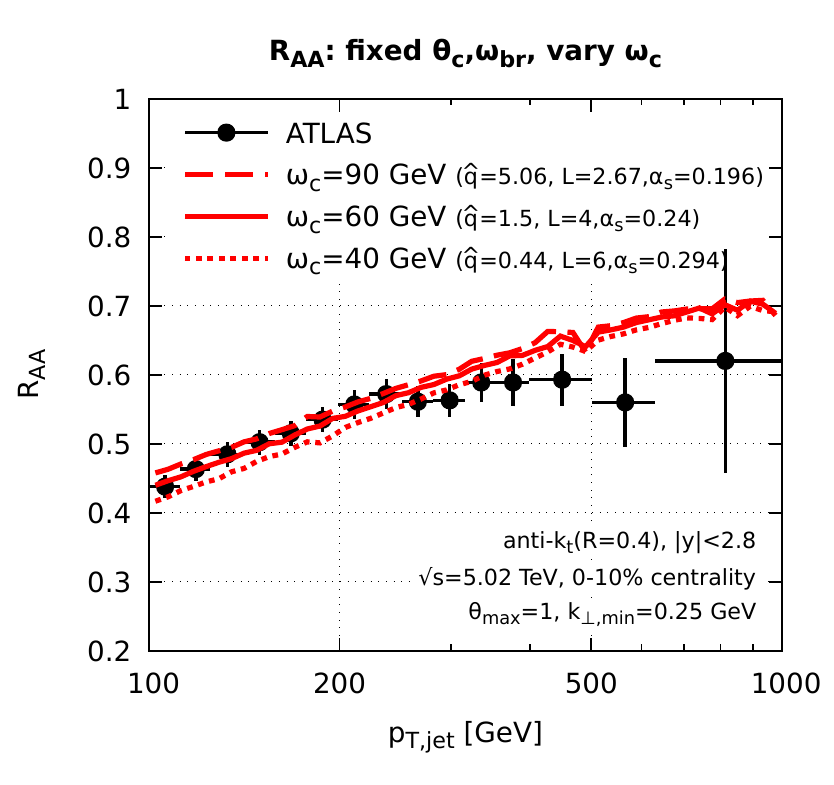}\hfill
  \includegraphics[width=0.48\textwidth]{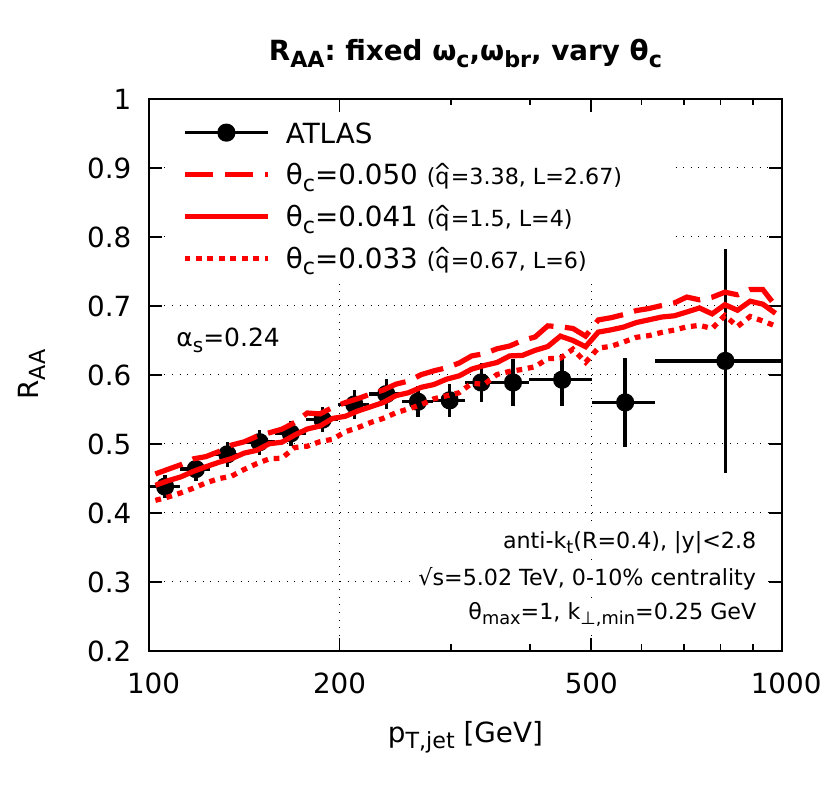}
  \caption{\small  The effects of varying the medium parameters $\hat q$, $L$ and $\amed $ in such a way to keep a constant value for $\obr$ (the same as the central value in  the right plot in Fig.~\ref{Fig:RAA}, i.e. $\obr=3.46$~GeV). Left: we vary $\oc$ by $\pm 50\%$ at fixed $\theta_c$. Right: we vary $\theta_c^2$ by $\pm 50\%$ at fixed $\oc$.}
  \label{Fig:variab}
\end{figure}

Figs.~\ref{Fig:RAA}--\ref{Fig:variab} show our predictions together
with the LHC (ATLAS) data \cite{Aaboud:2018twu} as a function of the transverse momentum
$p_{T,{\rm jet}}$ of the jet (that we shall simply denote as $p_T$ in this section).
As discussed in Sect.~\ref{sec:MC}, our calculation involves 5 free
parameters: the 3 ``physical'' parameters $\hat q$, $L$ and $\amed$
which characterise the medium properties and 2 ``unphysical''
parameters $\theta_{\rm max}$ and $k_{\perp,\text{min}}$ which
specify the boundaries of the phase-space for the perturbative parton
shower. Our aim is to study the dependence of our results under
changes of these parameters.

The first observation, visible in Fig.~\ref{Fig:RAA}~(left) is that the
$R_{AA}$ ratio appears to be very little sensitive to variations of
the ``unphysical'' parameters $\theta_{\rm max}$ and $k_{\perp,\text{min}}$.
Although our results for the inclusive jet spectrum do depend on these
parameters,\footnote{In particular on $\theta_\text{max}$ as the parton
  shower would generate collinear logarithms of $\theta_\text{max}/R$
  to all orders.} the impact on $R_\text{AA}$ remains well within the
experimental error bars when changing $\theta_{\rm max}$ by a factor
2 and  $k_{\perp,\text{min}}$ by a factor larger than 3.
 
In Fig.~\ref{Fig:RAA} right and in the two plots of
Fig.~\ref{Fig:variab} we fix the ``unphysical'' parameters and vary
the medium ones.
The variations are done, following Table~\ref{tab:parameters}, so as
to keep two of the three physical scales $\obr$, $\oc$ and $\theta_c$
fixed while varying the third. It is obvious from the figures that
$R_\text{AA}$ is most sensitive to variations of $\obr$
(Fig.~\ref{Fig:RAA}, right) and shows only a small dependence on $\oc$
and $\theta_c$.
This is in perfect agreement with the expectations from
section~\ref{sec:energyloss-mies} that the jet energy loss is mostly
driven by the scale $\obr$.
Furthermore, the small variations of $R_\text{AA}$ with changes in $\oc$
and $\theta_c$ can be attributed to the slight change in the
phase-space for VLEs leading to a corresponding change in the number
of sources for energy loss (see section~\ref{sec:energyloss-full} and,
in particular, \eqn{logeloss}).

One remarkable fact about the LHC measurements is the
fact that  $R_\text{AA}$ increases very slowly with the jet
$p_T$. This implies that the jet energy loss $\mathcal{E}_{\rm{jet}}$
must itself increase with $p_T$ to avoid a fast approach of
$R_\text{AA}$ towards unity. In our picture, such an increase is indeed
present, as manifest in Fig.~\ref{Fig:jeteloss-v-pt}, and is associated with the steady 
rise of the phase-space for VLEs leading to an increase in the 
number of sources for MIEs (cf.~\eqn{logeloss}).

\section{$z_g$ distribution for monochromatic jets}
\label{sec:zgmed}

We now turn to a discussion of the $z_g$-distribution in the medium.
Our purpose is not only to present Monte Carlo simulations, but also
to identify the main mechanisms responsible for the various features
seen in the simulations.
We follow the same strategy as in the previous section, namely
we start with ``monochromatic'' jets initiated by a parton of fixed
flavour and $p_{T0}$, a case which is easier to discuss analytically,
before we turn (in Sect.~\ref{sec:data}) to the full $z_g$ distribution including the hard
process, for  which it makes sense to compare our results with the LHC data.

\subsection{General definitions and $z_g$ in the vacuum}

For completeness, we first recall the definition of the soft drop (SD)
procedure~\cite{Larkoski:2014wba}. For a given jet of radius $R$, SD
first reclusters the constituents of the jet using the
Cambridge/Aachen (C/A)
algorithm~\cite{Dokshitzer:1997in,Wobisch:1998wt}. The ensuing jet is
then iteratively declustered, i.e.\ the last step of the pairwise
clustering is undone, yielding two subjets of transverse momenta
$p_{T1}$ and $p_{T2}$ separated by a distance
$\Delta R_{12}=\sqrt{\Delta y_{12}^2+\Delta \phi_{12}^2}$ in
rapidity-azimuth.
This procedure stops when the SD condition is met, that is when
\beq\label{zgdef}
z_{12}\equiv \frac{\mathrm{min}(p_{T1},p_{T2})}{p_{T1}+p_{T2}} > \zc\left( \frac{\Delta R_{12}}{R} \right)^{\beta},
\eeq
where $\zc$ and $\beta$ are the SD parameters.
If the condition is not satisfied, the subjet with the smaller $p_T$
is discarded and the declustering procedure continues with the harder.
For $\beta=0$, which is what we adopt from now on, the SD procedure
coincides with the modified MassDrop Tagger~\cite{Dasgupta:2013ihk}.

With the above procedure, $\theta_g$ and $z_g$ are defined
respectively as $\Delta R_{12}$ and $z_{12}$ for the declustering
which satisfied the SD condition. When the declustering procedure
reaches a single parton, we set $\theta_g$ and $z_g$ to zero.
Furthermore, one can impose a lower cutoff
$\theta_g>\theta_\text{cut}$. This is commonly used for PbPb
collisions at the LHC and is thus our default as well.
We then study the differential $z_g$ distribution for a
jet initiated by a parton of flavour $i$ (quark or gluon).
We can consider two possible normalisation for the $z_g$
distribution: the ``self-normalised'' distribution, $p_i(z_g)$, and
the ``$N_\text{jets}$-normalised'' distribution, $f_i(z_g)$. The former
defined such that
\begin{equation}\label{pinorm}
\int_{\zc}^{1/2}\rmd z_g \,p_{i}(z_g) = 1.
\end{equation}
which is equivalent to normalising the $z_g$ distribution to the
number of jets which pass the SD condition and the optional cut on
$\theta_g$.
The latter is instead normalised to the total number of jets,
i.e.\ the normalisation includes jets which fail either the SD
condition or cut on the $\theta_g$.

We first recall the basic result for the $z_g$-distribution in the vacuum~\cite{Larkoski:2015lea}.  The double differential probability for bremsstrahlung starting with a parton of type $i\in\{\text{q,\,g}\}$   reads
\begin{equation}
\label{Pvac}
\rmd^2 \mathcal{P}_{i, \text{vac}}(z,\theta)= \frac{2C_i\alpha_s(zp_{T}\theta_g)}{\pi}\,\bar{P}_{i}(z)\,\rmd z\,
\frac{\rmd\theta}{\theta}\,\equiv \,\mathcal{P}_{i,\text{vac}}(z,\theta)\, \rmd\theta\rmd z,
\end{equation}
where  $\bar{P}_{i}(z)$ is the
symmetrised splitting function of a parton of type $i$.
The argument of the coupling is the relative transverse momentum of
the emission w.r.t.\ the parent parton.
We can also introduce the ``Sudakov factor'' $\Delta_i(R,\theta_g)$, which is the probability to have no emission at any angle between $\theta_g$ and $R$ and with any splitting fraction $z\ge \zc$:
\begin{equation}
\label{Ddef}
\Delta_i(R,\theta_g)=\exp\left(-\int_{\theta_g}^{R}{\rmd\theta}\int_{\zc}^{1/2}\rmd z\,\,\mathcal{P}_{i,\, \text{vac}}(z,\theta)\right)\,.
\end{equation}
The $z_g$-distribution is obtained by considering the probability for
both $z_g$ and $\theta_g$, marginalised over $\theta_g$. The former is
simply expressed as the probability to have no branching between
$\theta_g$ and $R$ times the probability of a branching with
$\theta=\theta_g$ and $z=z_g$, so that
\begin{equation}\label{pzgthetag}
  p_{i, \text{vac}}(z_g)=\frac{\Theta(z_g-z_{\text{cut}})}
  {1-\Delta_i(R,\thetacut)}\int_{\thetacut}^{R}\rmd\theta_g
\,\mathcal{P}_{i, \text{vac}}(z_g,\theta_g) \Delta_i(R,\theta_g),
\end{equation}
where we have included an optional cut $\theta_g>\thetacut$. The
overall factor $(1-\Delta_i)^{-1}$ enforces the normalisation condition
\eqref{pinorm}. It would be equal to one in the absence of the minimal
angle $\thetacut$.
This also means that $f_i(z_g)$ coincides with $p_i(z_g)$ in the limit $\thetacut\to 0$.

In this context, it is worth pointing out that, in the limit
$\theta_\text{cut}\to 0$, $z_g$ is a peculiar observable from the point
of view of perturbative QCD. Indeed, while Eq.~\eqref{pzgthetag} is
overall finite, its expansion at any finite order of perturbation
theory is collinearly divergent, due to the fact that
 $\mathcal{P}_{i, \text{vac}}(z_g,\theta_g)$
diverges when $\theta_g\to 0$. It is only after an all-order
resummation that the exponential form of the Sudakov regulates the
divergence. In other words, although $z_g$ is collinear unsafe, it is
Sudakov safe~\cite{Larkoski:2015lea}.

To discuss the physics underlying the $z_g$ distribution, it is sometimes helpful
to consider the fixed-coupling approximation. One can then easily
perform the angular integration in~\eqn{pzgthetag} and get
\begin{equation}
\label{pzg_fixed}
 p_{i,{ \text{vac}}}(z_g)=\frac{\bar{P}_{i}(z_g)}{\int_{z_{\text{cut}}}^{1/2}\bar{P}_{i}(z)dz}\Theta(z_g-z_{\text{cut}}).
\end{equation}
This makes it clear that the $z_g$-distribution provides a direct measurement of the splitting function.

\subsection{In-medium $z_g$ distribution: Monte Carlo results}

We now present our Monte Carlo results for the $z_g$-distribution
created by ``monochromatic'' jets which propagates through the
quark-gluon plasma. We focus on the \njets-normalised distribution
$f_{i, \text{med}}(z_g)$, which carries more information. We define
the corresponding nuclear modification factor,
$\mathcal{R}_i(z_g)\equiv f_{i, \text{med}}(z_g)/f_{i,
  \text{vac}}(z_g)$. Similarly we define
$\mathcal{R}_i^\text{(norm)}(z_g)\equiv p_{i,
  \text{med}}(z_g)/p_{i,\text{vac}}(z_g)$ as the nuclear modification
factor of the self-normalised $z_g$ distributions.
We study four different values for the initial $p_{T0}$ spanning a
wide range in $p_{T0}$, from $100$~GeV to $1$~TeV.  We use the same SD
parameters as in the CMS analysis~\cite{Sirunyan:2017bsd}, namely
$\beta=0$ and $\zc=0.1$, together with a cut $\theta_g>\thetacut=0.1$.
In this section we mostly highlight the main features of our Monte
Carlo simulations and provide a brief physical interpretation. More
detailed analytic calculations are postponed to
sections~\ref{sec:high} for high-energy jets and~\ref{sec:lowpt} for
lower-energy jets.

\begin{figure}[t] 
  \centering
  \includegraphics[page=2,width=0.48\textwidth]{./FIGS/zg-parton-ptdep}\hfill
  \includegraphics[page=1,width=0.48\textwidth]{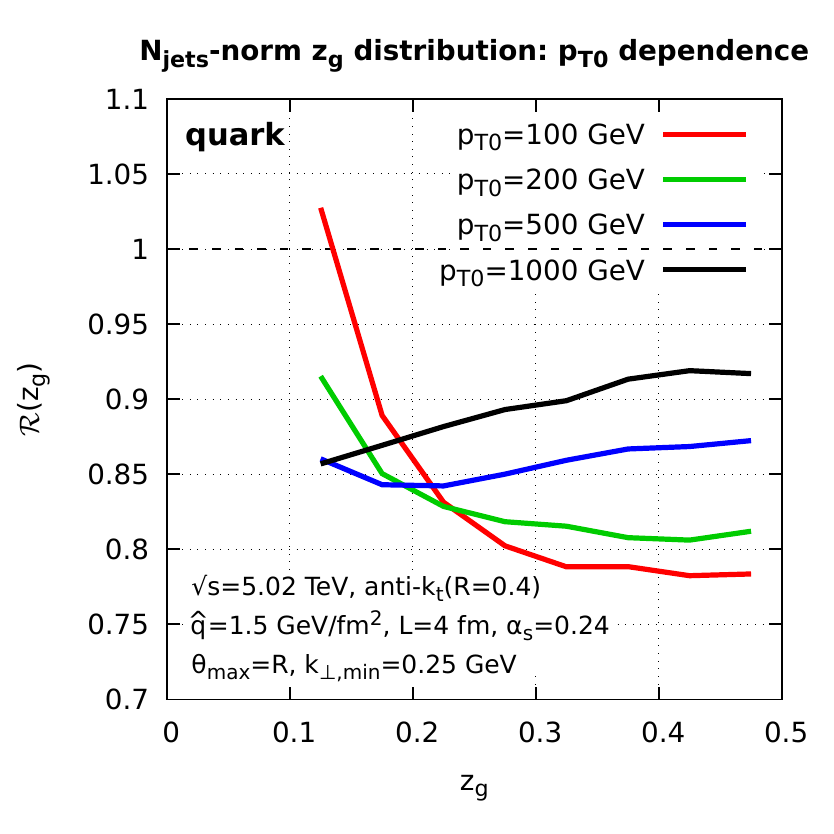}
  \caption{\small A summary of our MC results  for the medium/vacuum ratio $\mathcal{R}_i(z_g)$ of the 
$N_\text{jets}$-normalised distributions, for monochromatic jets initiated either by a gluon (left figure), or by a quark (right figure), and for 4  different values for the initial transverse momentum $p_{T0}$. }
  \label{Fig:zgvar}
\end{figure}

Our results are shown in Fig.~\ref{Fig:zgvar}, separately for jets
initiated by a gluon (left plot) and by a quark (right plot), using
our default MC parameters (cf.\ Table~\ref{tab:parameters}).
As for our study of energy loss for monochromatic jets in
section~\ref{sec:eloss}, we set the angular cutoff scale of our Monte
Carlo to $\theta_\text{max}=R$ with $R=0.4$ the jet radius.
Each of the plots in Fig.~\ref{Fig:zgvar} show qualitatively different
behaviours between our lowest $p_{T0}$ (100~GeV) value and the largest
one (1~TeV).
More precisely, for the highest energy jets, $p_{T0}=1$~TeV, the ratio
$\mathcal{R}(z_g)$ is always smaller than one, indicating a nuclear
suppression, and it increases monotonously with $z_g$, meaning that
the nuclear suppression is larger at small $z_g$.
Conversely, for lower $p_{T0}$, while the nuclear suppression becomes
stronger at large $z_g$, a peak develops at small $z_g$ where
$\mathcal{R}(z_g)$ can even become larger than one, indicating a
nuclear {\em enhancement}.

\begin{figure}[t] 
  \centering
  \includegraphics[width=0.45\textwidth]{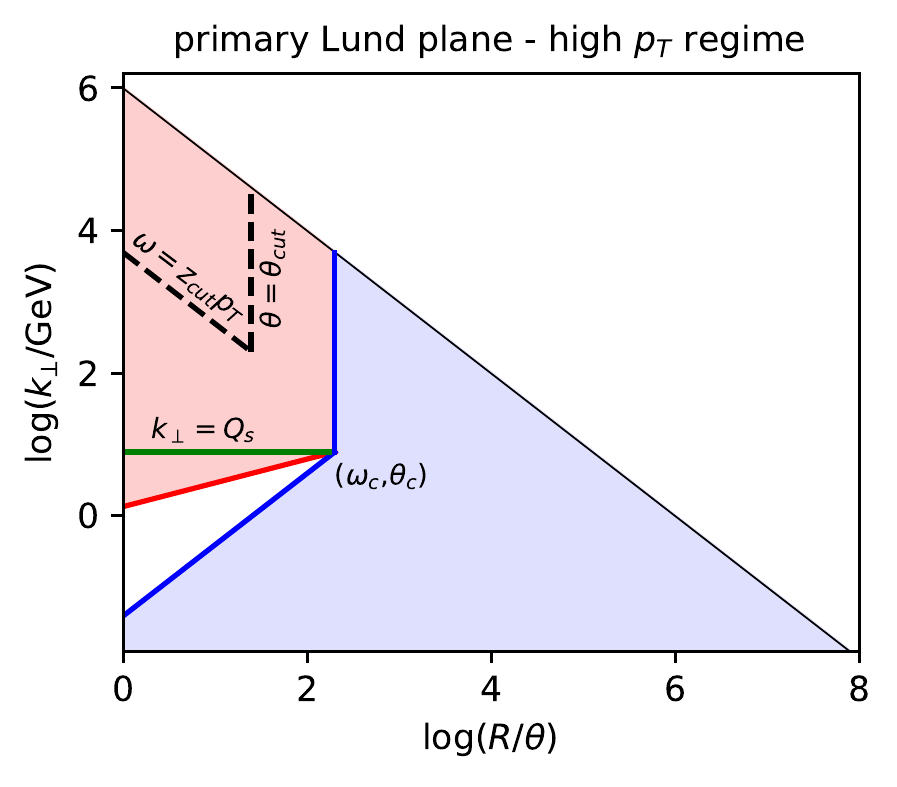}\hfill
  \includegraphics[width=0.45\textwidth]{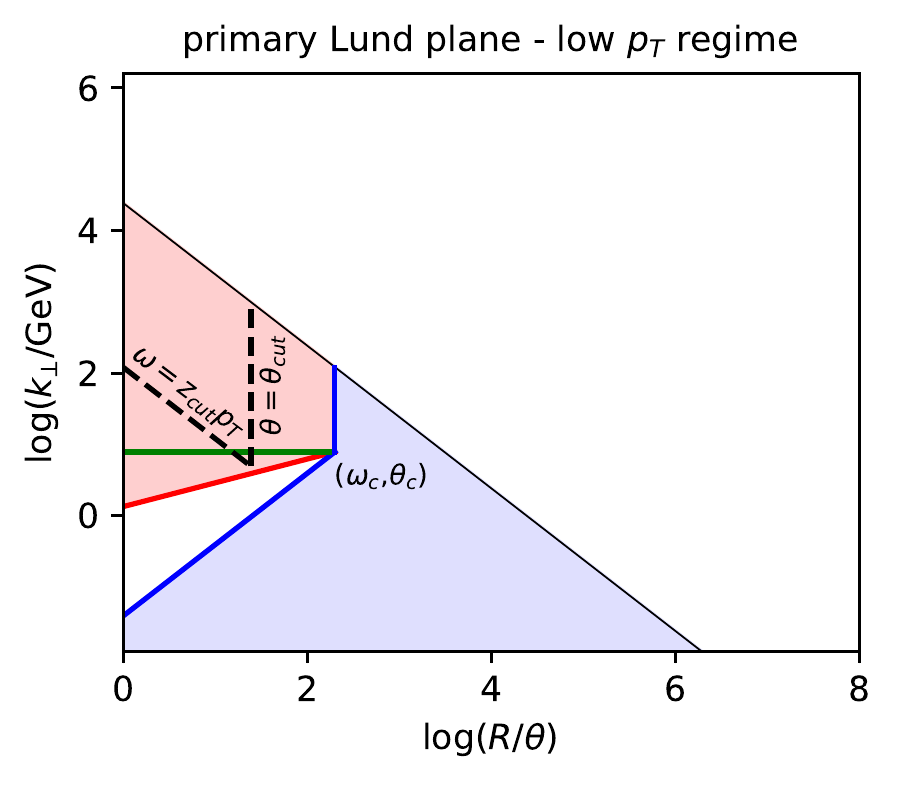}
  \caption{\small  The kinematic regions in the Lund plane that are covered by the SD algorithm
  in the case of a high-energy jet ($\zc p_T >\oc$) in the left figure and of a low-energy jet ($\zc p_T <\oc$) in the
  right figure. As suggested by the pictorial representation in the right figure, the most interesting situation for ``low-energy jets'' is such that there is only little overlap
  between the kinematic region for SD and the phase-space for MIEs.}
  \label{Fig:zgLund}
\end{figure}

Let us first discuss the behaviour at large $p_{T0}$, focusing on
$p_{T0}=1$~TeV.
In this case, the softest radiation that can be captured by Soft Drop
has an energy $\zc p_{T0}=100$~GeV which is still larger than the
hardest medium-induced emissions which have energies
$\omega\sim\oc=60$~GeV. Hence, for jets with high-enough $p_{T0}$, SD
can only select vacuum-like emissions. To illustrate this, we show in
the left plot of Fig.~\ref{Fig:zgLund} the phase-space selected by
SD. Under these circumstances, the only nuclear effect on the
$z_g$-distribution is the energy lost by the two (hard, $z_g > \zc$)
subjets passing the SD condition.
Due to this energy loss, the {\it
  effective} splitting fraction $z_g$ measured by SD turns out to be 
slightly smaller than the {\it physical} splitting fraction $z$ at the
branching vertex (see Sect.~\ref{sec:high} for details).
If we call for now this shift $\Delta z=z-z_g > 0$, we have
(cf.~(\ref{pzg_fixed}))
\beq
\mathcal{R}(z_g) \approx \frac{\bar{P}_g(z=z_g+\Delta z)}{\bar{P}_g(z_g)}\,\simeq\,\frac{z_g}{z_g+\Delta z}\,\simeq\,1-\frac{\Delta z}{z_g}\quad
\mbox{for \ $\Delta z\ll z_g\ll 1$}\,,
\eeq
which explains the pattern (smaller than one and increasing with
$z_g$) observed at large $p_{T0}$ in Fig.~\ref{Fig:zgvar}.

The above discussion also suggests what changes when moving to the
opposite regime of (relatively) low energy jets, say
$p_{T0}=100$~GeV. In this case, the energy interval covered by SD,
that is $\omega$ between $\zc p_{T0}=10\,{\rm GeV}$ and
$p_{T0}/2=50$~GeV, fully overlaps with the BDMPS-Z spectrum for
medium-induced radiation which has $\omega\lesssim \oc=60$~GeV (cf.\
the right plot of Fig.~\ref{Fig:zgLund}). Consequently, the SD
condition can now be triggered either by a vacuum-like splitting, or
by a medium-induced one. Since the BDMPS-Z spectrum increases rapidly
at small $z$ (faster than the vacuum splitting function), this
naturally explains the peak in the ratio  $\mathcal{R}(z_g)$ at small
$z_g$, visible in Fig.~\ref{Fig:zgvar}.
The nuclear suppression observe at large $z_g$ suggests that in this
region, the energy loss effects dominate over the BDMPS-Z emissions.

The above arguments show that the $z_g$ distribution is best discussed
separately for high-energy and low-energy jets, where the separation
between the two regimes is set by the ratio $\zc p_{T0}/\oc$. The
high-energy jets, for which $p_{T0}>\oc/\zc$, are conceptually simper
as the in-medium $z_g$ distribution is only affected by the energy
loss via MIEs. For low-energy jets, i.e.\ jets with $p_{T0}< \oc/\zc$,
the $z_g$ distribution is affected by the medium both {\it directly}
when the SD condition is triggered by a MIE, and {\it indirectly} via the
energy loss of the two subjets emerging from the hard splitting.
This second case is more complex for a series of reasons and notably
because MIEs do not obey angular ordering.

\begin{figure}[t] 
  \centering
  \includegraphics[width=0.48\textwidth]{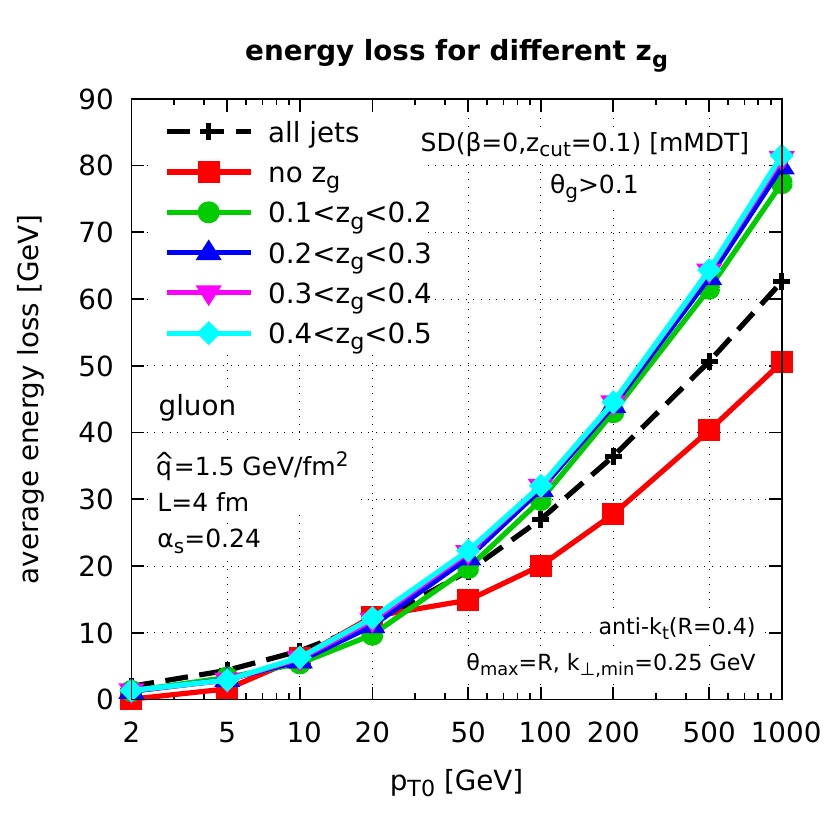}
 \caption{\small Our MC results for the average energy loss by a gluon-initiated jet are displayed
 as a function of the initial gluon energy $p_{T0}$ in bins of $z_g$. The inclusive (all jets) result  is also shown, by the dashed line.}
 \label{Fig:elosszg}
\end{figure}

Since $z_g$ is intrinsically tied to energy loss effects, it is
interesting to study how the average jet energy loss correlates with
$z_g$.
Our numerical results are presented in Fig.~\ref{Fig:elosszg}.
The dashed curve shows the MC results for the inclusive jets (all
values of $z_g$), the one denoted ``no $z_g$'' refers to jets which
did not pass the SD criterion or failed the cut $\theta_g>\thetacut$,
and the other curves correspond to different bins of $z_g$.
One clearly sees a distinction between the ``no $z_g$'' jets, which
lose much less energy than the average jet, and those which passed SD,
whose energy loss is larger than the average and quasi-independent of
$z_g$.
The main reason for this behaviour is that jets passing the SD
condition are effectively built of two relatively hard subjets. Since
the angular separation between these two hard subjets is larger than
$\thetacut=0.1 > \theta_c\simeq 0.04$, they lose energy (via MIEs) as
two independent jets, giving a larger-than-average energy loss. This
is mostly controlled by the geometry of the system, with only a
limited sensitivity to the precise sharing of the energy between the
subjets. On the other hand, the jets which did {\it not} pass SD are
typically narrow one-prong jets with either no hard substructure
or with some substructure at an angle smaller that $\thetacut=0.1$
(i.e.\ at an angle $\simeq \theta_c$). These jet therefore lose less
energy that an average jet.
The fact that the angular cutoff $\thetacut=0.1$ is close to the
critical value $\theta_c\simeq 0.04$ is clearly essential for the
above arguments.

A last comment concerns the difference between the $z_g$-distributions
for gluon- and quark-initiated jets, as shown in the left and right
plots of Fig.~\ref{Fig:zgvar}, respectively. The deviation of the
medium/vacuum ratio from unity appears to be larger for quark jets
than for gluon jets. This might look surprising at first sight given
that the average energy loss is known to be larger for the gluon jet
than for the quark one (cf.\ Figs.~\ref{Fig:jeteloss-v-pt}
and~\ref{Fig:jeteloss-v-R}). However we will show in
section~\ref{sec:high} that the $z_g$ distribution is mostly
controlled by the energy loss of the softest subjet, which is
typically a {\it gluon} jet even when the leading parton is a
quark. The difference between quark and gluon jets in
Fig.~\ref{Fig:zgvar} is in fact controlled by ``non-medium'' effects,
like the difference in their respective splitting functions.

\subsection{Analytic insight for high-energy jets: VLEs and energy loss}
 \label{sec:high}
 
We begin our analytic calculations of the nuclear effects on the
$z_g$ distribution with the case of a high-energy jet,
$p_{T0}>\oc/\zc$
In this case, the SD condition is triggered by an in-medium VLE that
we call the ``hard splitting'' in what follows. This splitting occurs
early (since $\tf\ll L$, cf. \eqn{tfvac}) and the daughter partons
propagate through the medium over a distance of order $L$. During
their propagation, they evolve into two subjets, both via VLEs (which
obey angular ordering, except possibly for the first emission outside
the medium) and via MIEs (which can be emitted at any angle).

Since the C/A algorithm is used by the SD procedure, both
subjets have an opening angle of order\footnote{On average, the two
  subjets have an (active) area $\approx 0.69\pi\theta_g^2$, while a
  single jet of radius $\theta_g$ has an area
  $\approx 0.81\pi\theta_g^2$, cf.~Fig~8 of
  Ref.~\cite{Cacciari:2008gn}. Each subjet thus have an effective
  radius of order $\sqrt{0.69/0.81}\theta_g\approx 0.92\theta_g$ which
  is very close to $\theta_g$.} $\theta_g$, with $\theta_g$ the angle
of the hard splitting.
Consequently, the emissions from the two subjets with angles larger
than $\theta_g$ --- either MIEs, or VLEs produced outside the medium
--- are {\it not} clustered within the two subjets. Accordingly, their
reconstructed transverse momenta $p_{T1}$ and $p_{T2}$ are generally
lower than the initial momenta, $\omega_1$ and $\omega_2$, of the
daughter partons produced by the hard splitting. This implies a
difference between the {\it reconstructed} splitting fraction
$z_g=p_{T,1}/(p_{T,1}+p_{T,2})$ and the {\it physical} one,
$z\equiv \omega_1/(\omega_1+\omega_2)$. This difference is controlled
by the energy lost by the two subjets.

Let us first mention that the energy loss via VLEs outside the medium
at angles $\theta >\theta_g$ can be neglected. Indeed, since these emissions
have $\tf \sim 1/(\omega\theta^2) > L$, they are soft and only
give very small contributions to the energy loss.
We have checked this explicitly with MC studies of the VLEs {\em
  alone}: we find that the effect on the $z_g$ distribution of the
vetoed region and of the violation of angular ordering for the first
emission outside the medium are much smaller than those associated
with the energy loss via MIEs.

We then discuss the role of {\it colour coherence} for the energy loss
via MIEs. If the splitting angle $\theta_g$ is smaller than $\theta_c$
the daughter partons are not discriminated by the medium. This is a
case of {\em coherent} energy loss where the MIEs at angles
$\theta>\theta_g$ are effectively sourced by their parent
parton~\cite{MehtarTani:2010ma,MehtarTani:2011tz,CasalderreySolana:2011rz,CasalderreySolana:2012ef},
so that $z=z_g$.
On the other hand, for larger splitting angles
$\theta_g \gg \theta_c$, the colour coherence is rapidly washed out, so
the two daughter partons act as independent sources of MIEs. In this
case, one can write
$\omega_i=p_{Ti}+ \mathcal{E}_i(\omega_i, \theta_g)$, where $i=1,2$
and $\mathcal{E}_i(\omega_i, \theta_g)$ is the average energy loss for a
jet of flavour $i$, initial energy $\omega_i$ and opening angle
$\theta_g$ (cf.\ e.g.~\eqn{logeloss}).
It would be relatively straightforward to deal with generic values of
$\theta_g$, both coherent and incoherent. In practice, all existing
measurements at the LHC imposes a minimal angle
$\theta_g \ge \thetacut$, with $\thetacut=0.1$. Since
$\theta_\text{cut}$ is larger than $\theta_c$ for all our choices of
parameters, we only consider the incoherent case
$\theta_g>\theta_\text{cut}$ in what follows.

That said, the relation between the measured $z_g$ and the physical
splitting fraction $z$ can be written as (assuming $p_{T1}<p_{T2}$)
\begin{equation}\label{zinzg}
  z_g
  =\frac{p_{T1}}{p_{T1}+p_{T2}}
  = \frac{zp_T-\mathcal{E}_1(zp_T, \theta_g)}{p_T-\mathcal{E}_1(zp_T, \theta_g)
    -\mathcal{E}_2((1-z)p_T, \theta_g)}
  \,\equiv\, \mathcal{Z}_g(z, \theta_g)\,,
\end{equation}
where $p_{T}\equiv \omega_1+\omega_2$ is the energy (or transverse
momentum) of the parent parton at the time of the ``hard''
branching. In what follows we approximate $p_{T}\simeq p_{T,0}$. This
is valid as long as one can neglect two effects: \texttt{(i)} the
transverse momentum of the partons which have been groomed away during
previous iterations of the SD procedure, and \texttt{(ii)} MIEs prior
to the hard branching. The former is indeed negligible as long as we
work in the standard limit $\zc\ll 1$, and the latter is also
negligible based on our short formation time arguments in
section~\ref{sec:factorisation-vles}.

For a given average energy loss $\mathcal{E}(p_{T}, \theta_g)$ one
can, at least numerically, invert \eqn{zinzg} to obtain the physical
splitting fraction $z=\mathcal{Z}(z_g, \theta_g)$ corresponding to the
measured $z_g$. The kinematic constraint $z_g>\zc$ thus implies a
constraint on $z$,
$z\ge \mathcal{Z}(\zc, \theta_g)$,\footnote{Here we assume
  that $\mathcal{Z}_g(z, \theta_g)$ is a monotonously increasing
  function of $z$.} and the in-medium $z_g$ distribution in this
high-energy regime becomes a straightforward generalisation of
\eqn{pzgthetag}
\begin{equation}
 \label{zghighpt}
  p_{i, \text{med}}(z_g)= \mathcal{N}
 \int_{\thetacut}^{R}\rmd\theta_g\,
\Delta_i^{\text{VLE}}  (R,\theta_g)
  \int_0^{1/2}\rmd z\, \mathcal{P}_{i,\text{vac}}(z,\theta_g)\,\delta\big(z_g\minus \mathcal{Z}_g(z, \theta_g)\big)
   \Theta(z_g\minus\zc),
\end{equation}
where the Sudakov factor is formally the same as in the vacuum,
\eqn{Ddef}, but with the new, medium-dependent, lower limit
$\mathcal{Z}(\zc, \theta)$ on $z$:
\begin{equation}\label{DeltaVLE}
\Delta_i^{\text{VLE}}
(R,\theta_g)=\exp\left(-\int_{\theta_g}^{R}{\rmd\theta}\int_{0}^{1/2}\rmd
  z\,\,\mathcal{P}_{i,\text{vac}}(z,\theta)\,
  \Theta\big(z-\mathcal{Z}(\zc, \theta)\big)\right)\,.
\end{equation}
The normalisation factor $\mathcal{N}$ in~(\ref{zghighpt}) is given by
$(1-\Delta_i^{\text{VLE}})^{-1}$. The \njets-normalised distribution
$f_{i, \text{med}}(z_g)$ is obtained by simply removing this factor
$\mathcal{N}$.

In practice we will replace $\theta_g$ by $R$ in the argument of the
energy loss. This is motivated by two facts.
Firstly, due to the SD procedure, we know that the jet is free of
emissions with $\omega>\zc p_{T0}$ at angles between $\theta_g$ and
$R$, simply because such an emission would have triggered the
SD condition. The remaining emissions between $\theta_g$ and $R$ are
therefore soft and we neglect them.
Secondly, the angular phase-space $\thetacut < \theta < R$ is
relatively small and $\mathcal{E}$ is slowly varying over this domain.
With this approximation, both $\mathcal{Z}_g$ and $\mathcal{Z}$
becomes independent of $\theta_g$ and the Sudakov factor
\eqref{DeltaVLE} simplifies to the vacuum one, \eqn{Ddef}, evaluated
at $\zc\to\mathcal{Z}(\zc)$.

The above picture can be further simplified by noticing that the
energy losses are typically much smaller than $zp_T$ and
$(1-z)p_T$. This means that the difference between $z$ and $z_g$ is
parametrically small and we can replace $z$ by $z_g$ in the arguments
of the energy loss in~(\ref{zinzg}) so that 
  \begin{equation}\label{zvszg}
z \equiv \mathcal{Z}(z_g, \theta_g) \simeq\,z_g+\,\frac{\mathcal{E}_1-z_g(\mathcal{E}_1+\mathcal{E}_2)}{p_T}\,,
\end{equation}
with $\mathcal{E}_1\equiv \mathcal{E}_1(z_gp_T)$ and $\mathcal{E}_2\equiv \mathcal{E}_2((1-z_g)p_T)$. 
Since $z_g<1/2$, this shows that the physical $z$ is
typically\footnote{Small deviations from this behaviour can happen close to
  $z_g=1/2$ when $\mathcal{E}_2\neq \mathcal{E}_1$. In this limit, our
  assumption that the softer physical parton ($z<1/2$) matches with
  the softer measured subjet ($z_g<1/2$) has to be reconsidered
  anyway.} larger than $z_g$.

As before, it is useful to consider the fixed-coupling scenario where  
the $z_g$-dependence of \eqn{zghighpt} factorises from the integral
over $\theta_g$. After dividing out by the vacuum distribution
$\propto \bar{P}_{i}(z_g)$, we find
\begin{equation}\label{Rdef}
  \mathcal{R}(z_g)
  \equiv \frac{f_\text{med}(z_g)}{f_\text{vac}(z_g)}
  \simeq \mathcal{J}(z_g)\,
  \frac{\bar{P}_g(\mathcal{Z}(z_g))}{\bar{P}_g(z_g)}\,,\qquad\mbox{with}\quad
\mathcal{J}(z_g)\equiv \bigg |\frac{\rmd  \mathcal{Z}(z_g)}{\rmd z_g}
\bigg |
\simeq 1-\frac{\mathcal{E}_1+\mathcal{E}_2}{p_T}\,,
\end{equation}
where $\mathcal{J}$ is a Jacobian and the last equality
in~\eqref{Rdef} is obtained using the simplified expression~(\ref{zvszg}).

At this level, it becomes necessary to specify the energy lost by a
subjet.
At high energy, both $zp_T$ and $(1-z)p_T$ are large and the energy
lost by the subjets is sensitive to the increase in the number of
partonic sources for MIEs (cf.\ section~\ref{sec:energyloss-full} and
Fig.~\ref{Fig:jeteloss-v-pt}).
To test this picture, we consider two energy loss scenarios.
First, the case of an energy loss which captures the increase in the
number of sources for MIEs and increases with the jet $p_T$, as in
Eq.~(\ref{logeloss}).
Since Eq.~(\ref{logeloss}) is not very accurate we will instead use
$\mathcal{E}_j=\mathcal{E}_{j,\text{fit}}$ corresponding to the fit of
the Monte Carlo result shown in Fig.~\ref{Fig:jeteloss-v-pt}.
The second scenario corresponds to what would happen in the absence of
VLEs, i.e.\ when only MIEs from the leading parton in each subjet are
included. This gives an energy loss which saturates to a constant
$\mathcal{E}_j=\varepsilon_j$ at large $p_T$ (see again
Fig.~\ref{Fig:jeteloss-v-pt}).
Clearly, the first scenario is the most physically realistic.

\begin{figure}[t] 
  \centering
  \includegraphics[page=1,width=0.48\textwidth]{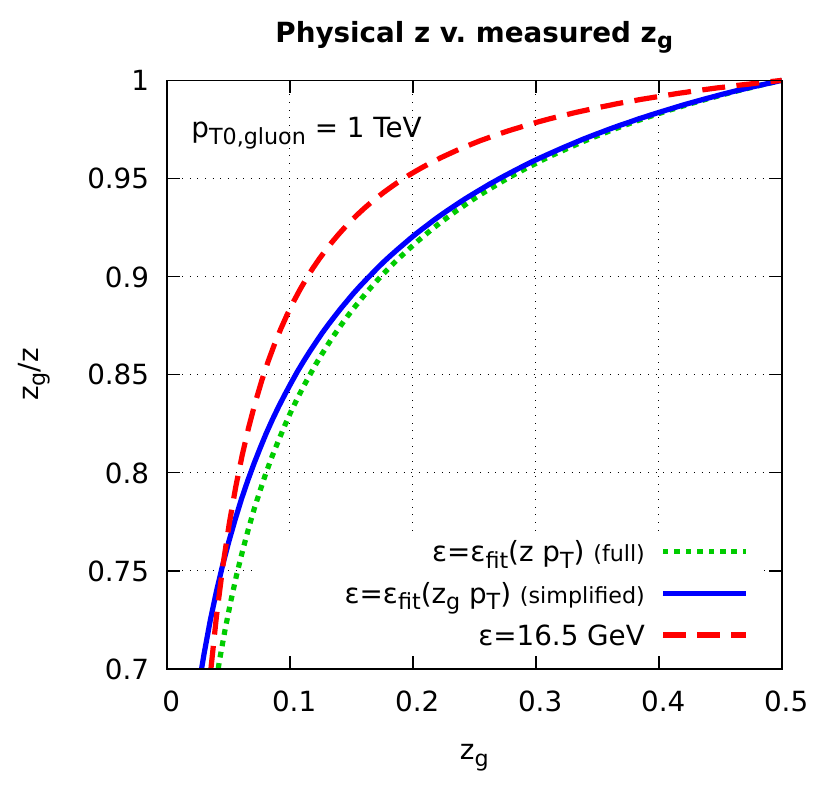}
  \hfill
  \includegraphics[page=2,width=0.48\textwidth]{./FIGS/delta_zg}
  \caption{\small The ratio $z_g/z$ (left figure) and the difference $z-z_g$ (right figure) between the
  splitting fraction $z_g$ measured by SD and the physical splitting fraction $z$ for the hard 
  splitting.  This difference is given by Eq.~(\ref{zinzg}) that we evaluate with
   two scenarios for the energy loss by the subjets: a constant energy loss and a $p_T$-dependent
  one; for the second scenario, we also show the predictions of the simplified relation~(\ref{zvszg}).
   }
\label{Fig:Deltazg} 
\end{figure}

For definiteness, let us first consider the case of a 1-TeV
gluon-initiated jet.\footnote{We only included the dominant partonic
  channel $g\to gg$ in our analytic calculation.}
Fig.~\ref{Fig:Deltazg} shows the relation between the physical
splitting fraction $z$ and the measured $z_g$, with the ratio $z_g/z$
plotted on the left panel and the difference $z-z_g$ on the right
panel. We see that $z$ is larger than $z_g$ in both energy-loss
scenarios. The difference $z-z_g$ decreases when increasing $z$ (at
least for $z>\zc$), while the ratio $z_g/z$ gets close to 1. The
effects are roughly twice as large for the full energy-loss scenario
than for a constant energy loss.
The dotted (green) curve shows the result obtained using the ``full''
relation~(\ref{zinzg}) while the solid (blue) line uses the simplified
version, Eq~(\ref{zvszg}). As expected, they both lead to very similar
results and we therefore make the simplified version our default from
now on.

\begin{figure}[t] 
  \centering
  \includegraphics[page=1,width=0.48\textwidth]{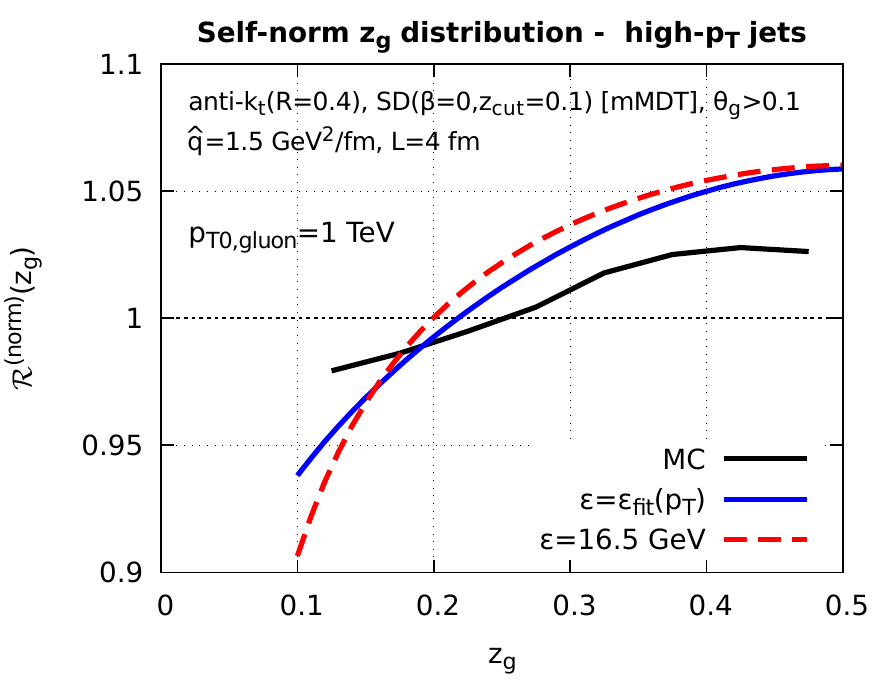}\hfill
  \includegraphics[page=2,width=0.48\textwidth]{./FIGS/plot-highpt-ratio}
 \caption{\small Our MC results for the medium/vacuum ratio of the $z_g$ distributions associated with
 gluon-initiated jets with initial energy $p_{T0}=1$~GeV are compared to our analytic
 predictions for the two scenarios of energy loss described in the text. The left plot refers to the
 self-normalised distributions, cf. \eqn{zghighpt}, and the right plot to the $N_\text{jets}$-normalised ones.} \label{Fig:zghighpt}
\end{figure}

The nuclear modification factor for the $z_g$ distribution obtained
from our analytic calculation~(\ref{zghighpt}), including
running-coupling corrections, is shown in Fig.~\ref{Fig:zghighpt} for
both the self-normalised distribution $p_\text{med}(z_g)$ (left) and the
$N_\text{jets}$-normalised one $f_\text{med}(z_g)$ (right).
We see that $\mathcal{R}(z_g)$ increases with $z_g$. This is expected
since, at small $z_g$, $\bar{P}_g(z)/\bar{P}_g(z_g)\simeq z_g/z$ which
increases with $z_g$ (see e.g.\ Fig.~\ref{Fig:Deltazg}, left).
Furthermore, the medium/vacuum ratio of the $N_\text{jets}$-normalised distributions
(Fig.~\ref{Fig:zghighpt}, right) is always smaller than one.
With reference to the fixed-coupling estimate in Eq.~(\ref{Rdef}), 
this is a combined effect of $z$ being larger than $z_g$, hence
$\bar{P}_g(z)<\bar{P}_g(z_g)$, and of the extra Jacobian in front of
Eq.~(\ref{Rdef}). 

Regarding the comparison between the two scenarios for the energy loss, we see that
although they produce similar results for the self-normalised $z_g$
distribution, the (physical) ``full jet'' scenario predicts a larger
suppression than the ``constant'' one for the medium/vacuum ratio of the 
$N_\text{jets}$-normalised distributions.
In particular, the former predicts a value for $\mathcal{R}(z_g)$ 
which remains significantly smaller
than one even at $z_g$ close to 1/2.
This behaviour is in also in better agreement with our Monte Carlo simulations.
Generally speaking, it is worth keeping in mind that the
\njets-normalised ratio is better suited to disentangle between different
energy-loss models than the self-normalised ratio which is bound to cross one
by construction.

Since the nuclear modification of the $z_g$ distribution appears to be
so sensitive to the energy loss, it is interesting to check whether
this observable follows the Casimir scaling of the jet energy loss.
We show that this is not the case and that the nuclear modification
is even slightly larger for quark than for gluon jets.
In practice, the $z_g$ distribution is controlled by the energy loss
of the softest among the two subjets created by the hard splitting,
which is typically a gluon independently of the flavour of the initial
parton. Let us then consider \eqn{zvszg} in which
we take $\mathcal{E}_1= \mathcal{E}_{\text g}$ and
$\mathcal{E}_2=\mathcal{E}_R$, with $R={\text q}$ or ${\text g}$ depending on the
colour representation of the leading parton. Simple algebra yields
\beq\label{zGszQ}
z^{(\text{g-jet})}(z_g)\,\simeq\,z^{(\text{q-jet})}(z_g)+z_g\,\frac{\mathcal{E}_{\text g}-\mathcal{E}_{\text q}}{p_T}
\eeq
where $z^{(R\text{-jet})}(z_g)$ is the physical splitting fraction $z$
corresponding to a measured fraction $z_g$ for the case of a leading
parton of flavour $R$, and the energy loss functions $\mathcal{E}_R$
are evaluated at $(1-z_g)p_T$.
Since $\mathcal{E}_{\text g}\simeq  2\mathcal{E}_{\text q}$ the second term
in~\eqref{zGszQ} is positive and thus $z^{(\text{g-jet})}(z_g) >z^{(\text{q-jet})}(z_g)$
as expected on physical grounds. Yet, the difference between
$z^{(\text{g-jet})}(z_g)$ and $z^{(\text{q-jet})}(z_g)$ is weighted by $z_g$,
hence it suppressed at
    small $z_g$, where the energy loss effects should be more important.
Furthermore, the effects of the difference $z^{(\text{g-jet})}(z_g)-
z^{(\text{q-jet})}(z_g)$ are difficult to distinguish in practice
since there are other sources of differences between the  $z_g$
distributions of quark and  gluon jets like the non-singular terms in
the splitting functions and the different Sudakov factors. In practice
these effects appear to dominate over difference between
$z^{(\text{g-jet})}(z_g)$ and $z^{(\text{q-jet})}(z_g)$

\begin{figure}[t] 
  \centering
  \includegraphics[width=0.48\textwidth,page=2]{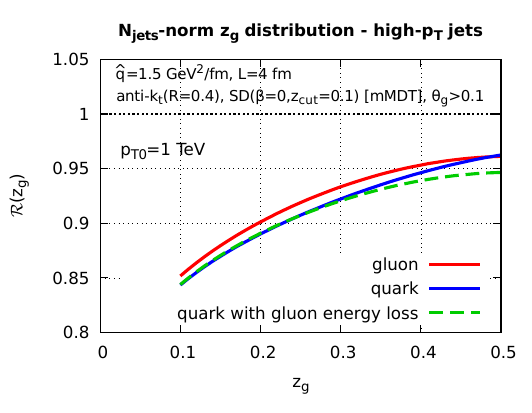}
 \caption{\small The results of the analytic calculations for the medium/vacuum ratio 
 $\mathcal{R}(z_g)$ for both gluon- and quark-initiated jets, together with the fictitious
 case where a quark-initiated  subjet loses the same amount of energy as a gluon-initiated one.}
 \label{Fig:QvsG}
\end{figure}

This is confirmed by our analytic calculations in
Fig.~\ref{Fig:QvsG}. Together with our previous results for a
gluon jet, we show two scenarios for quark jets: \texttt{(a)} a realistic
scenario which takes into account the different quark and gluon energy
losses (cf.\ Fig.~\ref{Fig:jeteloss-v-pt}), and  \texttt{(b)} a
fictitious case, which assumes that a quark subjet loses the same
energy as a gluon one, i.e.\ $\mathcal{E}_{\text q}=\mathcal{E}_{\text g}$.
In both cases, the nuclear suppression of the $z_g$ distribution
appears larger for quark-initiated jets than for gluon-initiated
jets, in qualitative agreement with our Monte Carlo findings (recall 
 Fig.~\ref{Fig:zgvar}).
This is clearly driven by effects beyond the energy loss difference
between quarks and gluons (cf our case \texttt{(b)}), even though this
difference has indeed the effect of slightly increasing
$\mathcal{R}(z_g)$, especially close to $z_g=1/2$, as visible by comparing
the curves corresponding to the cases \texttt{(a)} and \texttt{(b)}.

In summary, the main lessons one draws from our study of high-energy
jets are as follows:
\texttt{(i)} the incoherent energy loss by the
two subjets created by the hard splitting leads to a suppression in
the nuclear $z_g$ distribution which is larger at small $z_g$;
\texttt{(ii)} the MC results are sensitive to the evolution of
the subjets multiplicity via VLEs which leads to an energy loss
increasing with the subjet $p_T$;
\texttt{(iii)} this last effect may be hidden when studying the
self-normalised $z_g$ distribution; in that respect, the \njets-normalised
distribution is better suited to disentangle between different
energy-loss models.

\subsection{Analytic insight for low-energy jets: MIEs and energy loss}
\label{sec:lowpt}

We now turn to more phenomenologically-relevant case of ``low energy''
jets, $p_{T0} <  \omega_c/\zc$ (with $\omega_c/\zc=600$~GeV for 
our default choice of medium parameters), for which the ``hard'' 
emission that triggers the SD condition can be either vacuum-like or medium-induced.
In both cases, the  two ensuing subjets lose energy via MIEs, which, as 
explained in the previous section, implies 
that the measured value $z_g$ is different (typically slightly
smaller) than the physical value $z$.

Our main goal in this section is to develop analytic approximations
which qualitatively and even semi-quantitatively capture this complex
dynamics.
For definiteness, we focus on gluon-initiated jets with
$p_{T0}=200$~GeV. This value is at the same time low enough to be
representative for the low-energy regime and large enough to justify
some convenient approximations, like the single emission approximation
for the MIEs captured by SD.
For pedagogical reasons it is convenient to first consider two
simplified situations --- a jet built with MIEs alone and a jet in
which the SD condition can only be triggered by a VLE (as in the
``high-energy'' case) ---, before addressing the full picture in
Sect.~\ref{sec:full}.

\subsubsection{Low-energy jets: medium-induced emissions only}
\label{sec:low}

To study the case where the SD condition is triggered by a MIE, we
consider jets generated via MIEs only, disabling VLEs.
Since the emission angles of MIEs are controlled by their transverse
momentum broadening (cf. Sect.~\ref{sec:TMB}), they are not ordered in
angle.
Hence, the reclustering of the jet constituents with the C/A algorithm
does not necessarily respects the physical ordering of the MIEs in
time.
In particular, the branching selected by the SD procedure may not be a
{\it primary} emission, i.e.\ a direct emission by the leading
parton.
However, as long as $\zc p_{T0}$ is sufficiently large compared to the
characteristic scale $\obr=3.46$~GeV for multiple branching --- which
is definitely the case for our 200~GeV jets, ---  the
probability to select a non-primary branching is suppressed by
$\amed$. From now on we can therefore assume that SD selects a {\it primary} MIE.

Next, we can argue that the MIEs captured by the SD algorithm are soft and
located in a small corner near $z=\zc$ and $\theta=\thetacut$.
Indeed, the bulk of the MIEs lies below the line $\kt=Q_s=2.4$~GeV and
the smallest value of $\kt$ that can be selected by SD, namely
$\zc p_{T0}\thetacut= 2$~GeV, is only slightly smaller than
$Q_s$. This is visible in the phase-space diagram of
Fig.~\ref{Fig:zgLund} (right).
Together with the fact that the BDMPS-Z rate~\eqref{BDMPSZrate} grows
quickly as $z\to 0$, this means that MIEs contribute to the $z_g$
distribution only at small $z_g$.
After SD, one therefore has a soft subjet of transverse momentum
$p_{T1}$ corresponding to the MIE and a harder subjet of
momentum $p_{T2}$ corresponding to the leading parton.

\begin{figure}[t] 
  \centering
    \includegraphics[width=0.48\textwidth]{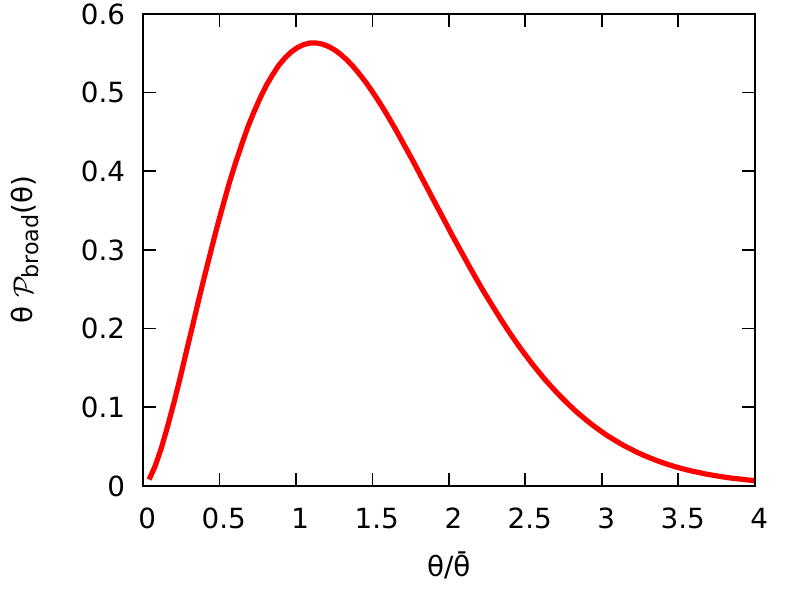}\hfill%
    \includegraphics[width=0.48\textwidth]{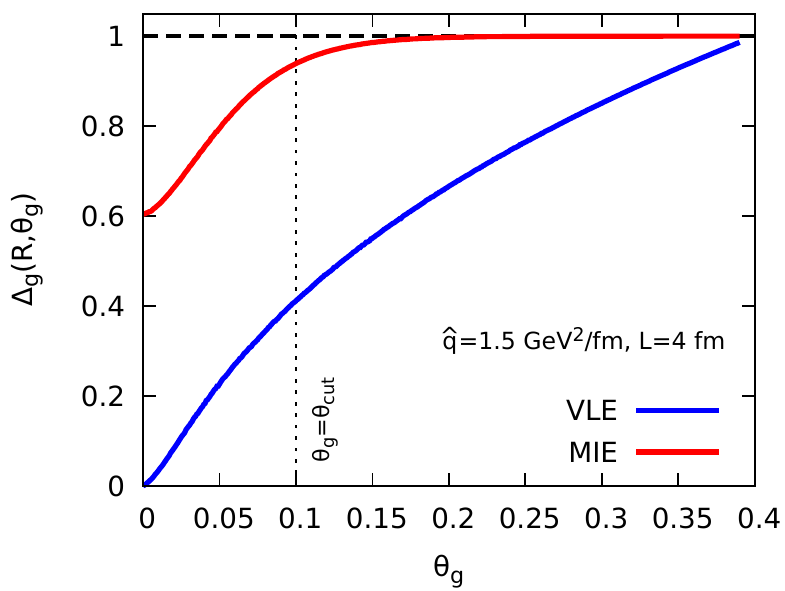}
 \caption{\small  Left: after multiplication by $\theta$, the angular distribution $\mathcal{P}_{\text{broad}}$ in \eqn{pbroad} scales as a function of the ratio $\theta/\bar\theta$, with  $\bar\theta=(\sqrt{\pi}/3)Q_s/\omega$. Right: the Sudakov factors $\Delta_{\text g}^{ \text{MIE}} (R,\theta_g)$ for MIEs, cf.~\eqn{Deltamed}, and 
 $\Delta_{\text g}^{\text{VLE}}(R,\theta_g)$ for VLEs, cf.~\eqn{DeltaVLE}, for a gluon-initiated jet with
 $ p_{T0}=200$~GeV and $R=0.4$.  }
 \label{Fig:broad}
\end{figure}

The differential probability for the emission of a primary MIE with
$\omega_1\equiv\omega \ll p_{T0}$ is given by the BDMPS-Z
spectrum~\eqref{BDMPSZspec} multiplied by the angular distribution
produced via transverse momentum broadening after emission,
\eqn{thetabroad}. The latter depends on the distance $\Delta t=L-t$,
with $t$ the emission time, travelled by the two
subjets through the medium.
In principle one should therefore work differentially in $t$. Since
this would be a serious complication, we rather use a picture in which
we average over all the emission times $t$, distributed with uniform
probability over the interval $0<t<L$. The differential probability
for the ``hard'' splitting then takes the form
\begin{equation}
\label{Pmed}
\rmd^2 \mathcal{P}_{i, \text{med}}(\omega,\theta)= \frac{C_i\amed}{\pi}\Theta\left({\oc}-\omega\right)
\sqrt{\frac{2\oc}{\omega^{3/2}}}\,\mathcal{P}_{\text{broad}}(\omega,\theta)\,{\rmd \omega}
\rmd\theta
 \,\equiv \,\mathcal{P}_{i,\text{med}}(\omega,\theta) \rmd \omega \rmd\theta,
\end{equation}
where~\cite{Mehtar-Tani:2016aco}
  \begin{equation}
\label{pbroad}
 \mathcal{P}_{\text{broad}}(\omega,\theta)\equiv \frac{1}{L}\int_0^L\rmd t\,\frac{2\omega^2\theta}{\hat{q}(L-t)}\exp\left\{-\frac{\omega^2\theta^2}{\hat{q}(L-t)}\right\}=
 2\theta\,\frac{\omega^2}{Q_s^2}\,\Gamma\left(0,\frac{\omega^2\theta^2}{Q_s^2}\right),
\end{equation}
with $\Gamma(0,x)$ the incomplete Gamma function.
This distribution predicts an average value
$\bar k_\perp=\tfrac{\sqrt{\pi}}{3}Q_s$ for $k_\perp\equiv \omega\theta$. It
shows a peak near $\bar k_\perp$, and a rather wide tail at
larger values $\kt > \bar k_\perp$ (see Fig.~\ref{Fig:broad}, left).
Since we have just argued that SD selects emissions in a narrow range
in $k_\perp$, close to $Q_s$, the tail of this distribution plays an
important role in our discussion. This is amplified by the fact that
$\bar k_\perp$ is slightly smaller than $Q_s$.
Note that in terms of the emission angle, this argument means
that the emissions selected by SD will need to acquire a $\theta$
larger than the peak value $\bar\theta(\omega)=\bar k_\perp/\omega$ 
from broadening in order to pass the
$\theta_\text{cut}$ condition.
In future work, it will be interesting to study how a description of
broadening beyond the Gaussian approximation affects quantitatively
our results.

Additionally, we need to account for the fact that both the MIE that
triggers the SD condition and the leading parton lose energy. 
The situation is mostly the same as for our earlier high-energy case
except that now the medium-induced gluon emission can occur anywhere
inside the medium, i.e.\ at any time $t$ with $0< t<L$.
For an emitted gluon of energy $\omega$, one can write
$p_{T1}=\omega-\varepsilon_{\text g}(\omega,\theta_g, \Delta t)$, where the
energy loss depends explicitly on the distance $\Delta t=L-t$
travelled by the subjet through the medium.
In our kinematic range, this energy loss is relatively small,
$\varepsilon_{\rm g}\ll \omega$, and therefore varies slowly with
$\omega$ (cf.\ \eqn{ElossHigh} so we can neglect the $\omega$
dependence.
It depends however quadratically on $\Delta t$.
For simplicity, we use a time-averaged picture in which
$\langle t\rangle \simeq \langle\Delta t\rangle \simeq L/2$ and
therefore
$\varepsilon_{\text g}(\Delta t)\approx\varepsilon_{\text g}(L/2)\simeq
\tfrac{1}{4}\varepsilon_{\text g}(L)\equiv \bar\varepsilon_{\text g}$, with
$\varepsilon_{\text g}(L)\propto L^2$ the energy loss corresponding to a
distance $L$.

The situation for the harder subjet matching with the leading parton, is more complex, owing to
differences between the time ordering of MIEs and the
(angular-ordered) C/A clustering used by SD.
Indeed, MIEs from the leading parton at times smaller than $t$ and
angles smaller than $\theta_g$ will be clustered by the C/A algorithm
in the harder subjet.
These emissions can carry a substantial amount of energy so that,
without a full picture of the time evolution of the jet, it is
delicate to even define a physical transverse momentum, $\omega_2$,
for the harder subjet at the vertex where the MIE triggering the SD
condition is emitted.
We can however take a different approach and realise that, by
definition, the difference $p_{T0}-(p_{T1}+p_{T2})$ corresponds to the
energy $\varepsilon_i(p_{T0},\theta_g)$ lost by the initial parton at
angles larger than $\theta_g$. As for the high-energy case, we can
neglect the energy lost between $\theta_g$ and $R$ and hence write
$p_{T1}+p_{T2}\simeq p_{T0}-\varepsilon_i(p_{T0},R)$.

In fine, the measured value of $z_g$ is related to the initial energy
$\omega$ of the gluonic MIE subjet via
\beq\label{zgMIE}
z_g\,\simeq\,\frac{\omega-\bar\varepsilon_{\text g}(\omega,\theta_g)}{p_{T0}-\varepsilon_{i}(p_{T0},R)}\,
\qquad\text{ with }\,\bar\varepsilon_{\text g}(\omega,\theta_g)=\frac{1}{4}\varepsilon_{\text g}(\omega,\theta_g).
\eeq
Since both energy losses in~\eqref{zgMIE} are small, one can ignore
their $p_T$ dependence and use the fits to the MC results shown in Figs.~\ref{Fig:jeteloss-v-R}.
The $z_g$ distribution created via MIEs can then be computed using a formula similar to that used for VLEs in the previous subsections, cf. \eqn{zghighpt}, namely \begin{equation}
 \label{zglowpt}
  f_{i,\text{med}}(z_g)= \int_{\thetacut}^{R}\rmd\theta_g\,
\Delta_i^{ \text{MIE}} (R,\theta_g)
  \int\rmd \omega\, \mathcal{P}_{i,\text{med}}(\omega,\theta_g)\delta\left(z_g-\frac{\omega-\bar\varepsilon_{\rm g}(\theta_g)}{p_{T0}-\varepsilon_{i}(R)}\right)
   \Theta(z_g-\zc),
\end{equation}
with the Sudakov factor $\Delta_i^{ \text{MIE}} (R,\theta_g)
$ accounting for the probability to have no primary MIEs with
$\omega>\omega_{\text{cut}}\equiv \bar\varepsilon_{\rm
  g}+\zc(p_{T0}-\varepsilon_{i}(R))$ and $\theta >\theta_g$ at any
point  inside the medium:
\begin{equation}\label{Deltamed}
\Delta_i^{ \text{MIE}} (R,\theta_g)=\exp\left(-\int_{\theta_g}^{R}{\rmd\theta}\int\rmd \omega\,\,\mathcal{P}_{i,\text{med}}(\omega,\theta)\, \Theta\left(\omega-\omega_{\text{cut}}\right)\right)\,.
\end{equation}
The self-normalised distribution can be computed as
$p_{i, \text{med}}(z_g)= \mathcal{N} f_{i, \text{med}}(z_g)$ with
$\mathcal{N}=\tfrac{1}{1-\Delta_i^{ \text{MIE}} (R,\thetacut)}$.
The Sudakov factor is plotted for both MIEs, \eqn{Deltamed}, and VLEs,
\eqn{DeltaVLE}, as a function of $\theta_g$ in the right plot of
Fig.~\ref{Fig:broad}. In both plots, we use $p_{T0}=200$~GeV and a gluon-initiated
jet.
While this factor is clearly important for VLEs at all $\theta_g$, it
remains very close to one for MIEs. It is mostly irrelevant for the
shape of the $z_g$ distribution and only has a small impact on its
overall normalisation.

\begin{figure}[t] 
  \centering
  \includegraphics[width=0.48\textwidth]{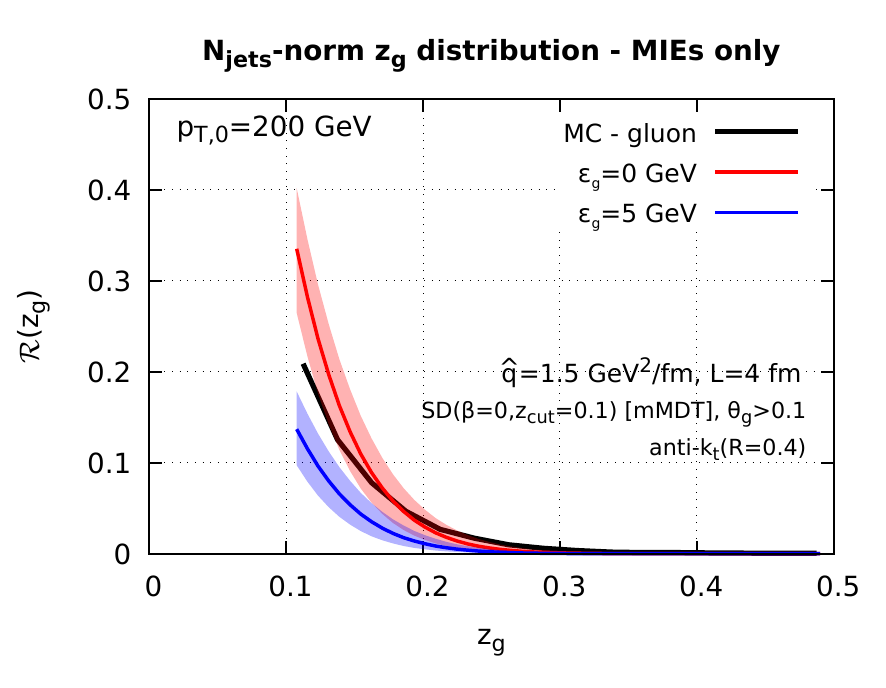}\hfill%
  \includegraphics[width=0.48\textwidth]{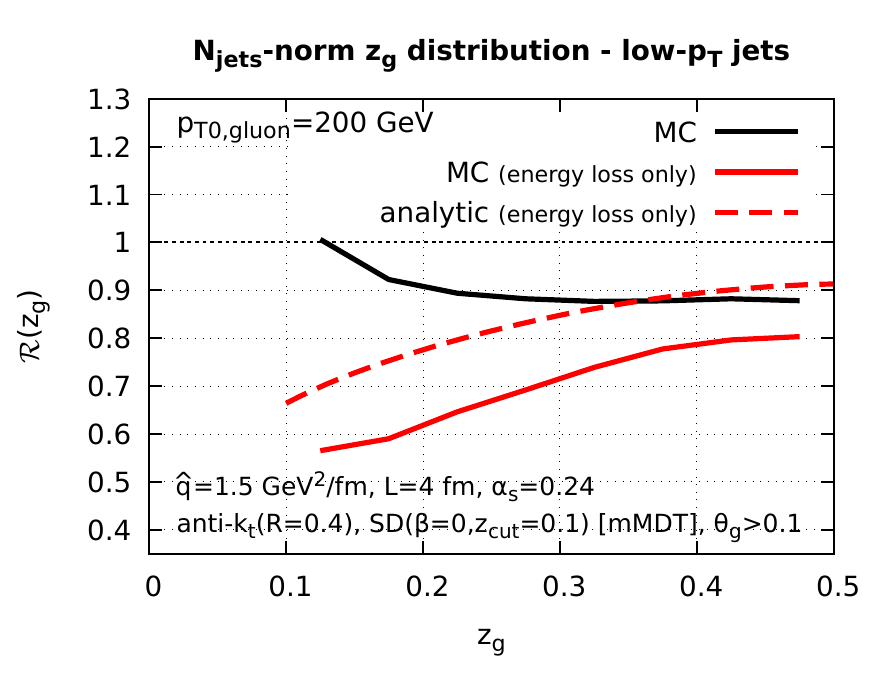}
  \caption{\small Separation of the nuclear effects on the $z_g$
    distribution of low-energy jets into a contribution due to the
    MIEs (left) and a contribution due to the VLEs with energy loss
    (right). Left: MIEs only (MC vs.\ analytic calculations).
    Right: Monte Carlo results for the full parton shower (black) vs.\
    the case where only VLEs with energy loss are included (red). We
    also show an analytic result for the second case (dashed line).}
  \label{Fig:lowptMIEVLE}
\end{figure}

In Fig.~\ref{Fig:lowptMIEVLE}(left) we compare our analytic approximation for
the ratio\footnote{Here, the medium/vacuum ratio is not
  a genuine nuclear modification factor. For example, in the absence of medium effects, it would be equal to zero, not to one.} $f_{\text{med}}(z_g)/f_{\text{vac}}(z_g)$ of the
\njets-normalised $z_g$ distributions corresponding the a gluon-initiated
jet to the MC results obtained by ``switching off'' the VLEs from the
general numerical code.
The energy losses are estimated from the 
 fit to the MC results shown in the left plot of Fig.~\ref{Fig:jeteloss-v-R}, which yields
$\varepsilon_{\rm g}\simeq 16.5$~GeV for the whole jet with $R=0.4$ and $\bar\varepsilon_{\rm g}\simeq 5$~GeV for the subjet with $\theta_g\simeq\thetacut=0.1$. Given the large uncertainty in the calculation of $\bar\varepsilon_{\rm g}$, we present  two sets of results, one corresponding to $\varepsilon_{\rm g}=0$~GeV and the other one to $\varepsilon_{\rm g}=5$~GeV.
For each of these 2 choices we indicate by a band the uncertainty associated with $\pm 10\%$ variations in the saturation scale $Q_s$ around its central value $Q_s=\sqrt{\hat q L}=2.4$~GeV. This variation corresponds to the fact that the relation \eqref{zgMIE} between $z_g$ and $\omega$ is only approximate and the associated uncertainty in the value of $\omega$ has consequences, via \eqn{pbroad}, on the angular distribution; this uncertainty was mimicked by varying $Q_s$.
 
Fig.~\ref{Fig:lowptMIEVLE}(left) shows a qualitative agreement between the MC
and the analytic calculations: all the curves have a visible rise
at small $z_g$, reflecting the fact that the BDMPS-Z
spectrum behaves like $z^{-3/2}$ which is more singular than the
vacuum spectrum $\propto z^{-1}$.
This being said, our analytic study is still too poor to
quantitatively reproduce the MC results, or to discriminate between
various scenarios for the energy loss. In particular, the ``zero
energy loss'' scenario is not in clear disagreement with the MC
results. This may be related to the fact that the angular distribution
in \eqn{pbroad} favours small values $z\sim\zc$ which biases the
distribution towards events with a smaller-than-average energy loss.

\subsubsection{Low-energy jets: energy loss only} 
\label{sec:lowpteloss}
 
In this section, we consider the situation (opposite to the previous
section) where the SD condition is triggered by a VLE.
In this case, we turn off the direct contribution of the MIEs to SD,
but only keep their (indirect) effect associated with incoherent
energy loss of the subjets found by the SD procedure.
The physical situation is similar to the high-energy case studied in
Sect.~\ref{sec:high} where the ``direct'' contribution of the MIEs to
SD was negligible by definition.

To artificially remove the direct contribution of the MIEs from the MC
simulations, we have enforced that all the partons generated via MIEs
propagate at angles $\theta \gg R$.
This obviously overestimates the jet energy loss, simply because some
of the partons which would have remained within the jet are
artificially moved outside. This is fine as long as we only focus on
illustrating the qualitative effects of the energy loss on the $z_g$
distribution.

In Fig.~\ref{Fig:lowptMIEVLE}(right), we compare Monte Carlo
results obtained with this artificial removal of the direct
contribution of MIEs (``energy loss only'', red,  curve) to the
full simulation (``full'', black, curve), where both VLEs and MIEs
contribute directly.
Fig.~\ref{Fig:lowptMIEVLE}(right) also shows the prediction of an analytic
calculation which ignores the direct contribution of the MIEs to SD.
This calculation is the same as the one presented in
Sect.~\ref{sec:high} for the case of a high-energy jet, i.e.\ it is
based on Eqs.~\eqref{zinzg}, \eqref{Pvac} and~\eqref{DeltaVLE}, now
applied to $p_{T}\simeq p_{T,0}=200$~GeV.  The respective Sudakov
factor is plotted in the right plot of Fig.~\ref{Fig:broad}.
By inspection of these curves, we first notice that the effect of energy loss alone is the same for
low-energy jets as it was for high-energy jets: it leads to a strong
nuclear suppression\footnote{This suppression appears to be larger for the respective MC calculation than for the analytical one 
because, for the former, the energy loss is artificially amplified.} of the $z_g$ distribution with larger effects at small $z_g$. 
Second, adding the direct contribution of the MIEs changes
the picture significantly: the medium/vacuum ratio is now decreasing
with $z_g$ and the ratio even becomes larger than one at small $z_g$.
The difference between the two curves is, at least qualitatively,
consistent with an additional peak at small $z_g$ from MIEs (see e.g.\
Fig.~\ref{Fig:lowptMIEVLE}, left).

\subsubsection{Low-energy jets: full parton showers} 
\label{sec:full}

Now that we have studied both effects separately, we can provide an
analytic calculation for the complete $z_g$ distribution for a
low-energy jet, including both VLEs and MIEs.

Due to angular ordering, VLEs selected by the SD procedure are
necessarily primary gluon emissions from the leading parton.
However, whenever SD selects a MIE with energy fraction $z$ and
emission angle $\theta_g$, this emission can be emitted by any of the
partonic sources created via VLEs with energy $\omega>zp_{T0}$. Since,
by definition, SD selects the largest-angle emission with $z_g$ above $\zc$,
these sources of MIE can have any angle in the range
$\theta_c<\theta<\theta_g$. Such emissions are formally clustered by
the C/A algorithm together with the subjet corresponding to the leading
parton.

The $z_g$ distribution for a full parton shower generated by a parton
of type $i=(\text{q,\,g})$  is obtained by incoherently summing up the
probabilities for SD to select either a VLE or a MIE:
\begin{align}  \label{full}
 f_i(z_g)& = \int_{\thetacut}^{R}\rmd\theta_g\,
\Delta_i^{\text{VLE}}  (R,\theta_g) \,\Delta_i^{\text{MIE}} (R,\theta_g) \\
 &\times
 \int_{0}^{1/2}\rmd
   z\Big[\mathcal{P}_{i,\text{vac}}(z,\theta_g)\delta\big(
   \mathcal{Z}_{g,\text{vac}}(z,\theta_g)-z_g\big)+\mathcal{P}_{i,\text{med}}(z,\theta_g)\delta\big(
   \mathcal{Z}_{g,\text{med}}(z,\theta_g)-z_g\big)\Big]\Theta(z_g-z_{\text{cut}}).
   \nonumber
 \end{align}
Here, we have used different functions, $ \mathcal{Z}_{g,\text{vac}}(z,\theta_g)$ and $ \mathcal{Z}_{g,\text{med}}(z,\theta_g)$, for the relation between the measured splitting fraction $z_g$ and the physical one $z$, to take into account the fact that the energy loss is generally different for the subjets produced by a VLE or a MIE.  
The function $ \mathcal{Z}_{g,\text{vac}}(z,\theta_g)$ is given by
\eqn{zinzg}, with the $p_T$ of the parent gluon identified with the
$p_{T0}$ of the leading parton. As before, we replace $\theta_g$ by $R$ in
the energy loss and take $\mathcal{E}_{1}(zp_T,R)$ and
$\mathcal{E}_{2}((1-z)p_T,R)$ from the fits in Fig.~\ref{Fig:jeteloss-v-pt}.
In the case of a medium-induced splitting, we use a generalisation of \eqn{zgMIE}, that is
\beq\label{zgfull}
z_g\,\simeq\,\frac{zp_{T0}-\bar\varepsilon_{\rm g}(\theta_g)}{p_{T0}-\mathcal{E}_{i}(p_{T0},R,z_g)}
\,\equiv  \mathcal{Z}_{g,\text{med}}(z,\theta_g)\,.
\eeq
The main difference w.r.t.\ \eqn{zgMIE} refers to the energy loss by
the whole jet, i.e.\ the function $\mathcal{E}_{i}(p_{T0},R,z_g)$ in
the denominator: not only this has now a strong dependence upon
$p_{T0}$, due to the rise in the number of partonic sources via VLEs,
but this must be evaluated for the special jets which include a hard
splitting with a given value $z_g>0.1$ and with any
$\theta_g>\thetacut=0.1$. From our MC calculation illustrated in
Fig.~\ref{Fig:elosszg}, we know that it is larger than the
average energy loss and largely independent of $z_g$. We use
$\mathcal{E}_{\text g}=43$~GeV for  $p_{T0}=200$~GeV and $R=0.4$. As for
$\bar\varepsilon_{\text g}(\theta_g)$, we use 5~GeV  as in Sect.~\ref{sec:low}
and study the sensitivity of our results to variations around this value.

The vacuum splitting probability density $\mathcal{P}_{i,\text{vac}}$
and the associated Sudakov factor $\Delta_i^{\text{VLE}}$ take the
same form as in Sect.~\ref{sec:high}, Eqs.~\eqref{Pvac}
and~\eqref{DeltaVLE}.
The corresponding medium-induced probability density
$\mathcal{P}_{i,\text{med}}$ takes a form similar to Eq.~\eqref{Pmed}
modified to account for the fact that each VLE produced in the medium
can act as a source of MIE. We therefore write
\begin{align}
\label{Pfullmed}
  \mathcal{P}_{i,\text{med}}(z,\theta_g)&=\nu(z,\theta_g)\,\frac{\alpha_{s,\text{med}} C_i}{\pi}
  \sqrt{\frac{2\oc}{p_{T0}}}\,z^{-3/2}\,\mathcal{P}_{\text{broad}}(z,\theta_g).
\end{align}
The number of MIE sources $\nu(z,\theta_g)$ is obtained from the
density $\tfrac{\rmd^2N_{\text{VLE}}^{\text{(in)}}}{\rmd\omega
  \rmd\theta}$ of VLEs produced inside the medium:
\beq
\label{nudef}
\nu(z,\theta_g)\equiv 1+ \int_{z p_{T0}}^{p_{T0}}\rmd\omega\int_{\theta_c}^{\theta_g}\rmd\theta \frac{\rmd^2N_{\text{VLE}}^{\text{(in)}}}{\rmd\omega \rmd\theta}\,.
\eeq
In this last expression, the first term corresponds to the leading
parton and the integration boundaries in the second term impose that
an MIE which triggers the SD condition has to come from a source of
larger energy at an angle between $\theta_c$ and $\theta_g$. 
The associated Sudakov factor $\Delta_i^{\text{MIE}}$ is then constructed in terms of $\mathcal{P}_{i,\text{med}}$ as in \eqn{Deltamed}.
We note however that, although for the case with only MIEs, the
exponentiation of the emission probability in the Sudakov factor
$\Delta_i^{\text{MIE}}$ is relatively straightforward, this
does not obviously hold in the presence of multiple sources of
MIEs.
However, since the Sudakov factor $\Delta_i^{\text{MIE}}$ only introduces a
small correction (see Fig.~\ref{Fig:broad}, right), 
we have kept the exponential form,
Eq.~(\ref{Deltamed}), for simplicity and as an easy way to maintain
the conservation of probability for MIEs.

\begin{figure}[t] 
  \centering
  \begin{minipage}[t]{0.45\textwidth}
    \mbox{ }\\
    \includegraphics[page=1,width=\textwidth]{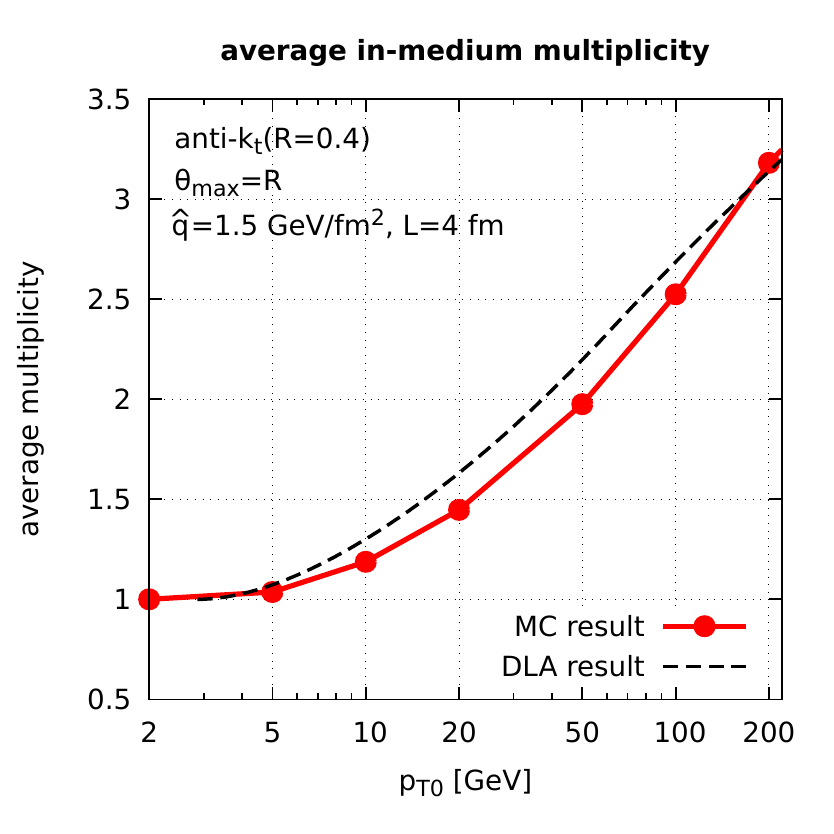}
  \end{minipage}
  \hfill%
  \begin{minipage}[t]{0.43\textwidth}
    \mbox{ }\\
    \vspace*{1.0cm}

    \caption{\small The average multiplicity of partons created inside
      the medium via VLEs by a leading gluon with $p_{T0}=200$~GeV: the
      full Monte Carlo results shown by (red) points are compared with a
      running-coupling extension of the DLA estimate in \eqn{DLA},
      integrated over the ``inside'' region of the phase-space of
      Fig.~\ref{Fig:LundPS}.} \label{Fig:multi}
  \end{minipage}
\end{figure}

\begin{figure}[t] 
  \centering
  \includegraphics[page=1,width=0.48\textwidth]{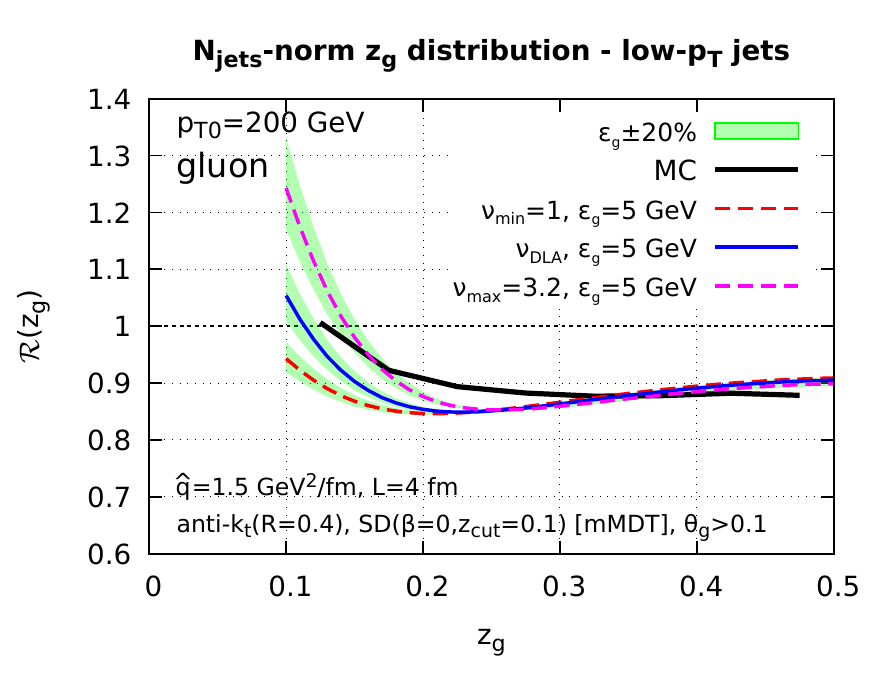}\hfill%
  \includegraphics[page=2,width=0.48\textwidth]{./FIGS/plot_zg-low_pT}\hfill%
  \caption{\small
    $\mathcal{R}(z_g)$ for a leading parton (left: gluon, right:
    quark) with initial transverse momentum $ p_{T0}=200$~GeV. The
    Monte Carlo results are compared to analytic calculations
    corresponding to 3 different approximations for the number $\nu$
    of sources emitting MIEs (see the text for details). }
  \label{Fig:zglowpt}
\end{figure}

In practice, we test three different approximations\footnote{Note also that the formal
limit $\nu\to 0$, in which one keeps only the ``direct'' contribution of the VLEs to \eqn{Pfullmed},
corresponds to the analytic result shown with dashed line in Fig.~\ref{Fig:lowptMIEVLE}(right).}
for $\nu$: \texttt{(i)} $\nu\equiv\nu_{\rm min}=1$ includes only the leading
parton, \texttt{(ii)} $\nu\equiv\nu_{\rm max}=3.2$ is our MC estimate
for the average multiplicity of partons created inside the medium via
VLEs by a leading gluon with $p_{T0}=200$~GeV,
cf. Fig~\ref{Fig:multi}. This has to be seen as a maximal value since
it ignores the kinematic limits in \eqn{nudef}). \texttt{(iii)}
$\nu\equiv\nu_{\rm DLA}$ obtained by evaluating \eqn{nudef} with a DLA
estimate for the gluon multiplicity, corresponding to \eqn{DLA} with
$\theta_\text{max}\to\theta_g$ and taking the coupling $\abar$ at the
scale $k_\perp=E\theta_g$. We see in Fig.~\ref{Fig:multi} that this DLA
approximation gives a reasonable description of the VLE multiplicity in a
jet (setting $\theta_g=R$).
For the case $\nu=\nu_{\rm DLA}$ we show no
variation band in Fig.~\ref{Fig:zglowpt} to avoid overlapping bands 
in an already complicated plot, but it is quite clear
what should be the effects of varying $\bar\varepsilon_{\rm g}$ and
$Q_s^2$.

In Fig.~\ref{Fig:zglowpt}, we show our MC results for a full shower
generated by a leading parton, gluon (left) or quark (right), with
$ p_{T0}=200$~GeV, together with our analytic approximation based on
\eqn{full} using the three different approximations for $\nu$.
In each case, the central curve corresponds to the average values
$\bar\varepsilon_{\text g}=5$~GeV for the subjet energy loss in \eqn{zgfull}  and $Q_s^2=\hat q
L=6$~GeV$^2$ for the saturation momentum squared
in~\eqref{pbroad}. The bands around these central curves correspond to
variations by 20\% of $\varepsilon_{\rm g}$. Note that varying $Q_s^2$
by 20\% has a smaller effect.
The unphysical case $\nu_{\rm min}=1$ is disfavoured by the comparison
with the MC results as it underestimates the peak associated with the
direct MIE contribution. The other two cases are at least
qualitatively consistent with the MC results.

It is also interesting to notice the dependence of the results upon the flavour
of the leading parton. The rise of the nuclear distribution at small $z_g$ appears
to be stronger  for the gluon-initiated jet than for the quark-initiated one and this
difference is rather well captured by our analytic approximations, where it is due
to a change in the number  $\nu$ of partonic sources for  a ``hard'' MIE: one has
indeed $\nu-1\propto C_R$ (recall e.g. the discussion in footnote~\ref{foot:CR}). 
 At large $z_g$ on the other hand, the analytic approximation
appears to be less satisfactory for the quark-initiated jet --- most likely, because
it underestimates the energy loss by the subjets resulting from a hard VLE. As a matter
of facts, a similar difficulty occurs in the high-energy case, as can be seen by comparing
the quark-jet results for $p_{T0}=1$~TeV  in Figs.~\ref{Fig:zgvar} and \ref{Fig:QvsG} respectively.

The main conclusions that we can draw from in Fig.~\ref{Fig:zglowpt}
and from the overall discussion in this section is that $z_g$
distribution for ``low energy jets'' is a superposition of two main
effects: \texttt{(i)} incoherent energy loss for the subjets created
by a vacuum-like splitting; this controls the $z_g$ distribution at
moderate and large values of $z_g$, where it yields a nuclear
modification factor which slowly increase with $z_g$, and
\texttt{(ii)} sufficiently hard medium-induced emissions, with
$z\gtrsim \zc$, which leads to a significant growth of the $z_g$
distribution at small values $z_g\sim\zc$.
This behaviour is qualitatively reproduced by our simple analytic
calculations which shows, for example, that including multiple (VLE)
sources of MIEs is important. It is however more delicate to draw more
quantitative conclusions as several effects entering the calculation
would require a more involve treatment.

\section{$z_g$ distribution with realistic initial jet spectra}
\label{sec:data}

Even if studying monochromatic jets is helpful to understand
the dominant physical effects at play, any realistic measurement would
instead impose cuts on the $p_{T,{\rm jet}}\equiv p_T$ of the final jet.
For this we need the full $p_{T0}$ spectrum of the hard scattering.
Here, we follow our prescription from Section~\ref{sec:RAA} and use a
LO dijet spectrum where both final partons are showered using our Monte
Carlo. One can then cluster and analyse the resulting event.

In the case of the $z_g$ distribution, it is interesting to note that
we expect a competition between two effects.
On one side, due to the steeply-falling underlying $p_{T0}$ spectrum,
cutting on the jet $p_T$ tends to select jets which lose less energy
than average.
On the other side, we have seen from Fig.~\ref{Fig:elosszg} that jets
with $z_g>\zc$ and $\theta_g>\thetacut$ lose more energy than average.

Below, we first study the case of the $N_\text{jets}$-normalised $z_g$
distribution which is best suited to discuss the underlying physical
details highlighted in section~\ref{sec:zgmed}.
Our distributions are also qualitatively compared to a recent experimental 
analysis by ALICE~\cite{Acharya:2019djg}.
We then consider the self-normalised $z_g$ distribution which is more
easily compared to the CMS measurements~\cite{Sirunyan:2017bsd}.

\subsection{Phenomenology with the \njets-normalised $z_g$ }

\begin{figure}[t] 
  \centering
  \includegraphics[width=0.48\textwidth]{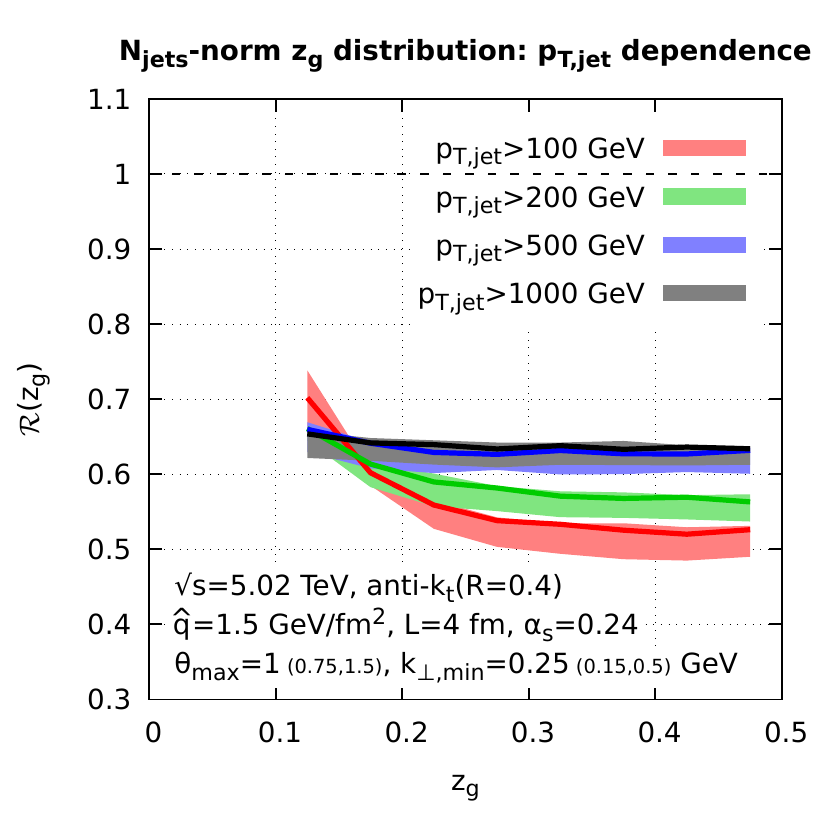}\hfill%
  \includegraphics[width=0.48\textwidth]{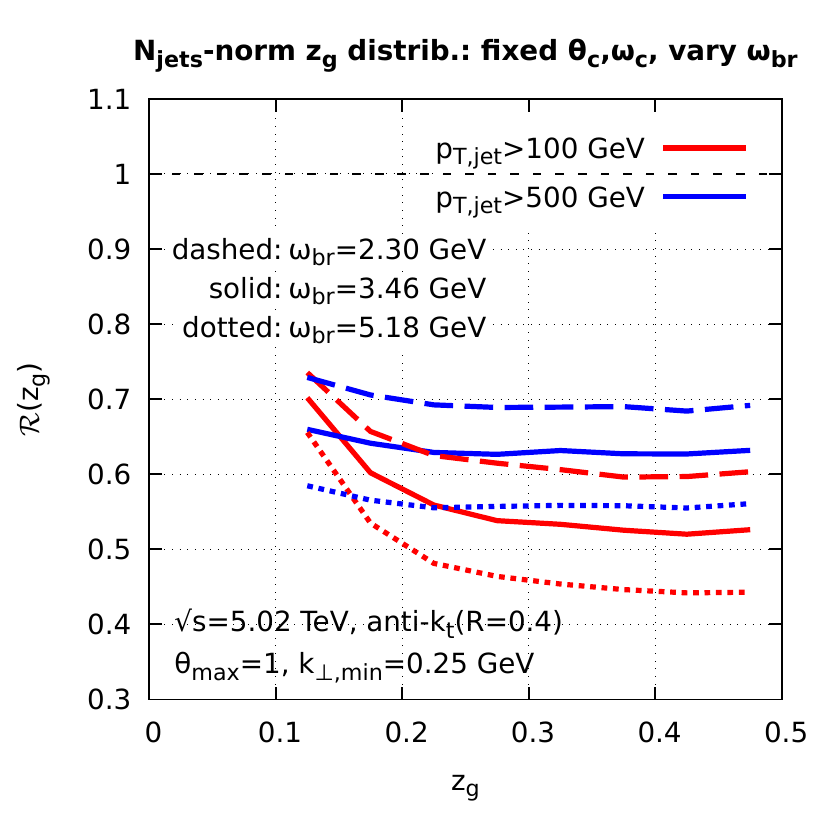}
\caption{\small Our full MC predictions for the medium/vacuum ratio $\mathcal{R}(z_g)$ of the 
$N_\text{jets}$-normalised $z_g$ distributions, including the convolution with the initial jet spectrum.   Left: the sensitivity of our results to changes in the kinematic cuts $\theta_{\rm max}$ and $k_{\perp,\text{min}}$. Right: the effect of varying $\obr$ (by $\pm 50\%$) at fixed values for $\oc$ and $\theta_c$.}
\label{Fig:zgvar1}
\end{figure}

\begin{figure}[t] 
  \centering
  \includegraphics[width=0.48\textwidth]{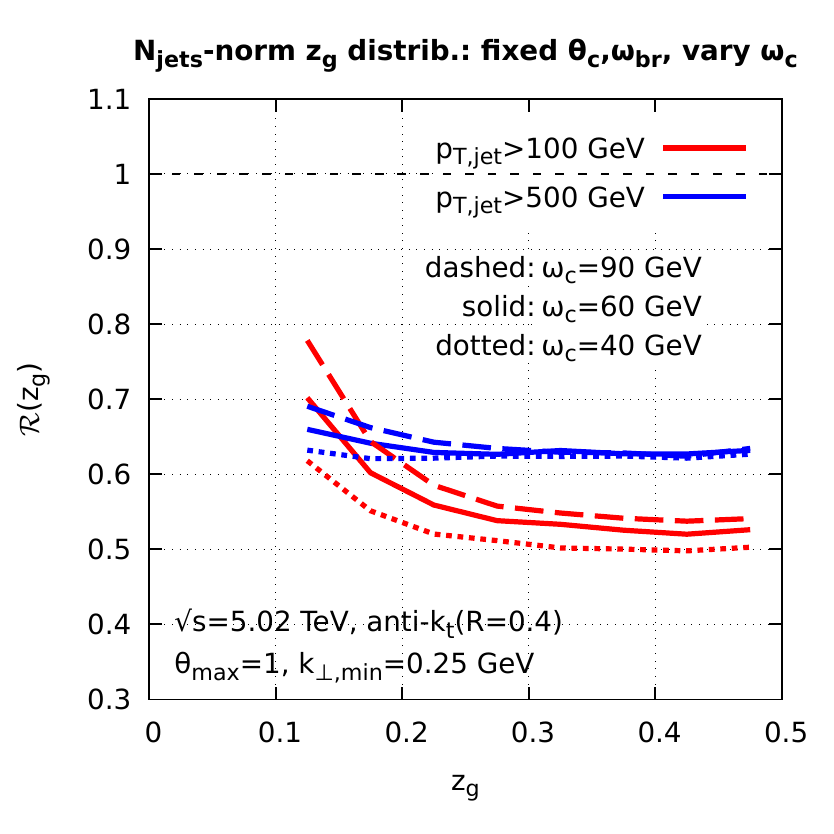}\hfill%
  \includegraphics[width=0.48\textwidth]{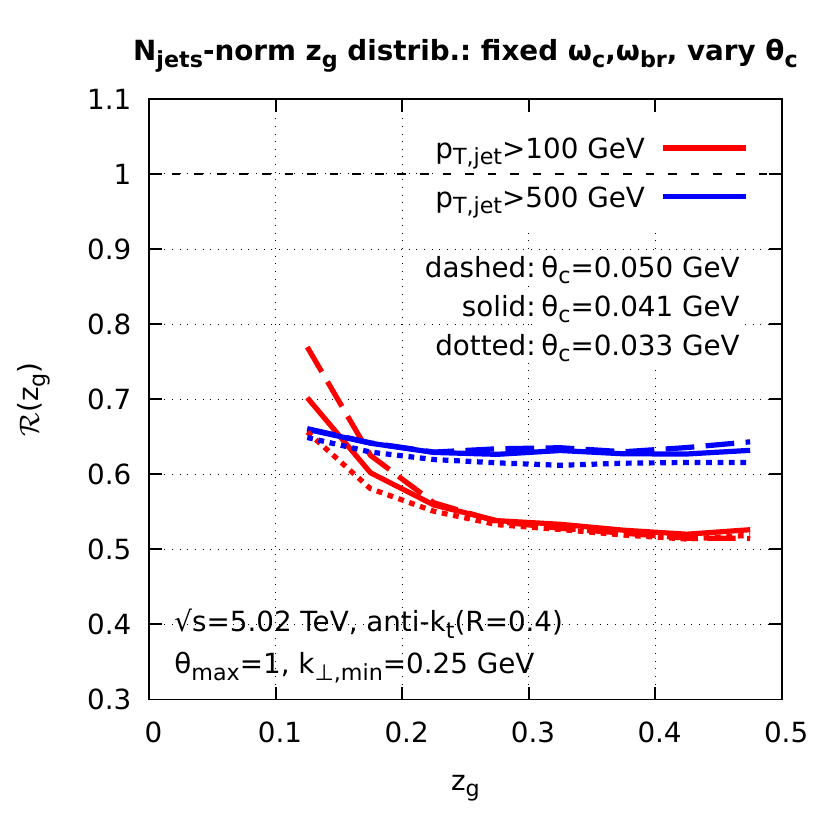}
  \caption{\small The effects of varying $\hat q$, $L$ and $\amed $
    keeping $\obr=3.46$~GeV fixed at its central value (cf.\ the right
    plot of Fig.~\ref{Fig:RAA}). Left: we vary $\oc$ by $\pm 50\%$ at
    fixed $\theta_c$. Right: we vary $\theta_c^2$ by $\pm 50\%$ at
    fixed $\oc$.}
\label{Fig:zgvar2}
\end{figure}

Our main results for the $N_\text{jets}$-normalised $z_g$ distribution
are plotted in Figs.~\ref{Fig:zgvar1}--\ref{Fig:zgvar2}. They are the
analog of the results for $R_{AA}$ shown in
Figs.~\ref{Fig:RAA}--\ref{Fig:variab}: they highlight the
$p_T$-dependence of our predictions together as well as their
sensitivity to changes in the physical ($\hat q$, $L$ and $\amed$) and
unphysical ($\theta_{\rm max}$ and $k_{\perp,\text{min}}$)
parameters.
The various curves shown in these figures have been obtained by integrating
the $z_g$ distribution over all the values of $p_{T}$ above a lower cutoff 
$p_{T,{\rm min}}$ (explicitly shown for each curve). 
In practice, our choices for this cutoff are the same
as the values taken for $p_{T0}$ in Fig.~\ref{Fig:zgvar}.
Since the jet $p_T$ spectrum falls rapidly with $p_T$ and the jet
energy loss is relatively small compared to the jet $p_T$, it makes
sense to compare the respective results.

First of all, we see from Fig.~\ref{Fig:zgvar1}, left, that our
predictions are robust w.r.t.\ variations of the unphysical parameters in
our Monte Carlo.
Then, based on the analyses from section~\ref{sec:zgmed}, we expect
the $z_g$ distribution to be mostly sensitive to changes in the
multiple-branching energy scale $\obr$ which controls both the energy
loss and the rate for SD to be triggered by a MIE.
When varying $\obr$ by 50\% around its central value, keeping $\oc$
and $\theta_c$ fixed, the $z_g$ distribution is indeed strongly
affected, see Fig.~\ref{Fig:zgvar1}, right.
The effects of changing either $\oc$ or $\theta_c$, at fixed $\obr$,
are much less pronounced as seen in Fig.~\ref{Fig:zgvar2}.
The residual variations observed when varying $\oc$ or $\theta_c$ can be
mainly attributed to variations in the phase-space for VLEs, which
affect the multiplicity of sources for MIEs and the energy loss (cf.\
\eqn{Pfullmed}). Besides, for the low-$p_T$ jets, a change in $\oc$
can have a sizeable effect on the phase-space for MIEs that are
accessible to SD (cf.\ Fig.~\ref{Fig:zgLund}). This is indeed seen in
Fig.~\ref{Fig:zgvar2} which shows a stronger dependence on $\oc$ than
on $\theta_c$, especially for the peak at low $p_T$ and small $z_g$.

Comparing now to the results for monochromatic jets shown in
Fig.~\ref{Fig:zgvar}, we observe important differences that can be
understood as follows. When studying the \njets-normalised ratio, the
deviation of
$\mathcal{R}(z_g)= f_{\text{med}}(z_g)/f_{\text{vac}}(z_g)$ from unity
is proportional to the ratio, $N_{\rm SDjets}/N_{\rm jets}$, between
the number of jets which passed the SD condition and the total number
of jets. This ratio is considerably smaller when using a realistic jet
spectrum, Fig.~\ref{Fig:zgvar1} (left), than for monochromatic jets,
Fig.~\ref{Fig:zgvar}.
This is explained by the fact that jets passing the SD condition lose
more energy than average jets and therefore have a more suppressed
production rate.
Moreover, among the jets which have passed SD with a given $z_g>\zc$,
the initial cross-section favours those where the subjets have lost
less energy,
leading to a flattening in the shape of the ratio $\mathcal{R}(z_g)$
at large $z_g$, in agreement with Fig.~\ref{Fig:zgvar1} left.
Finally, imposing a lower $p_T$ cut on jets introduces a bias towards quark
jets, which lose less energy than gluon jets. Since the former have a
smaller $\mathcal{R}(z_g)$ than the latter (cf.  Fig.~\ref{Fig:zgvar}), this further reduces
$\mathcal{R}(z_g)$ for jets.

Another interesting feature of Fig.~\ref{Fig:zgvar1} left is the fact
that the ratio $\mathcal{R}(z_g)$ is almost identical for
$p_T=500$~GeV and $p_T=1$~TeV. We believe that this purely
fortuitous. First, the normalisation factor
$N_{\rm SDjets}/N_{\rm jets}$ penalises the jets with
$p_T=1$~TeV more than those with $p_T=500$~GeV, thus reducing an
initially-small difference between the respective results in
Fig.~\ref{Fig:zgvar}.
Second, as $p_T$ increases so does the fraction of quark-initiated
jets, thus contributing to a reduction of $\mathcal{R}(z_g)$.

\begin{figure}
  \centering
  \begin{minipage}[t]{0.50\textwidth}
    \mbox{ }\\
    \includegraphics[width=\textwidth]{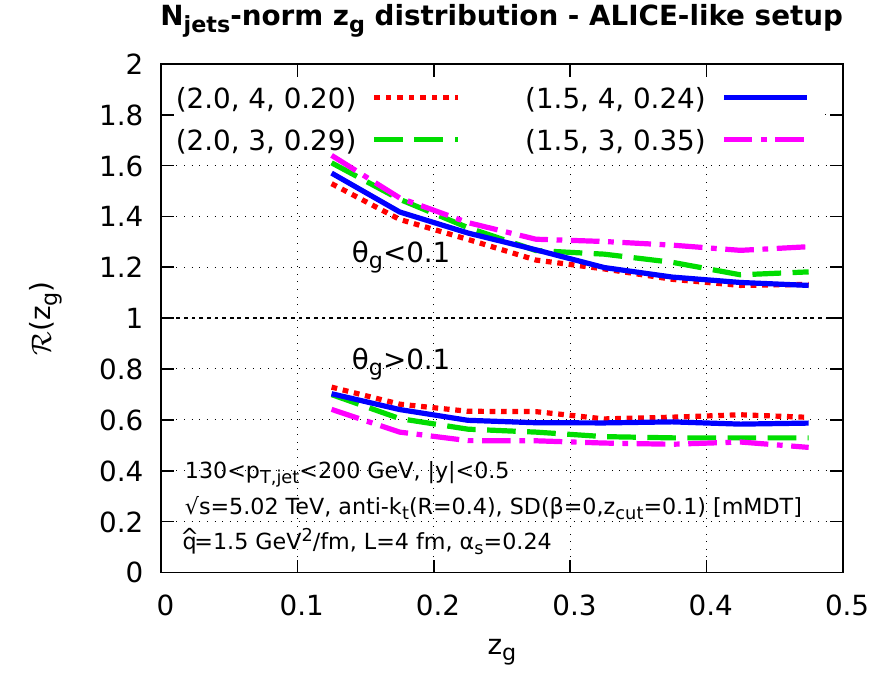}%
  \end{minipage}
  \hfill%
  \begin{minipage}[t]{0.43\textwidth}
    \caption{\small Predictions of our Monte Carlo generator for the $z_g$
      distribution obtained with a setup similar to the one used in the
      ALICE measurement of Ref.\cite{Acharya:2019djg}. We included the
      distribution obtained with either $\theta_g<0.1$ (bottom set of
      curves), or $\theta_g>0.1$ (top set of curves). In each case we
      show the result for different sets of medium parameters,
      $\hat{q}$, $L$ and $\alpha_s$ as indicated in the
      legend.}\label{Fig:ALICE}
  \end{minipage}
\end{figure}

At this point, it is interesting to compare our predictions with the
measurements by the ALICE collaboration~\cite{Acharya:2019djg} at the
LHC.
This is not immediately straightforward as the ALICE measurement is at
a different collider energy than what we have considered so far, uses
only charged tracks which are not accessible in our parton-level
shower, and is not unfolded for the detector effects  and residual
background fluctuations.
For simplicity, we keep the collider energy at 5.02~TeV. Since the
charged and full transverse momenta of jets are roughly proportional
to one another, we scale
the acceptance region for the jet $p_T$ from $[80,120]$~GeV to
$[130,200]$~GeV and work with all the particles.
The discussion below should therefore, at best, be considered as
qualitative.

Our findings are presented in Fig.~\ref{Fig:ALICE} where, following
Ref.~\cite{Acharya:2019djg} (see the first and third plots in Fig.~3),
we have considered both the case $\theta_g > 0.1$ and the case
$\theta_g < 0.1$.
Our predictions are shown for a range of medium parameters (see also
Table~\ref{tab:parameters} of Fig.~\ref{Fig:zgnorm}).
In all cases, our results are qualitatively similar to those of the
experimental analysis: the ratio $\mathcal{R}(z_g)$ is decreasing with
$z_g$, it shows nuclear suppression ($\mathcal{R}(z_g)<1$) for the
large-angle case $\theta_g > 0.1$ and nuclear enhancement
($\mathcal{R}(z_g) > 1$) for the small-angle case $\theta_g < 0.1$.
Within our framework, the enhancement observed for $\theta_g < 0.1$
and the rise at small $z_g$ are both associated with medium-induced
emissions\footnote{Since with our parameters the minimal angle
  for MIEs is $\theta_c\approx 0.04$, MIEs can pass the SD condition
  even for $\theta_g < 0.1$.} being captured by SD.
The suppression visible for $\theta_g > 0.1$ is a
consequence of incoherent energy loss as seen in Sect.~\ref{sec:zgmed}.

\subsection{Self-normalised $z_g$ distribution and CMS data}

We want to compare the predictions of our Monte Carlo generator to the
measurement of the self-normalised $z_g$ distribution by the CMS
collaboration in Ref.~\cite{Sirunyan:2017bsd}.
This comparison should however be taken with care since the CMS
results are not unfolded for detector (and residual Underlying Event)
effects and are instead presented under the form of ``PbPb/smeared
pp'' ratios. 
Without a proper dedicated study, it is delicate to assess the precise
effects of this smearing on $\mathcal{R}^\text{(norm)}(z_g)$.

\begin{figure}[t] 
  \centering
  \begin{minipage}[b]{0.48\textwidth}
    \centering
    {\small 
      \begin{tabular}{|ccc|ccc|}
        \hline
        $\hat{q}$ & $L$ & $\alpha_s$
        & $\theta_c$ & $\omega_c$ & $\omega_\text{br}$ \\
        \hline
        1.5   & 3     & 0.35  & 0.0629 & 33.75 & 4.134 \\
        1.5   & 4     & 0.24  & 0.0408 &  60 & 3.456 \\
        2     & 3     & 0.29  & 0.0544 & 45    & 3.784 \\
        2     & 4     & 0.2   & 0.0354 & 80    & 3.200 \\
        \hline
      \end{tabular}}\\
    \vspace*{0.2cm}
    \includegraphics[width=\textwidth]{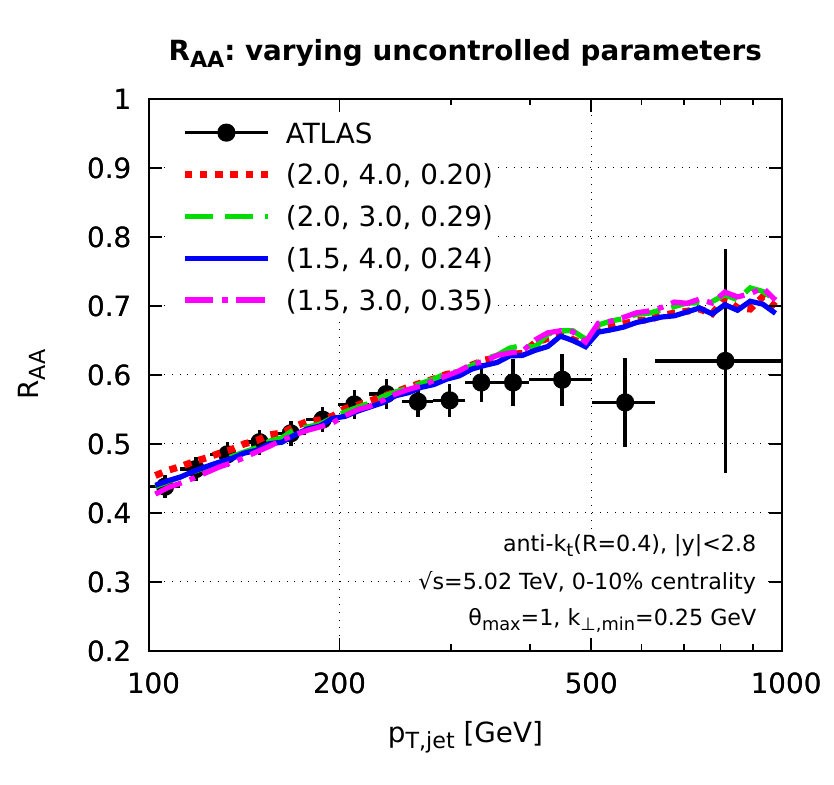}
  \end{minipage}\hfill%
  \includegraphics[width=0.48\textwidth]{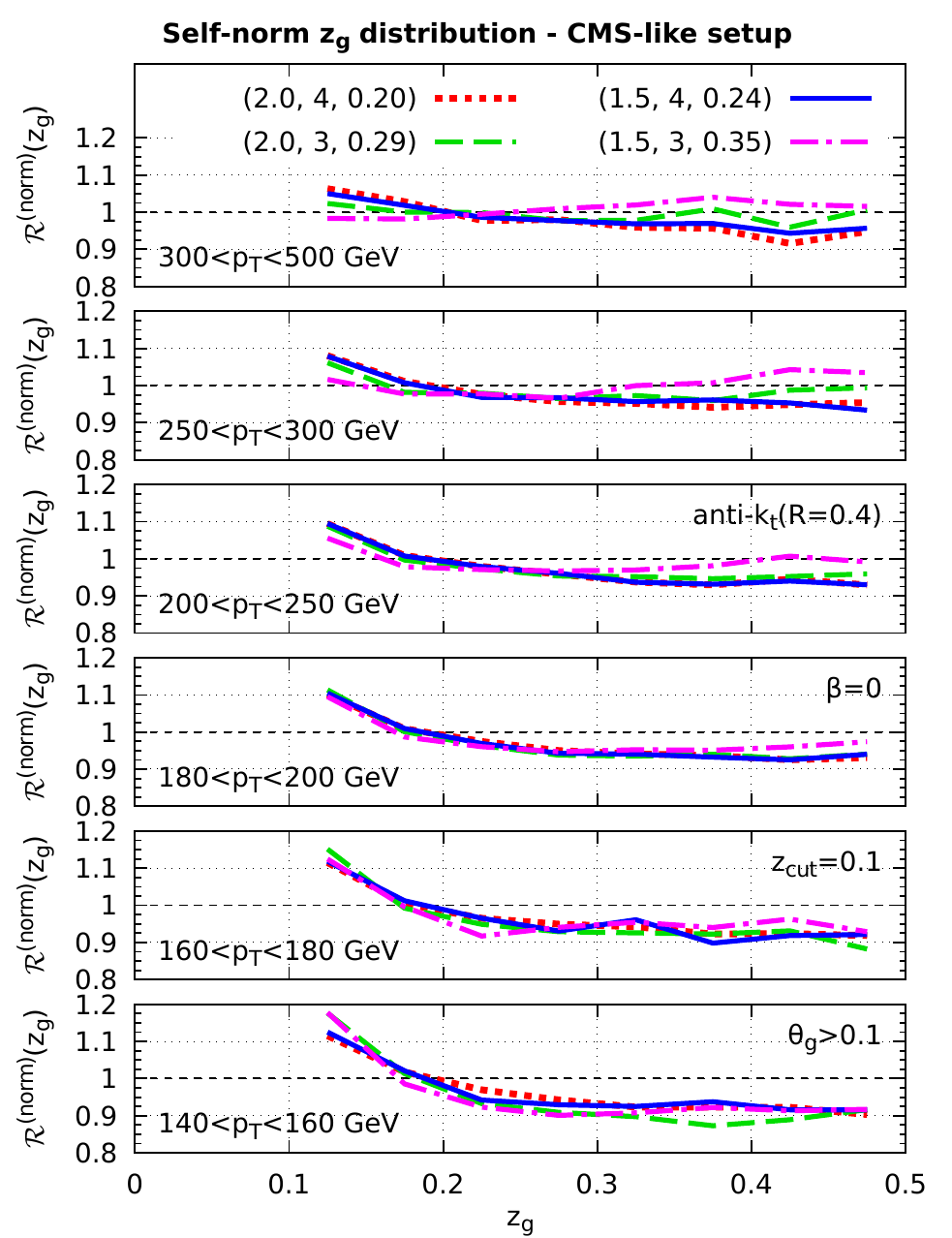}
 \caption{\small  Left: our MC results for jet $R_{AA}$ and for 4 sets of
 medium parameters which give quasi-identical predictions are compared to
 the ATLAS data \cite{Aaboud:2018twu} (black dots with error bars). Right:
 the MC predictions for the medium/vacuum ratio $\mathcal{R}^\text{(norm)}(z_g)$ of the 
self-normalised $z_g$ distributions are presented in bins of $p_T$ for
the same 4 sets of medium parameters as in the left figure.}
 \label{Fig:zgnorm}
\end{figure}

Our findings are shown in Fig.~\ref{Fig:zgnorm}(right).
In the left plot, we show a selection of 4 sets of medium parameters,
$\hat q$, $L$ and $\amed$ (reported from Table~\ref{tab:parameters}
for readability) which provide a good description of the LHC data
\cite{Aaboud:2018twu} for the jet $R_{AA}$ ratio.
In the right plot, we show the corresponding predictions for the $z_g$
nuclear modification factor, $\mathcal{R}^\text{(norm)}(z_g)$, using
the same bins and cuts as in the CMS analysis.

We see that our 4
choices of medium parameters correspond to
somewhat different values for the physical medium scales $\obr$, $\oc$
and $\theta_c$. They therefore lead to different
predictions both for the average energy lost {\it by a single parton}
at large angles, dominated by $\obr$, and for the number and
distribution of sources, which is controlled via the phase-space
boundaries for VLEs by $\oc$ and $\theta_c$.
While the $R_{AA}$ ratio is most sensitive to variations in $\obr$,
small variations in $\obr$ ($\sim 30$\% between our extreme values)
can be compensated by larger variations of $\oc$ and $\theta_c$ (a
factor $\sim 2$ between our extreme values).
Since the interplay between the 3 scales $\obr$,  $\oc$ and $\theta_c$
is different for $R_{AA}$ and $\mathcal{R}^\text{(norm)}(z_g)$, our 4
sets of parameters predict different behaviours for the latter.
However, both observables are predominantly controlled by the energy loss
of the jet, so the spread in $\mathcal{R}^\text{(norm)}(z_g)$ remains
limited.
Some differences are nonetheless observable, in particular for the two
bins with the largest $p_T$. The predictions obtained with a larger
$\obr$ --- i.e.\ larger single-parton energy loss but smaller
phase-space for VLEs --- show a pattern dominated by energy loss,
similar to what was seen in Sect.~\ref{sec:zgmed} for high-$p_T$
jets. Conversely, the predictions obtained with a smaller $\obr$ ---
i.e.\ smaller single-parton energy loss but larger phase-space for
VLEs --- show an enhancement of the small-$z_g$ peak associated to
MIEs.

If we compare these results with the CMS measurements (see e.g.\
Fig.~4 of Ref.~\cite{Sirunyan:2017bsd}), we see that the two agree
within the error bars for both the pattern and the magnitude of the
deviation from one.
In particular, the CMS data too indicate that
$\mathcal{R}^\text{(norm)}$ decreases quasi-monotonously with $z_g$ at
low $p_T$ and become flatter and flatter, approaching unity, when
increasing $p_T$.
This supports our main picture where the nuclear effects on the $z_g$
distribution are a combination of incoherent energy loss affecting a
vacuum-like splitting and a small-$z_g$ peak associated with the SD
condition being triggered by a MIE.
With increasing $p_T$ the first mechanism dominates over the over,
yielding a flatter distribution, in agreement with the CMS data. 
That being said, the current experimental uncertainty does not allow one
to distinguish between different sets of medium parameters.

\subsection{Substructure observables beyond the $z_g$ distribution}\label{sec:beyond-zg}

\begin{figure}[t] 
  \centering
  \includegraphics[page=1,width=0.48\textwidth]{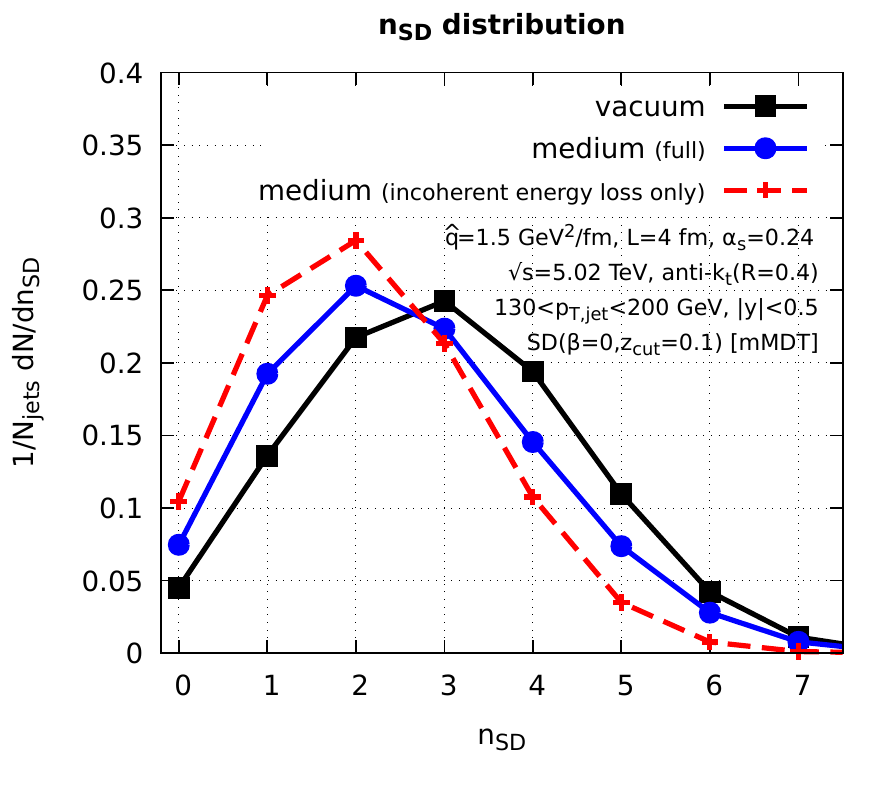}\hfill%
  \includegraphics[page=2,width=0.48\textwidth]{./FIGS/nsd-ALICE.pdf}
 \caption{\small $n_\text{SD}$ distributions emerging from our Monte Carlo simulations. Left: the 
 distributions themselves, for the vacuum shower, for the full in-medium parton shower, and also
 for the case where the MIEs contribute only to the energy loss (but not directly to SD). Right: the
 medium/vacuum ratios.}
 \label{Fig:nsd}
\end{figure}

Our final section discusses two substructure observables related to the $z_g$ distribution.

\paragraph{Iterated SD multiplicity.}
The first observable we consider is the Iterated SD
multiplicity~\cite{Frye:2017yrw}, $n_\text{SD}$, which has also been
measured on track-jets by the ALICE
collaboration~\cite{Acharya:2019djg}.\footnote{Our comparison to this
  measurement is subject to the same caveats that for the $z_g$
  distribution in the same paper.}
Iterated SD proceeds by iterating the Soft-Drop procedure, still
following the hardest branch in the jet, until all declusterings have
been exhausted. $n_\text{SD}$ is then defined as the number of
declusterings passing the SD condition.

Our results for the $n_\text{SD}$ distribution are presented in
Fig.~\ref{Fig:nsd} and show the same trend as the ALICE measurements
(Fig.~4 of Ref.~\cite{Acharya:2019djg}).
In particular, the $n_{\rm SD}$ distribution is shifted to smaller
values for jets created in PbPb collisions compared to pp collisions.
This might seem puzzling at first sight since in the
low-$p_{T,\text{jet}}$ range probed by the measurement, one could
naively expect an enhancement of $n_{\rm SD}$ due to the additional MIEs
passing the SD condition.
However, we believe that the dominant mechanism at play is the
incoherent energy loss which, as discussed in Sect.~\ref{sec:high},
results in an effective $z_g$ fraction smaller than the actual
momentum fraction $z$ at the splitting. This effect lowers the number
of measured hard splittings.
To support this argument, we have run a variant of our MC simulations
where all the partons created via MIEs are moved outside the jet and
hence only they only contribute to the energy loss.
The corresponding results, shown as crosses in Fig.~\ref{Fig:nsd}
demonstrate as expected an even stronger reduction in the average
value of $n_{\rm SD}$, which is only partially compensated by MIEs
captured by the Iterated SD procedure.

\begin{figure}[t] 
  \centering
  \includegraphics[page=1,width=0.48\textwidth]{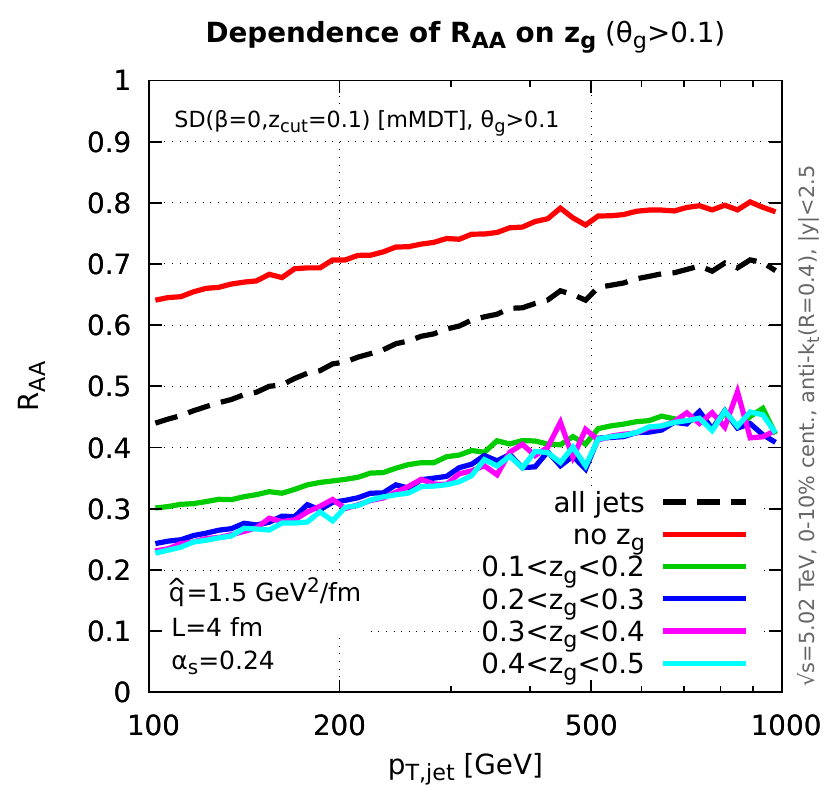}\hfill%
  \includegraphics[page=2,width=0.48\textwidth]{./FIGS/RAA-v-zg}
 \caption{\small  Our MC results for the jet  $R_{AA}$ as a function of $p_{T,{\rm jet}}$ are shown
 in bins of $z_g$ (left figure) and in bins of $\theta_g$ (right figure). The inclusive (all-$z_g$, respectively
 all-$\theta_g$) results are shown with dashed lines.}
 \label{Fig:zgRAA} 
\end{figure}

\paragraph{Correlation between $R_{AA}$ and $z_g$.}
Given that both $R_{AA}$  and the $z_g$ distribution are primarily
controlled by the jet energy loss, it is interesting to study the
correlation between these 2 variables (similarly to
Fig.~\ref{Fig:elosszg} for the energy loss of monochromatic jets). To
that aim,  we show in Fig.~\ref{Fig:zgRAA} the ratio $R_{AA}$ as a
function of $p_T\equiv p_{T,{\rm jet}}$ for different bins in $z_g$
(imposing $\theta_g>0.1$) (left plot) and for different bins in
$\theta_g$ (right plot). For reference, the inclusive $R_{AA}$ ratio is shown by
the ``all jets'' curve.
The curve labelled as ``no $z_g$'' in the left plot includes both the
events which did not pass the SD criteria and the events which failed
the $\theta_g>0.1$ constraint.
Correspondingly, the curve labelled ``$\theta_g<0.03$'' in the right
plot includes both the events with a genuine splitting passing the SD
condition with $\theta_g<0.03$ and the events which did not pass the
SD condition.

The remarkable feature in both plots is the striking difference
between the events which passed SD and those which did not.
As explained when we discussed Fig.~\ref{Fig:elosszg}, this difference
reflects the fact that, on average, two-prong jets lose more energy
than single-prong ones.
These results also reveal the role played by colour (de)coherence and
the emergence of a critical angle $\theta_c$. With our choice of
parameters, $\theta_c\simeq 0.04$ corresponds to the region in
$\theta_g$ where $R_{AA}$ changes significantly. For example, the
curve corresponding to $\theta_g<0.03 < \theta_c$ receives almost
exclusively contributions from single-prong jets --- even more so
than the ``no $z_g$ curve in the left plot --- and thus shows a nuclear
factor $R_{AA}$ close to unity.
This suggests that measuring $R_{AA}$ in bins of $\theta_g$ can be
interesting to better characterise the propagation of jets in the
quark-gluon plasma.

\section{Conclusions and perspectives}
\label{sec:conc}

In this paper, we have presented a new picture, emerging from
perturbative QCD, for the parton shower created by an energetic parton
propagating through a dense quark-gluon plasma. This picture is
factorised in time with vacuum-like emissions occurring first and
creating sources for subsequent medium-induced radiation.
Both types of emission are Markovian processes, yielding a modular
Monte Carlo implementation of our picture. This allows us to study
separately various aspects of the dynamics of jet quenching and assert
their relative importance.
In practice, we have focused on two observables for which we believe
our approximations to be robust: the jet nuclear modification factor
for $R_{AA}$ and the nuclear effects on the $z_g$ distribution given
by the Soft Drop procedure.

For both observables, we obtained good qualitative and
semi-quantitative descriptions of the respective LHC data and
we discussed the physical interpretation of the various trends seen in
the data.
To make our physical discussions more convincing, we supplemented the
numerical calculations of the $z_g$ distribution with suitable
analytic calculations, which were helpful to pinpoint the different mechanisms
at play and compare their effects.
Our
formalism involves a few free parameters, notably $\hat q$ and $L$,
but we have checked that the quality of our description of the data
depends on these parameters only via microscopic scales built with
these parameters.
In particular, the energy scale $\obr\sim \alpha_s^2\hat q L^2$
controls the energy loss via soft medium-induced emissions at large
angles.

We found that these two observables are to a large
extent controlled by the jet energy loss.
They are therefore very sensitive both to vacuum-like emissions (which
drive the number of sources for medium-induced radiation) and to
medium-induced emissions.
In particular, the increase of the number of vacuum-like emissions
with the jet $p_T$ causes an increase in the jet energy loss.

We showed that the nuclear $z_g$ distribution is affected by both a
{\it direct} contribution of the MIEs and their {\it indirect}
contribution, via the incoherent energy loss of the two subjets
selected by the SD procedure.
The former leads to a pronounced rise of the $z_g$ distribution at
small $z_g$, as seen in the data.
The latter causes a nuclear suppression ($\mathcal{R}(z_g)<1$ and slowly increasing with
$z_g$), which is best seen when normalising the $z_g$ distribution to
the total number of jets (the \njets-normalised distributions in our
nomenclature).
The interplay between the two effects depends on the jet $p_T$.
At low $p_T$, $p_T\lesssim 300$~GeV, corresponding to the current
range covered by the LHC analyses, both effects
contribute.
As the jet $p_T$ increases, MIEs become too soft to trigger the SD
condition and only the indirect effect of (incoherent) energy loss
survives.
As a consequence, we predict that for $p_T\gtrsim 500$~GeV, the
(\njets-normalised) nuclear modification factor $\mathcal{R}(z_g)$
should be systematically smaller than one and slowly increasing with
$z_g$.
The onset of such a transition is consistent with the largest $p_T$
bin of the CMS measurement.

Some of our results have been anticipated by previous studies in the
literature, with somewhat different conclusions.
When studying the consequences of the incoherent energy loss on the
$z_g$ distribution, Refs.~\cite{Mehtar-Tani:2016aco,Chang:2017gkt}
obtained results which are qualitatively similar to the curve denoted
as ``energy loss only'' in our Fig.~\ref{Fig:lowptMIEVLE}, right, i.e.\
$\mathcal{R}(z_g)<1$ with an increase with $z_g$.
This trend is opposite to the one seen in the LHC data at
$p_T\lesssim 250$~GeV \cite{Sirunyan:2017bsd}, which led
Refs.~\cite{Mehtar-Tani:2016aco,Chang:2017gkt} to argue that the LHC
data favours a scenario of {\it coherent} energy loss by the two
subjets.
Such a conclusion seems difficult to reconcile with the
fact that, in the CMS analysis~\cite{Sirunyan:2017bsd}, the angular
separation $\theta_g$ between the two subjets is constrained to values
$\theta_g\ge \thetacut=0.1$ which, with our current estimates,
are considerably larger than the critical angle $\theta_c$ for the
onset of colour decoherence.
In our picture, we instead conclude that the rise of the $z_g$
distribution at small $z_g$ is due to the relatively hard MIEs that
can be captured by SD and which more than compensate the
suppression due to the incoherent energy loss. 
This is indeed visible for our full Monte Carlo
results in Fig.~\ref{Fig:lowptMIEVLE}, right.
This shows the importance of having a complete physical picture
and the usefulness of the corresponding numerical implementation.

In the remaining part of these conclusions, we emphasise the
limitations of our current approach and thus outline some of our
projects for the future. Even though we describe the parton showers
from first principles, our current treatment of the medium --- a
homogeneous ``brick'' of quark-gluon plasma --- is insufficient for
more detailed phenomenological studies. Besides our implementation of
an angular-ordered final-state parton shower can be significantly
refined.

A first step towards a more realistic description of the medium is to
include its longitudinal expansion, e.g.\ by giving a suitable
time-dependence to the jet quenching parameter $\hat q$
\cite{Baier:1998yf,Zakharov:1998wq, Arnold:2008iy,Iancu:2018trm}. One
would also need a more realistic geometry (say, an expanding cylinder
for central collisions) together with a probability distribution for
the location of the hard process inside the medium.
Further refinements would also include the radial expansion and a
fully dynamical description of the plasma, including its response to
the jet.
The current belief is that the inclusion of the medium backreaction is
important for observables like the jet shapes
\cite{Chatrchyan:2013kwa} and the geometrical distribution of the
energy lost by the jet, e.g.\ as measured by the dijet asymmetry
\cite{Aad:2010bu,Chatrchyan:2011sx,Khachatryan:2016erx}.

Several other improvements of the medium-induced cascade can be
implemented.
First, we should relax the fixed coupling approximation in the
treatment of the medium-induced radiation. Second, one should use a
dynamic treatment of the transverse momentum broadening, which
explicitly includes the elastic collisions and thus goes beyond our
current Gaussian approximation.
In practice, hard collisions would generate a power-law tail
$\propto \kt^{-4}$ at large $\kt$.
This can have a sizeable impact on
the $z_g$ distribution which, as discussed in Sect.~\ref{sec:low}, is
sensitive to the large-$\kt$ tail of the $k_\perp$ distribution.
A Monte Carlo implementation of the elastic collisions (in the
diffusion approximation) has recently been presented
\cite{Kutak:2018dim} for a jet evolving via (BDMPS-Z) medium-induced
radiation alone. In the future, we plan to extend the method in
\cite{Kutak:2018dim} to the full parton shower, including VLEs. More
generally, one could also add the effects of elastic collisions in
terms of energy loss (the ``drag force'') and longitudinal momentum
broadening. This would be important for the in-medium dynamics of the
softest quanta from the jet, in particular for the possibility of
their
thermalisation~\cite{Baier:2000sb,Kurkela:2015qoa,Iancu:2015uja}.

Finally, our vacuum parton shower should be extended to include the
single-logarithmic effects of soft gluon emissions beyond the
collinear limit, and to include both final-state and initial-state
radiation.
This can e.g.\ be done using a dipole shower which would also have the
advantage of facilitating the interface of our parton shower with
hadronisation models.

Instead of going through the complex task of completing our Monte
Carlo with a detailed description of the medium and of its
interactions with the jet, one can alternatively think about using our
parton showers as an input for the recently developed
JETSCAPE~\cite{Cao:2017zih,Putschke:2019yrg} framework, which offers various
approaches for treating the interactions between the parton shower and
the medium.
Last but not least, it would be both important and instructive to
understand in detail the relation between our approach and the
related approaches in the literature, notably those used by the
Monte Carlo event generators MARTINI
\cite{Schenke:2009gb} and JEWEL~\cite{Zapp:2011ya,Zapp:2012ak}, which
share with us the fact that both the parton showers and the in-medium
interactions are treated in perturbative QCD.

\smallskip

\section*{Acknowledgements}

We thank Al Mueller for inspiring remarks as well as Yacine
Mehtar-Tani and Konrad Tywoniuk for useful discussions.
The work of E.I. and G.S. is supported in part by the Agence Nationale de la Recherche project 
 ANR-16-CE31-0019-01.

\bigskip
\bibliographystyle{utcaps}
\bibliography{refs}

\providecommand{\href}[2]{#2}\begingroup\raggedright\begin{thebibliography}{10}

\bibitem{Mehtar-Tani:2013pia}
Y.~Mehtar-Tani, J.~G. Milhano, and K.~Tywoniuk, ``{Jet physics in heavy-ion
  collisions},'' \href{http://dx.doi.org/10.1142/S0217751X13400137}{{\em
  Int.J.Mod.Phys.} {\bfseries A28} (2013) 1340013},
\href{http://arxiv.org/abs/1302.2579}{{\ttfamily arXiv:1302.2579 [hep-ph]}}.

\bibitem{Blaizot:2015lma}
J.-P. Blaizot and Y.~Mehtar-Tani, ``{Jet Structure in Heavy Ion Collisions},''
  \href{http://dx.doi.org/10.1142/S021830131530012X}{{\em Int. J. Mod. Phys.}
  {\bfseries E24} no.~11, (2015) 1530012},
\href{http://arxiv.org/abs/1503.05958}{{\ttfamily arXiv:1503.05958 [hep-ph]}}.

\bibitem{Qin:2015srf}
G.-Y. Qin and X.-N. Wang, ``{Jet quenching in high-energy heavy-ion
  collisions},'' \href{http://dx.doi.org/10.1142/S0218301315300143,
  10.1142/9789814663717_0007}{{\em Int. J. Mod. Phys.} {\bfseries E24} no.~11,
  (2015) 1530014}, \href{http://arxiv.org/abs/1511.00790}{{\ttfamily
  arXiv:1511.00790 [hep-ph]}}.
[,309(2016)].

\bibitem{Andrews:2018jcm}
H.~A. Andrews {\em et al.}, ``{Novel tools and observables for jet physics in
  heavy-ion collisions},''
\href{http://arxiv.org/abs/1808.03689}{{\ttfamily arXiv:1808.03689 [hep-ph]}}.

\bibitem{Larkoski:2015lea}
A.~J. Larkoski, S.~Marzani, and J.~Thaler, ``{Sudakov Safety in Perturbative
  QCD},'' \href{http://dx.doi.org/10.1103/PhysRevD.91.111501}{{\em Phys. Rev.}
  {\bfseries D91} no.~11, (2015) 111501},
\href{http://arxiv.org/abs/1502.01719}{{\ttfamily arXiv:1502.01719 [hep-ph]}}.

\bibitem{Larkoski:2014wba}
A.~J. Larkoski, S.~Marzani, G.~Soyez, and J.~Thaler, ``{Soft Drop},''
  \href{http://dx.doi.org/10.1007/JHEP05(2014)146}{{\em JHEP} {\bfseries 05}
  (2014) 146},
\href{http://arxiv.org/abs/1402.2657}{{\ttfamily arXiv:1402.2657 [hep-ph]}}.

\bibitem{Sirunyan:2017bsd}
{\bfseries CMS} Collaboration, A.~M. Sirunyan {\em et al.}, ``{Measurement of
  the Splitting Function in $pp$ and Pb-Pb Collisions at
  $\sqrt{s_{_{\mathrm{NN}}}} =$ 5.02 TeV},''
  \href{http://dx.doi.org/10.1103/PhysRevLett.120.142302}{{\em Phys. Rev.
  Lett.} {\bfseries 120} no.~14, (2018) 142302},
\href{http://arxiv.org/abs/1708.09429}{{\ttfamily arXiv:1708.09429 [nucl-ex]}}.

\bibitem{Acharya:2019djg}
{\bfseries ALICE} Collaboration, S.~Acharya {\em et al.}, ``{Exploration of jet
  substructure using iterative declustering in pp and Pb-Pb collisions at LHC
  energies},''
\href{http://arxiv.org/abs/1905.02512}{{\ttfamily arXiv:1905.02512 [nucl-ex]}}.

\bibitem{Chien:2016led}
Y.-T. Chien and I.~Vitev, ``{Probing the Hardest Branching within Jets in
  Heavy-Ion Collisions},''
  \href{http://dx.doi.org/10.1103/PhysRevLett.119.112301}{{\em Phys. Rev.
  Lett.} {\bfseries 119} no.~11, (2017) 112301},
\href{http://arxiv.org/abs/1608.07283}{{\ttfamily arXiv:1608.07283 [hep-ph]}}.

\bibitem{Mehtar-Tani:2016aco}
Y.~Mehtar-Tani and K.~Tywoniuk, ``{Groomed jets in heavy-ion collisions:
  sensitivity to medium-induced bremsstrahlung},''
  \href{http://dx.doi.org/10.1007/JHEP04(2017)125}{{\em JHEP} {\bfseries 04}
  (2017) 125},
\href{http://arxiv.org/abs/1610.08930}{{\ttfamily arXiv:1610.08930 [hep-ph]}}.

\bibitem{Chang:2017gkt}
N.-B. Chang, S.~Cao, and G.-Y. Qin, ``{Probing medium-induced jet splitting and
  energy loss in heavy-ion collisions},''
  \href{http://dx.doi.org/10.1016/j.physletb.2018.04.019}{{\em Phys. Lett.}
  {\bfseries B781} (2018) 423--432},
\href{http://arxiv.org/abs/1707.03767}{{\ttfamily arXiv:1707.03767 [hep-ph]}}.

\bibitem{Milhano:2017nzm}
G.~Milhano, U.~A. Wiedemann, and K.~C. Zapp, ``{Sensitivity of jet substructure
  to jet-induced medium response},''
  \href{http://dx.doi.org/10.1016/j.physletb.2018.01.029}{{\em Phys. Lett.}
  {\bfseries B779} (2018) 409--413},
\href{http://arxiv.org/abs/1707.04142}{{\ttfamily arXiv:1707.04142 [hep-ph]}}.

\bibitem{Dokshitzer:1997in}
Y.~L. Dokshitzer, G.~D. Leder, S.~Moretti, and B.~R. Webber, ``{Better jet
  clustering algorithms},''
  \href{http://dx.doi.org/10.1088/1126-6708/1997/08/001}{{\em JHEP} {\bfseries
  08} (1997) 001},
\href{http://arxiv.org/abs/hep-ph/9707323}{{\ttfamily arXiv:hep-ph/9707323
  [hep-ph]}}.

\bibitem{Wobisch:1998wt}
M.~Wobisch and T.~Wengler, ``{Hadronization corrections to jet cross-sections
  in deep inelastic scattering},'' in {\em {Monte Carlo generators for HERA
  physics. Proceedings, Workshop, Hamburg, Germany, 1998-1999}}, pp.~270--279.
\newblock 1998.
\newblock
\href{http://arxiv.org/abs/hep-ph/9907280}{{\ttfamily arXiv:hep-ph/9907280
  [hep-ph]}}.
\newblock

\bibitem{Dokshitzer:1991wu}
Y.~L. Dokshitzer, V.~A. Khoze, A.~H. Mueller, and S.~I. Troian, ``{Basics of
  perturbative QCD},''. Gif-sur-Yvette, France. Ed. Frontieres (1991) 274 p.

\bibitem{Baier:1996kr}
R.~Baier, Y.~L. Dokshitzer, A.~H. Mueller, S.~Peigne, and D.~Schiff,
  ``{Radiative Energy Loss of High Energy Quarks and Gluons in a Finite-Volume
  Quark-Gluon Plasma},''
  \href{http://dx.doi.org/10.1016/S0550-3213(96)00553-6}{{\em Nucl. Phys.}
  {\bfseries B483} (1997) 291--320},
\href{http://arxiv.org/abs/hep-ph/9607355}{{\ttfamily arXiv:hep-ph/9607355}}.

\bibitem{Baier:1996sk}
R.~Baier, Y.~L. Dokshitzer, A.~H. Mueller, S.~Peigne, and D.~Schiff,
  ``{Radiative Energy Loss and P(T)-Broadening of High Energy Partons in
  Nuclei},'' \href{http://dx.doi.org/10.1016/S0550-3213(96)00581-0}{{\em Nucl.
  Phys.} {\bfseries B484} (1997) 265--282},
\href{http://arxiv.org/abs/hep-ph/9608322}{{\ttfamily arXiv:hep-ph/9608322}}.

\bibitem{Zakharov:1996fv}
B.~G. Zakharov, ``{Fully Quantum Treatment of the Landau-Pomeranchuk-Migdal
  Effect in QED and QCD},'' \href{http://dx.doi.org/10.1134/1.567126}{{\em JETP
  Lett.} {\bfseries 63} (1996) 952--957},
\href{http://arxiv.org/abs/hep-ph/9607440}{{\ttfamily arXiv:hep-ph/9607440}}.

\bibitem{Zakharov:1997uu}
B.~G. Zakharov, ``{Radiative Energy Loss of High Energy Quarks in Finite-Size
  Nuclear Matter and Quark-Gluon Plasma},''
  \href{http://dx.doi.org/10.1134/1.567389}{{\em JETP Lett.} {\bfseries 65}
  (1997) 615--620},
\href{http://arxiv.org/abs/hep-ph/9704255}{{\ttfamily arXiv:hep-ph/9704255}}.

\bibitem{Baier:1998kq}
R.~Baier, Y.~L. Dokshitzer, A.~H. Mueller, and D.~Schiff, ``{Medium-Induced
  Radiative Energy Loss: Equivalence Between the Bdmps and Zakharov
  Formalisms},'' \href{http://dx.doi.org/10.1016/S0550-3213(98)00546-X}{{\em
  Nucl. Phys.} {\bfseries B531} (1998) 403--425},
\href{http://arxiv.org/abs/hep-ph/9804212}{{\ttfamily arXiv:hep-ph/9804212}}.

\bibitem{MehtarTani:2010ma}
Y.~Mehtar-Tani, C.~A. Salgado, and K.~Tywoniuk, ``{Antiangular Ordering of
  Gluon Radiation in QCD Media},''
  \href{http://dx.doi.org/10.1103/PhysRevLett.106.122002}{{\em Phys. Rev.
  Lett.} {\bfseries 106} (2011) 122002},
\href{http://arxiv.org/abs/1009.2965}{{\ttfamily arXiv:1009.2965 [hep-ph]}}.

\bibitem{MehtarTani:2011tz}
Y.~Mehtar-Tani, C.~A. Salgado, and K.~Tywoniuk, ``{Jets in QCD Media: from
  Color Coherence to Decoherence},''
  \href{http://dx.doi.org/10.1016/j.physletb.2011.12.042}{{\em Phys. Lett.}
  {\bfseries B707} (2012) 156--159},
\href{http://arxiv.org/abs/1102.4317}{{\ttfamily arXiv:1102.4317 [hep-ph]}}.

\bibitem{CasalderreySolana:2011rz}
J.~Casalderrey-Solana and E.~Iancu, ``{Interference Effects in Medium-Induced
  Gluon Radiation},'' \href{http://dx.doi.org/10.1007/JHEP08(2011)015}{{\em
  JHEP} {\bfseries 08} (2011) 015},
\href{http://arxiv.org/abs/1105.1760}{{\ttfamily arXiv:1105.1760 [hep-ph]}}.

\bibitem{Blaizot:2012fh}
J.-P. Blaizot, F.~Dominguez, E.~Iancu, and Y.~Mehtar-Tani, ``{Medium-induced
  gluon branching},'' \href{http://dx.doi.org/10.1007/JHEP01(2013)143}{{\em
  JHEP} {\bfseries 1301} (2013) 143},
\href{http://arxiv.org/abs/1209.4585}{{\ttfamily arXiv:1209.4585 [hep-ph]}}.

\bibitem{Blaizot:2013vha}
J.-P. Blaizot, F.~Dominguez, E.~Iancu, and Y.~Mehtar-Tani, ``{Probabilistic
  picture for medium-induced jet evolution},''
  \href{http://dx.doi.org/10.1007/JHEP06(2014)075}{{\em JHEP} {\bfseries 1406}
  (2014) 075},
\href{http://arxiv.org/abs/1311.5823}{{\ttfamily arXiv:1311.5823 [hep-ph]}}.

\bibitem{Apolinario:2014csa}
L.~Apolin\'ario, N.~Armesto, J.~G. Milhano, and C.~A. Salgado,
  ``{Medium-induced gluon radiation and colour decoherence beyond the soft
  approximation},'' \href{http://dx.doi.org/10.1007/JHEP02(2015)119}{{\em JHEP}
  {\bfseries 1502} (2015) 119},
\href{http://arxiv.org/abs/1407.0599}{{\ttfamily arXiv:1407.0599 [hep-ph]}}.

\bibitem{Caucal:2018dla}
P.~Caucal, E.~Iancu, A.~H. Mueller, and G.~Soyez, ``{Vacuum-like jet
  fragmentation in a dense QCD medium},''
  \href{http://dx.doi.org/10.1103/PhysRevLett.120.232001}{{\em Phys. Rev.
  Lett.} {\bfseries 120} (2018) 232001},
\href{http://arxiv.org/abs/1801.09703}{{\ttfamily arXiv:1801.09703 [hep-ph]}}.

\bibitem{Blaizot:2013hx}
J.-P. Blaizot, E.~Iancu, and Y.~Mehtar-Tani, ``{Medium-induced QCD cascade:
  democratic branching and wave turbulence},''
  \href{http://dx.doi.org/10.1103/PhysRevLett.111.052001}{{\em Phys.Rev.Lett.}
  {\bfseries 111} (2013) 052001},
\href{http://arxiv.org/abs/1301.6102}{{\ttfamily arXiv:1301.6102 [hep-ph]}}.

\bibitem{Fister:2014zxa}
L.~Fister and E.~Iancu, ``{Medium-induced jet evolution: wave turbulence and
  energy loss},'' \href{http://dx.doi.org/10.1007/JHEP03(2015)082}{{\em JHEP}
  {\bfseries 03} (2015) 082},
\href{http://arxiv.org/abs/1409.2010}{{\ttfamily arXiv:1409.2010 [hep-ph]}}.

\bibitem{Aaboud:2018twu}
{\bfseries ATLAS} Collaboration, M.~Aaboud {\em et al.}, ``{Measurement of the
  nuclear modification factor for inclusive jets in Pb+Pb collisions at
  $\sqrt{s_\mathrm{NN}}=5.02$ TeV with the ATLAS detector},''
  \href{http://dx.doi.org/10.1016/j.physletb.2018.10.076}{{\em Phys. Lett.}
  {\bfseries B790} (2019) 108--128},
\href{http://arxiv.org/abs/1805.05635}{{\ttfamily arXiv:1805.05635 [nucl-ex]}}.

\bibitem{Kauder:2017cvz}
{\bfseries STAR} Collaboration, K.~Kauder, ``{Measurement of the Shared
  Momentum Fraction $z_g$ using Jet Reconstruction in p+p and Au+Au Collisions
  with STAR},'' \href{http://dx.doi.org/10.1016/j.nuclphysbps.2017.05.028}{{\em
  Nucl. Part. Phys. Proc.} {\bfseries 289-290} (2017) 137--140},
\href{http://arxiv.org/abs/1703.10933}{{\ttfamily arXiv:1703.10933 [nucl-ex]}}.

\bibitem{Liou:2013qya}
T.~Liou, A.~Mueller, and B.~Wu, ``{Radiative $p_\bot$-broadening of high-energy
  quarks and gluons in QCD matter},''
  \href{http://dx.doi.org/10.1016/j.nuclphysa.2013.08.005}{{\em Nucl.Phys.}
  {\bfseries A916} (2013) 102--125},
\href{http://arxiv.org/abs/1304.7677}{{\ttfamily arXiv:1304.7677 [hep-ph]}}.

\bibitem{Blaizot:2014bha}
J.-P. Blaizot and Y.~Mehtar-Tani, ``{Renormalization of the jet-quenching
  parameter},'' \href{http://dx.doi.org/10.1016/j.nuclphysa.2014.05.018}{{\em
  Nucl.Phys.} {\bfseries A929} (2014) 202--229},
\href{http://arxiv.org/abs/1403.2323}{{\ttfamily arXiv:1403.2323 [hep-ph]}}.

\bibitem{Iancu:2014kga}
E.~Iancu, ``{The non-linear evolution of jet quenching},''
  \href{http://dx.doi.org/10.1007/JHEP10(2014)095}{{\em JHEP} {\bfseries 1410}
  (2014) 95},
\href{http://arxiv.org/abs/1403.1996}{{\ttfamily arXiv:1403.1996 [hep-ph]}}.

\bibitem{MehtarTani:2011gf}
Y.~Mehtar-Tani, C.~A. Salgado, and K.~Tywoniuk, ``{The Radiation Pattern of a
  QCD Antenna in a Dilute Medium},''
  \href{http://dx.doi.org/10.1007/JHEP04(2012)064}{{\em JHEP} {\bfseries 04}
  (2012) 064},
\href{http://arxiv.org/abs/1112.5031}{{\ttfamily arXiv:1112.5031 [hep-ph]}}.

\bibitem{Arnold:2015qya}
P.~Arnold and S.~Iqbal, ``{The LPM effect in sequential bremsstrahlung},''
  \href{http://dx.doi.org/10.1007/JHEP09(2016)072,
  10.1007/JHEP04(2015)070}{{\em JHEP} {\bfseries 04} (2015) 070},
  \href{http://arxiv.org/abs/1501.04964}{{\ttfamily arXiv:1501.04964
  [hep-ph]}}.
[Erratum: JHEP09,072(2016)].

\bibitem{Arnold:2016kek}
P.~Arnold, H.-C. Chang, and S.~Iqbal, ``{The LPM effect in sequential
  bremsstrahlung 2: factorization},''
  \href{http://dx.doi.org/10.1007/JHEP09(2016)078}{{\em JHEP} {\bfseries 09}
  (2016) 078},
\href{http://arxiv.org/abs/1605.07624}{{\ttfamily arXiv:1605.07624 [hep-ph]}}.

\bibitem{Wiedemann:2000za}
U.~A. Wiedemann, ``{Gluon Radiation Off Hard Quarks in a Nuclear Environment:
  Opacity Expansion},''
  \href{http://dx.doi.org/10.1016/S0550-3213(00)00457-0}{{\em Nucl. Phys.}
  {\bfseries B588} (2000) 303--344},
\href{http://arxiv.org/abs/hep-ph/0005129}{{\ttfamily arXiv:hep-ph/0005129}}.

\bibitem{Wiedemann:2000tf}
U.~A. Wiedemann, ``{Jet Quenching Versus Jet Enhancement: a Quantitative Study
  of the Bdmps-Z Gluon Radiation Spectrum},''
  \href{http://dx.doi.org/10.1016/S0375-9474(01)00362-1}{{\em Nucl. Phys.}
  {\bfseries A690} (2001) 731--751},
\href{http://arxiv.org/abs/hep-ph/0008241}{{\ttfamily arXiv:hep-ph/0008241}}.

\bibitem{Arnold:2001ba}
P.~B. Arnold, G.~D. Moore, and L.~G. Yaffe, ``{Photon Emission from
  Ultrarelativistic Plasmas},'' {\em JHEP} {\bfseries 11} (2001) 057,
\href{http://arxiv.org/abs/hep-ph/0109064}{{\ttfamily arXiv:hep-ph/0109064}}.

\bibitem{Arnold:2001ms}
P.~B. Arnold, G.~D. Moore, and L.~G. Yaffe, ``{Photon Emission from Quark Gluon
  Plasma: Complete Leading Order Results},'' {\em JHEP} {\bfseries 12} (2001)
  009,
\href{http://arxiv.org/abs/hep-ph/0111107}{{\ttfamily arXiv:hep-ph/0111107}}.

\bibitem{Arnold:2002ja}
P.~B. Arnold, G.~D. Moore, and L.~G. Yaffe, ``{Photon and Gluon Emission in
  Relativistic Plasmas},'' {\em JHEP} {\bfseries 06} (2002) 030,
\href{http://arxiv.org/abs/hep-ph/0204343}{{\ttfamily arXiv:hep-ph/0204343}}.

\bibitem{Mehtar-Tani:2018zba}
Y.~Mehtar-Tani and S.~Schlichting, ``{Universal quark to gluon ratio in
  medium-induced parton cascade},''
  \href{http://dx.doi.org/10.1007/JHEP09(2018)144}{{\em JHEP} {\bfseries 09}
  (2018) 144},
\href{http://arxiv.org/abs/1807.06181}{{\ttfamily arXiv:1807.06181 [hep-ph]}}.

\bibitem{Iancu:2015uja}
E.~Iancu and B.~Wu, ``{Thermalization of mini-jets in a quark-gluon plasma},''
  \href{http://dx.doi.org/10.1007/JHEP10(2015)155}{{\em JHEP} {\bfseries 10}
  (2015) 155},
\href{http://arxiv.org/abs/1506.07871}{{\ttfamily arXiv:1506.07871 [hep-ph]}}.

\bibitem{Blaizot:2014ula}
J.-P. Blaizot, Y.~Mehtar-Tani, and M.~A.~C. Torres, ``{Angular structure of the
  in-medium QCD cascade},''
  \href{http://dx.doi.org/10.1103/PhysRevLett.114.222002}{{\em Phys.Rev.Lett.}
  {\bfseries 114} no.~22, (2015) 222002},
\href{http://arxiv.org/abs/1407.0326}{{\ttfamily arXiv:1407.0326 [hep-ph]}}.

\bibitem{Blaizot:2014rla}
J.-P. Blaizot, L.~Fister, and Y.~Mehtar-Tani, ``{Angular distribution of
  medium-induced QCD cascades},''
  \href{http://dx.doi.org/10.1016/j.nuclphysa.2015.03.014}{{\em Nucl.Phys.}
  {\bfseries A940} (2015) 67--88},
\href{http://arxiv.org/abs/1409.6202}{{\ttfamily arXiv:1409.6202 [hep-ph]}}.

\bibitem{Escobedo:2016jbm}
M.~A. Escobedo and E.~Iancu, ``{Event-by-event fluctuations in the
  medium-induced jet evolution},''
  \href{http://dx.doi.org/10.1007/JHEP05(2016)008}{{\em JHEP} {\bfseries 05}
  (2016) 008},
\href{http://arxiv.org/abs/1601.03629}{{\ttfamily arXiv:1601.03629 [hep-ph]}}.

\bibitem{Escobedo:2016vba}
M.~A. Escobedo and E.~Iancu, ``{Multi-particle correlations and KNO scaling in
  the medium-induced jet evolution},''
  \href{http://dx.doi.org/10.1007/JHEP12(2016)104}{{\em JHEP} {\bfseries 12}
  (2016) 104},
\href{http://arxiv.org/abs/1609.06104}{{\ttfamily arXiv:1609.06104 [hep-ph]}}.

\bibitem{Baier:2000sb}
R.~Baier, A.~H. Mueller, D.~Schiff, and D.~Son, ``{'Bottom up' thermalization
  in heavy ion collisions},''
  \href{http://dx.doi.org/10.1016/S0370-2693(01)00191-5}{{\em Phys.Lett.}
  {\bfseries B502} (2001) 51--58},
\href{http://arxiv.org/abs/hep-ph/0009237}{{\ttfamily arXiv:hep-ph/0009237
  [hep-ph]}}.

\bibitem{Cacciari:2011ma}
M.~Cacciari, G.~P. Salam, and G.~Soyez, ``{FastJet User Manual},''
  \href{http://dx.doi.org/10.1140/epjc/s10052-012-1896-2}{{\em Eur. Phys. J.}
  {\bfseries C72} (2012) 1896},
\href{http://arxiv.org/abs/1111.6097}{{\ttfamily arXiv:1111.6097 [hep-ph]}}.

\bibitem{Cacciari:2008gp}
M.~Cacciari, G.~P. Salam, and G.~Soyez, ``{The anti-$k_t$ jet clustering
  algorithm},'' \href{http://dx.doi.org/10.1088/1126-6708/2008/04/063}{{\em
  JHEP} {\bfseries 04} (2008) 063},
\href{http://arxiv.org/abs/0802.1189}{{\ttfamily arXiv:0802.1189 [hep-ph]}}.

\bibitem{Dreyer:2018nbf}
F.~A. Dreyer, G.~P. Salam, and G.~Soyez, ``{The Lund Jet Plane},''
  \href{http://dx.doi.org/10.1007/JHEP12(2018)064}{{\em JHEP} {\bfseries 12}
  (2018) 064},
\href{http://arxiv.org/abs/1807.04758}{{\ttfamily arXiv:1807.04758 [hep-ph]}}.

\bibitem{Dasgupta:2013ihk}
M.~Dasgupta, A.~Fregoso, S.~Marzani, and G.~P. Salam, ``{Towards an
  understanding of jet substructure},''
  \href{http://dx.doi.org/10.1007/JHEP09(2013)029}{{\em JHEP} {\bfseries 09}
  (2013) 029},
\href{http://arxiv.org/abs/1307.0007}{{\ttfamily arXiv:1307.0007 [hep-ph]}}.

\bibitem{Cacciari:2008gn}
M.~Cacciari, G.~P. Salam, and G.~Soyez, ``{The Catchment Area of Jets},''
  \href{http://dx.doi.org/10.1088/1126-6708/2008/04/005}{{\em JHEP} {\bfseries
  04} (2008) 005},
\href{http://arxiv.org/abs/0802.1188}{{\ttfamily arXiv:0802.1188 [hep-ph]}}.

\bibitem{CasalderreySolana:2012ef}
J.~Casalderrey-Solana, Y.~Mehtar-Tani, C.~A. Salgado, and K.~Tywoniuk, ``{New
  picture of jet quenching dictated by color coherence},''
  \href{http://dx.doi.org/10.1016/j.physletb.2013.07.046}{{\em Phys.Lett.}
  {\bfseries B725} (2013) 357--360},
\href{http://arxiv.org/abs/1210.7765}{{\ttfamily arXiv:1210.7765 [hep-ph]}}.

\bibitem{Frye:2017yrw}
C.~Frye, A.~J. Larkoski, J.~Thaler, and K.~Zhou, ``{Casimir Meets Poisson:
  Improved Quark/Gluon Discrimination with Counting Observables},''
  \href{http://dx.doi.org/10.1007/JHEP09(2017)083}{{\em JHEP} {\bfseries 09}
  (2017) 083},
\href{http://arxiv.org/abs/1704.06266}{{\ttfamily arXiv:1704.06266 [hep-ph]}}.

\bibitem{Baier:1998yf}
R.~Baier, Y.~L. Dokshitzer, A.~H. Mueller, and D.~Schiff, ``{Radiative Energy
  Loss of High Energy Partons Traversing an Expanding {QCD} Plasma},''
  \href{http://dx.doi.org/10.1103/PhysRevC.58.1706}{{\em Phys. Rev.} {\bfseries
  C58} (1998) 1706--1713},
\href{http://arxiv.org/abs/hep-ph/9803473}{{\ttfamily arXiv:hep-ph/9803473}}.

\bibitem{Zakharov:1998wq}
B.~G. Zakharov, ``{Quark energy loss in an expanding quark gluon plasma},'' in
  {\em {QCD and high energy hadronic interactions. Proceedings, 33rd Rencontres
  de Moriond, Les Arcs, France, March 21-28, 1998}}, pp.~533--538.
\newblock 1998.
\newblock
\href{http://arxiv.org/abs/hep-ph/9807396}{{\ttfamily arXiv:hep-ph/9807396
  [hep-ph]}}.
\newblock

\bibitem{Arnold:2008iy}
P.~B. Arnold, ``{Simple Formula for High-Energy Gluon Bremsstrahlung in a
  Finite, Expanding Medium},''
  \href{http://dx.doi.org/10.1103/PhysRevD.79.065025}{{\em Phys. Rev.}
  {\bfseries D79} (2009) 065025},
\href{http://arxiv.org/abs/0808.2767}{{\ttfamily arXiv:0808.2767 [hep-ph]}}.

\bibitem{Iancu:2018trm}
E.~Iancu, P.~Taels, and B.~Wu, ``{Jet quenching parameter in an expanding QCD
  plasma},'' \href{http://dx.doi.org/10.1016/j.physletb.2018.10.007}{{\em Phys.
  Lett.} {\bfseries B786} (2018) 288--295},
\href{http://arxiv.org/abs/1806.07177}{{\ttfamily arXiv:1806.07177 [hep-ph]}}.

\bibitem{Chatrchyan:2013kwa}
{\bfseries CMS Collaboration} Collaboration, S.~Chatrchyan {\em et al.},
  ``{Modification of jet shapes in PbPb collisions at $\sqrt {s_{NN}} = 2.76$
  TeV},'' \href{http://dx.doi.org/10.1016/j.physletb.2014.01.042}{{\em
  Phys.Lett.} {\bfseries B730} (2014) 243--263},
\href{http://arxiv.org/abs/1310.0878}{{\ttfamily arXiv:1310.0878 [nucl-ex]}}.

\bibitem{Aad:2010bu}
{\bfseries Atlas} Collaboration, G.~Aad {\em et al.}, ``{Observation of a
  Centrality-Dependent Dijet Asymmetry in Lead-Lead Collisions at
  $\sqrt{s_{NN}}$= 2.76 TeV with the Atlas Detector at the LHC},''
  \href{http://dx.doi.org/10.1103/PhysRevLett.105.252303}{{\em Phys. Rev.
  Lett.} {\bfseries 105} (2010) 252303},
\href{http://arxiv.org/abs/1011.6182}{{\ttfamily arXiv:1011.6182 [hep-ex]}}.

\bibitem{Chatrchyan:2011sx}
{\bfseries CMS} Collaboration, S.~Chatrchyan {\em et al.}, ``{Observation and
  Studies of Jet Quenching in Pbpb Collisions at Nucleon-Nucleon Center-Of-Mass
  Energy = 2.76 TeV},''
  \href{http://dx.doi.org/10.1103/PhysRevC.84.024906}{{\em Phys. Rev.}
  {\bfseries C84} (2011) 024906},
\href{http://arxiv.org/abs/1102.1957}{{\ttfamily arXiv:1102.1957 [nucl-ex]}}.

\bibitem{Khachatryan:2016erx}
{\bfseries CMS} Collaboration, V.~Khachatryan {\em et al.}, ``{Correlations
  between jets and charged particles in PbPb and pp collisions at sqrt(s[NN])=
  2.76 TeV},''
\href{http://arxiv.org/abs/1601.00079}{{\ttfamily arXiv:1601.00079 [nucl-ex]}}.

\bibitem{Kutak:2018dim}
K.~Kutak, W.~P{\l}aczek, and R.~Straka, ``{Solutions of evolution equations for
  medium-induced QCD cascades},''
  \href{http://dx.doi.org/10.1140/epjc/s10052-019-6838-9}{{\em Eur. Phys. J.}
  {\bfseries C79} no.~4, (2019) 317},
\href{http://arxiv.org/abs/1811.06390}{{\ttfamily arXiv:1811.06390 [hep-ph]}}.

\bibitem{Kurkela:2015qoa}
A.~Kurkela and Y.~Zhu, ``{Isotropization and hydrodynamization in weakly
  coupled heavy-ion collisions},''
\href{http://arxiv.org/abs/1506.06647}{{\ttfamily arXiv:1506.06647 [hep-ph]}}.

\bibitem{Cao:2017zih}
{\bfseries JETSCAPE} Collaboration, S.~Cao {\em et al.}, ``{Multistage
  Monte-Carlo simulation of jet modification in a static medium},''
  \href{http://dx.doi.org/10.1103/PhysRevC.96.024909}{{\em Phys. Rev.}
  {\bfseries C96} no.~2, (2017) 024909},
\href{http://arxiv.org/abs/1705.00050}{{\ttfamily arXiv:1705.00050 [nucl-th]}}.

\bibitem{Putschke:2019yrg}
J.~H. Putschke {\em et al.}, ``{The JETSCAPE framework},''
\href{http://arxiv.org/abs/1903.07706}{{\ttfamily arXiv:1903.07706 [nucl-th]}}.

\bibitem{Schenke:2009gb}
B.~Schenke, C.~Gale, and S.~Jeon, ``{Martini: an Event Generator for
  Relativistic Heavy-Ion Collisions},''
  \href{http://dx.doi.org/10.1103/PhysRevC.80.054913}{{\em Phys. Rev.}
  {\bfseries C80} (2009) 054913},
\href{http://arxiv.org/abs/0909.2037}{{\ttfamily arXiv:0909.2037 [hep-ph]}}.

\bibitem{Zapp:2011ya}
K.~C. Zapp, J.~Stachel, and U.~A. Wiedemann, ``{A local Monte Carlo framework
  for coherent QCD parton energy loss},''
  \href{http://dx.doi.org/10.1007/JHEP07(2011)118}{{\em JHEP} {\bfseries 1107}
  (2011) 118},
\href{http://arxiv.org/abs/arXiv:1103.6252}{{\ttfamily arXiv:arXiv:1103.6252
  [hep-ph]}}.

\bibitem{Zapp:2012ak}
K.~C. Zapp, F.~Krauss, and U.~A. Wiedemann, ``{A perturbative framework for jet
  quenching},'' \href{http://dx.doi.org/10.1007/JHEP03(2013)080}{{\em JHEP}
  {\bfseries 03} (2013) 080},
\href{http://arxiv.org/abs/1212.1599}{{\ttfamily arXiv:1212.1599 [hep-ph]}}.

\end{thebibliography}\endgroup

\end{document}